\documentclass[aps,amsmath,twocolumn,amssymb,floatfix,showpacs,superscriptaddress,nofootinbib,longbibliography]{revtex4-1}

\usepackage{braket}
\usepackage[dvipsnames]{xcolor}
\usepackage{float}
\usepackage{subfigure}
\usepackage[colorlinks=true,linktoc=page,linkcolor=blue,citecolor=blue,urlcolor=blue]{hyperref}
\usepackage[toc,page]{appendix}

\newcommand{\vect}[1]{\boldsymbol{\mathrm{#1}}}
\mathchardef\mhyphen="2D 

\newcommand\bea{\begin{eqnarray}}
\newcommand\eea{\end{eqnarray}}
\newcommand\beq{\begin{equation}}  
\newcommand\eeq{\end{equation}}

\usepackage[normalem]{ulem}
\definecolor{lime}{HTML}{A6CE39}
\usepackage{sidecap,tikz}
\DeclareRobustCommand{\orcidicon}{\hspace{-1.0mm}
	\begin{tikzpicture}
		\draw[lime, fill=lime] (0.0,0.0) 
		circle [radius=0.15] 
		node[white] {{\fontfamily{qag}\selectfont \tiny \,ID}};
		\draw[white, fill=white] (-0.0525,0.095) 
		circle [radius=0.007];
	\end{tikzpicture}
	\hspace{-3.0mm}
}
\foreach \x in {A, ..., Z}{\expandafter\xdef\csname orcid\x\endcsname{\noexpand\href{https://orcid.org/\csname orcidauthor\x\endcsname}
		{\noexpand\orcidicon}}
}

\AtBeginDocument{%
	\newwrite\bibnotes
	\def\bibnotesext{Notes.bib}
	\immediate\openout\bibnotes=\jobname\bibnotesext
	\immediate\write\bibnotes{@CONTROL{REVTEX41Control}}
	\immediate\write\bibnotes{@CONTROL{%
			apsrev41Control,author="08",editor="1",pages="1",title="1",year="1"}}
	\if@filesw
	\immediate\write\@auxout{\string\citation{apsrev41Control}}%
	\fi
}%
\begin{document}


\title{ Finite temperature  dynamical quantum phase transition in a non-Hermitian system}

\author{Debashish Mondal\orcidA{}}
\email{debashish.m@iopb.res.in}
\affiliation{Institute of Physics, Sachivalaya Marg, Bhubaneswar-751005, India}
\affiliation{Homi Bhabha National Institute, Training School Complex, Anushakti Nagar, Mumbai 400094, India}

\author{Tanay Nag\orcidB{}}
\email{tanay.nag@physics.uu.se}
\affiliation{Department of Physics and Astronomy, Uppsala University, Box 516, 75120 Uppsala, Sweden}

\begin{abstract}
We investigate the interplay between the non-Hermiticity and finite temperature in the context of mixed state  
dynamical quantum phase transition (MSDQPT).  We consider a $p$-wave superconductor model, encompassing complex hopping and non-Hermiticity that can lead to  gapless phases in addition to gapped phases, to examine the MSDQPT and winding number via the intra-phase quench. We find that the MSDQPT is always present irrespective of the gap structure  of the underlying phase, however, the profile of Fisher zeros changes between the above phases. Such  occurrences of MSDQPT are in contrast to the zero-temperature case where DQPT does not  take place for the gapped phase. Surprisingly, the half-integer jumps in winding number at zero-temperature are washed away for finite temperature in the gapless phase. 
We study the evolution of the minimum time required by the system to experience MSDQPT with the inverse temperature such that gapped and gapless phases can be differentiated. Our study indicates that the minimum time shows monotonic (non-monotonic) behavior for the gapped (gapless) phase.

\end{abstract}

\maketitle

\section{Introduction}
\label{Sec:I}

The equilibrium phase transitions are associated with the nonanalyticities in the free-energy density that are marked by the zeros of the partition function namely Fisher zeros \cite{fisher1967theory,PhysRev.87.404,PhysRev.87.410}. In the non-equilibrium case, the dynamical free-energy density becomes singular at certain critical times in the  complex time plane where the dynamical quantum phase
transitions (DQPTs) takes place   \cite{Heyl13,PhysRevB.90.125106,PhysRevB.87.195104,Heyl15,Heyl_2018,Canovi_PRL, Utso_2017, Jafari_19feb,Vajna14,Dora2015,Somendra_PRL}.
When the time-evolved state under a sudden quench is 
orthogonal to the initial state,  the DQPTs occur
 corresponding to the vanishing Loschmidt amplitude
(LA) \cite{Nag16,Suzuki16,Sachdeva14,Nag12,PhysRevLett.96.140604,Cucchietti03,Jafari17}; 
this refers to the dynamical analogs of
equilibrium quantum phase transitions at quantum  critical points (QCPs). It has been shown that one can observe DQPT even without  quenching across the QCPs in sudden quench \cite{Uhrich20,Vajna14,Schmitt15,Halimeh17,Silva18,Halimeh20c,Hashizume2022,Lang_2018,Homrighausen_17,PhysRevB.106.045410,mishra2020disordered}. This list further extends to slow quenches \cite{Divakaran16,SS,PhysRevB.92.104306,Dutta17}, Floquet driving \cite{Zamani20,Jafari21,Jafari22,Naji22,Yang19,zhou2021floquet},
interacting  \cite{Halimeh17,Palmai_2015,Andraschko,Modak21}, bosonic \cite{Abdi19,Seyd_21,Stumper22} systems, time crystals~\cite{Kosior18,Kosior18b}, etc.

Thanks to the open
quantum systems \cite{Bergholtz19,Yang21}, and  quasiparticle systems with finite lifetime \cite{kozii2017non,Yoshida18,Shen18}, the Hermitian description of the problem expands to the non-Hermitian realm where exceptional points (EPs) 
appear instead of QCPs \cite{Bergholtz21,ghatak2019new,ashida2020non,Kawabata19,Shen18,Yoshida20,Yoshida19,Yoshida21}. As a result, the dynamical order parameter namely, winding number \cite{Vajna_15,Budich1}, characterizing the topological properties of the real-time dynamics can show intriguing jump profile as far as the non-unitary evolution of 
the non-Hermitian is concerned
\cite{Zhou1,Zhou_2021,Naji_PRA, Hamazaki_2021,Mondal22}.  The DQPTs are experimentally observed in trapped-ion  \cite{PhysRevLett.119.080501}, nuclear magnetic resonance \cite{Nie20}, optical lattice \cite{flaschner2018observation} systems. On the other hand, the non-Hermitian effects are practically realized in meta-materials such as  cold atom \cite{Gou20,li2019observation}, photonic \cite{Zeuner15,weimann2017topologically} and acoustic \cite{Weiwei18,Gao20} systems. Hence, it is important to study the interplay between DQPT and non-Hermiticity from the theoretical as well as experimental point of view.

The finite temperature extension of QPT has recently been examined in the context of LA \cite{Zanardi07,liang2019quantum,gu2010fidelity,Quan09}. Following the similar line of argument, the DQPT 
is investigated following an initial thermal distribution instead of a pure quantum state \cite{Utso_2017,Lang18,Lang18b,Mera18,Bandyopadhyay18,Hou20,Hou22}. This brings in the concept of density matrix,  characterised by an inverse temperature, leading to the mixed state DQPT (MSDQPT) where the 
quantum coherence is lost. For the open quantum systems in contact with the thermal bath can potentially lead to such MSDQPT \cite{Lang18c,zhou2021non,Sedlmayr18}.   
Given the fact that the DQPT persists in the finite temperature \cite{Utso_2017,Zhou_2021} and it can show anomaly in the non-Hermitian system \cite{Mondal22}, we here pose the following intriguing questions to understand the interplay between the non-Hermiticity and finite temperature: 
Can MSDQPT appear (disappear) when the DQPT in the 
the underlying non-Hermitian system at zero temperature
is absent (present)? Can we differentiate various non-Hermitian  phases by examining the MSDQPT? 
How do we understand the topology in the real-time dynamics of the non-Hermitian MSDQPT?

In this paper, we generalize the  framework  of DQPT and  winding number for  non-Hermitian finite-temperature cases such that the Hermitian and infinite temperature limits can be successfully extracted. Considering a one-dimensional (1D) $p$-wave superconductor with complex hopping and non-Hermiticity (see Fig.~\ref{Phase_Diagram}), we find sudden quench within the gapped phase exhibits MSDQPT unlike to  the  zero-temperature case \cite{Mondal22} (see Fig.~\ref{gapped}). Interestingly,  we only observe integer jumps as a signature of  MSDQPT for the 
sudden quenches within the gapless phases  contrasting to the zero-temperature profile of DQPT as observed previously (see Figs.  \ref{horizontal_gapless}, \ref{fig:non_hermi_gapless}). However, the winding number shows non-monotonic behavior in one of the gapless phases. The  threshold time, referred to as minimum critical time $t_{\rm cm}$, above which MSDQPT starts appearing can be different from the zero- as well as infinite-temperature limit (see Fig. \ref{temp_dep}). We can distinguish various gapless and the gapped phases by investigating the behavior of  $t_{\rm cm}$ with temperature. The above studies on lossy superconductivity are further extended to lossy chemical potential case for completeness.  Our study thus indicates that the temperature can non-trivially modify the dynamics of EPs as evident from the emergence of MSDQPT.

The structure of this paper is the  following. We present the framework of MSDQPT in Sec.~\ref{Sec:II} for finite temperature and non-Hermiticity. Next, we demonstrate the model under consideration in Sec.~\ref{Sec:III}. We examine the MSDQPT results for gapped phase in Sec.~\ref{topo_gapped_region} and for gapless phases in Secs.~\ref{horizontal_gapless_region} and \ref{vertical_gapless_region}. We differentiate among these phases with respect to their temperature profile in Sec.~\ref{temperature_dependence}. We provide plausible explanation behind our findings in Sec.~\ref{comments}. Finally, we conclude in Sec.~\ref{Sec:V}.

\section{MSDQPT framework}\label{Sec:II}

Let us consider a $2$-level system, described by Hamiltonian $H_{k}= \vect{h}_{k}\cdot \vect{\sigma}=h_{k}\hat{h}_{k}.\vec{\sigma}$, that is thermally attached to a heat bath at temperature $T=1/\beta$; here $\vect{h}_{k}=\{h_{k}^{x},h_{k}^{y},h_{k}^{z}\}$, and $\vect{\sigma}=\{\sigma_{x},\sigma_{y},\sigma_{z}\}$. Note that $H_{k}$ can be considered to be non-Hermitian without loss of generality where $h_{k}^{i}$ can be complex with $i=x,y,z$. 
The associated density matrix takes the form $\rho_{k}= \exp(- \beta H_{k})/ \operatorname{Tr} \left[\exp(- \beta H_{k})\right]=\big(\sigma_{0}- m (\hat{h}_{k}\cdot\vec{\sigma}\big)/2$,  where $m={\rm tanh}(\beta h_{k})$ (see Appendix.~ \ref{Sec:Initial_density_matrix} for more details). We start with this finite-temperature initial mixed state at time $t=0$ i.e., $\rho_{k}(0)$ corresponding to $H_{k,i}$ and suddenly quench to the final Hamiltonian $H_{k,f}$ such that 
the LA at a later time $t$ is given by~\cite{Utso_2017}  
$g_{k}(t)= \operatorname{Tr} \left[\rho_{k}(0) U_{k} (t) \right]=\cos(h_{k,f}t) - i ~ \sin(h_{k,f}t) B_{k}$
with $U_{k}(t)=e^{-i H_{k,f}t}$ and $B_{k}=-m (\vec{h}_{k,i}.\vec{h}_{k,f}/ h_{k,i} h_{k,f} )$ (see Appendix.~\ref{Sec:Loschmidt amplitude} for more details). 
\textcolor{black}{The dynamical analog of free energy is called rate function which is given by the logarithm of LA}~\cite{Budich1}
\begin{equation}
I(t)=-\frac{1}{2 \pi} \int_{\rm BZ} \ln\left(|g_{k}(t)|^{2}\right) \label{rate}.
\end{equation}

The nonanalyticities in the rate function, given by  $g_{k}(t)=0$, causes the Fisher zeros to appear in the complex time plane (see Appendix.~\ref{Sec:Fisher} for more details)
\begin{eqnarray}
	&& z_{n,k}=i \left(n+ \frac{1}{2}\right) \frac{\pi}{ h_{k,f}}  + \frac{1}{ h_{k,f}} \tanh^{-1}(B_{k}), \label{Fisher2}
\end{eqnarray}
where, $z_{n,k}=i t$, and $n \in Z$. \textcolor{black}{Note that the zeros of partition function is referred to as Fisher zeros.} As a result, MSDQPT occurs at the momentum $k=k_{c}$ and critical time $t_{c}=-i z_{n,k_{c}}$ where  $\operatorname{Re}[z_{n,k_{c}}]=0$ leading to
\begin{eqnarray}
&&\pi \left(n + 1/2\right) \operatorname{Im}[h_{k_{c},f}]  + \operatorname{Re}[h_{k_{c},f}] \operatorname{Re}[C_{k_{c}}]  \nonumber\\
&+& \operatorname{Im}[h_{k_{c},f}] \operatorname{Im}[C_{k_{c}}] =0  \label{kc}
\end{eqnarray}
and 
\begin{eqnarray}
t_{c}&=&\pi \left(n+\frac{1}{2}\right) \frac{\operatorname{Re}[h_{k_{c},f}]}{|h_{k_{c},f}|^{2}}  \nonumber \\
&+& \frac{\operatorname{Re}[h_{k_{c},f}] \operatorname{Im}[C_{k_{c}}] -\operatorname{Im}[h_{k_{c},f}] \operatorname{Re}[C_{k_{c}}]}{|h_{k_{c},f}|^{2}} \label{tc}
\end{eqnarray}
with $C_{k}= \tanh^{-1}(B_{k})$ (see Appendices ~\ref{Sec:kc} and \ref{Sec:tc} for more details). To be precise, MSDQPT occurs  when $z_n,k$ crosses the positive side of the imaginary axis such that positive $t_c$'s can only be the meaningful solutions of Eq.~(\ref{tc}).

On the other hand, the dynamical phase is given by (see Appendix.~\ref{Sec:dynamical_phase} for more details)
\begin{equation}
	\Phi_{k}^{\rm dyn}(t) = - \int_{0}^{t} dt^{\prime} \operatorname{Re}\left[h_{k,f} \frac{\tanh\left(2~ \operatorname{Im}[h_{k,f}]~ t^{\prime}\right)-m}{1-m~\tanh\left(2~ \operatorname{Im}[h_{k,f}]~ t^{\prime}\right)} \right].
	\label{dynamical}
\end{equation}
The winding number, \textcolor{black}{capturing the dynamical order parameter \cite{Erik00},} appear to be 
\begin{equation}
	\nu(t)= \oint_{\rm BZ}~dk~ \frac{\partial \Phi_{k}^{G}(t)}{\partial k} \label{wind},
\end{equation}
with $\Phi_{k}^{G}(t)=\Phi_{k}^{\rm tot}(t)-\Phi_{k}^{\rm dyn}(t)\label{geometrical main}$ and $\Phi_{k}^{\rm tot}(t)= - \ln \left(\frac{g_{k}(t)}{|g_{k}(t)|}\right)$. \textcolor{black}{ Hence the geometric phase is the net phase acquired by a non-equillibrium quantum system other than the dynamical phase.}\\

\textcolor{black}{The physical picture of MSDQPT refers to the quantum dynamics of a mixed state density matrix. The MSDQPT essentially captures the interference between the time evolved and initial density matrices. To be precise, where there is a complete destructive interference in real time i.e., $g_k(t)=0$, the rate function exhibits a singular behavior.   Using the concept of parallel transport it has been shown that noncyclic and unitary quantum evolutions of a pure quantum state are related to that of a mixed state \cite{Erik00}. Therefore,  the geometric phases for a mixed state can be thoroughly investigated with real time following the analysis of MSDQPT. The non-Hermiticity can effectively mimic the effect of  external  bath attached to a quantum system, and/or  interaction in the quantum system. In the present context, our study qualitatively tracks the evolution of geometric phase, associated with a mixed state, in an interacting system by considering a non-Hermitian system.}

\section{Model}\label{Sec:III}
We consider the non-Hermitian analog of a 1D p-wave superconductor with complex hopping as follows $H(\gamma_1,\gamma_2,\phi)=\Sigma_{k} \psi_{k} {\cal H} _{k}(\gamma_1,\gamma_2,\phi) \psi_{k}^{\dagger} $ with Nambu basis $\psi_k= (c_k, c^{\dagger}_{-k})$ ~\cite{kitaev2001unpaired,DeGo1,Manisha1,Rajak1,Mondal22},
\begin{eqnarray}
&&{\cal H} _{k}(\gamma_1,\gamma_2,\phi)= 2 w_{0} \sin \phi \sin k \hspace{1 mm} I + \left(2 \Delta \sin k + \frac{i \gamma_2}{2}\right) \sigma_{y} \nonumber \\
&&- \left(2 w_{0} \cos \phi \cos k +\mu + \frac{i \gamma_1}{2}\right) \sigma_{z} = \vect{h}_{k}.\vect{\sigma}, 
\label{Model_Hamiltonian}
\end{eqnarray}
where $w_{0}$, $\phi \in [0,\pi/2]$, $\Delta$, $\mu$ are the nearest neighbour hopping amplitude, phase of the hopping, superconducting gap, and chemical potential respectively.
The non-Hermiticity in $h_k^y$ for the $p$-wave superconductor gap function, referred to as the lossy superconductivity, might be caused by the spatially separated pairing processes \cite{shi2022topological}. On the other hand, the non-Hermiticity in $h_k^z$ can be originated by the non-reciprocal hopping \cite{li2020critical} or/and loss and gain in the chemical potential\cite{Yuce16}.

\begin{figure}[H]
	\centering
	\subfigure{\includegraphics[width=0.48\textwidth]{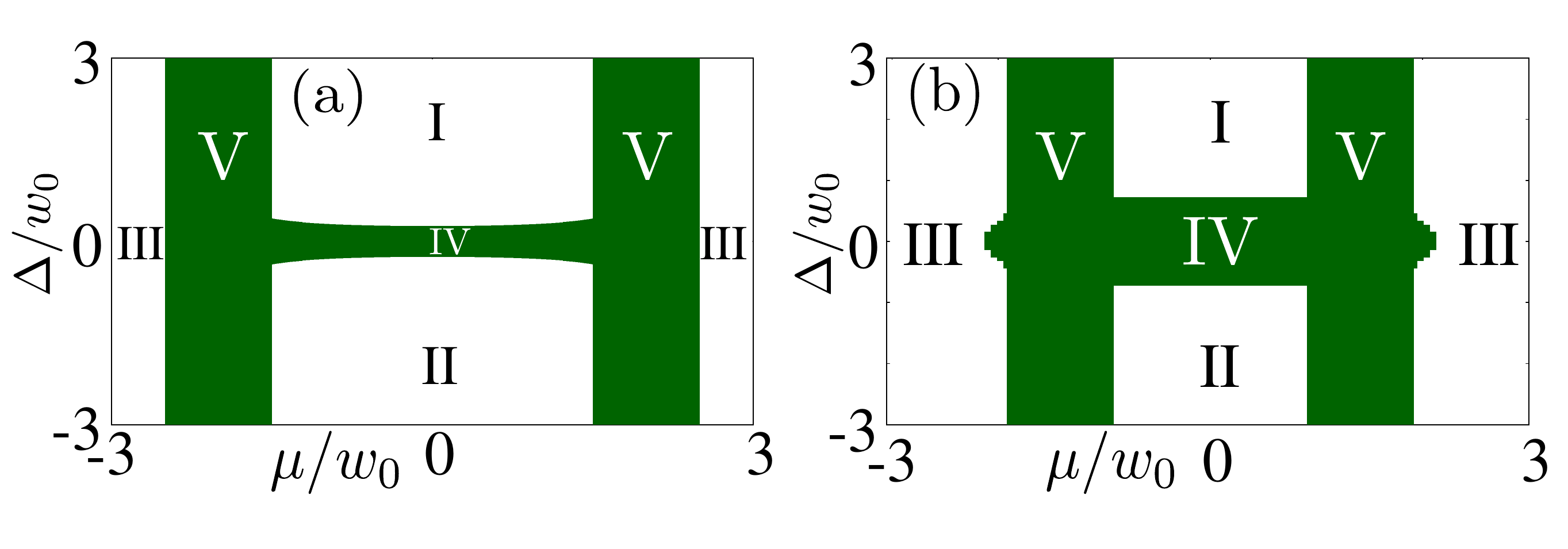}}
	\caption{ The phase diagram of the model Hamiltonian $H(0,1,0)$ [$H(0,1,\pi/4)$], given by Eq.~(\ref{Model_Hamiltonian}), is shown in (a) [(b)]. 
    The white regions I, II and III correspond to
    the gapped phase while green regions IV and V denote the gapless phases. With increasing $\phi$, the  phase IV increases in size while non-Hermiticity solely determines the width of phase V.
}
	\label{Phase_Diagram}
\end{figure}

The Hamiltonian Eq.~(\ref{Model_Hamiltonian}) becomes gapless for critical momentum $k_{*}$
when the real part of energies satisfies the following condition
\begin{eqnarray}
&&(2 w_{0} \cos \phi \cos k_{*} +\mu)^{2} - \frac{\gamma_1^{2}}{4} + 4 \Delta^{2} \sin^{2}k_{*}- \frac{\gamma_2^{2}}{4}\nonumber\\
&&=4 w_{0}^{2} \sin^{2} \phi \sin^{2} k_{*}
\label{gapless_condition}.
\end{eqnarray}
This allows us to chart out the phases diagram of the model Hamiltonian as shown in Fig.~\ref{Phase_Diagram}. For $\gamma_{1}=0$, and $\gamma_{2}\ne 0$, the gapless phase IV is bounded by horizontal lines $\Delta =\pm \sqrt{(4w^2_{0} \sin^2 \phi + \gamma_2^2/4- \mu^2 \sin^2\phi)/(4-\mu^2/w^2)}$ between the neighbouring gapless phases V that are  bounded by $[- 2 w_{0} \cos\phi - {\gamma_2}/{2}, - 2 w_{0} \cos\phi + {\gamma_2}/{2}]$ ($[2 w_{0} \cos\phi - {\gamma_2}/{2}, 2 w_{0} \cos\phi + {\gamma_2}/{2}]$) in the left (right) side. Notice that gapless phase IV is primarily caused by the phase of the complex hopping while  the non-Hermiticity term $\gamma_2$ is solely responsible for the other gapless phase V. 
The phases I and II are topological while III is trivial.  Notice that for the calculation of DQPT, we discard 
the identity term in Eq.~(\ref{Model_Hamiltonian})  as it does not alter non-equilibrium evolution.  

\section{Result}\label{Sec:IV}
We focus on the gapped and gapless phases in the above model in presence of lossy superconductivity only. The MSDQPT is studied for the intra-phase quench. For completeness,  we briefly discuss the fate of MSDQPT  for other inter-phase quench. We also discuss the MSDQPT in the above model with non-Hermitian chemical potential.   \textcolor{black}{The Hermitian counterpart of MSDQPT are demonstrated in Appendix. \ref{sec:hermitian case}.}

\subsection{Quench within gapped phase I}\label{topo_gapped_region}
\begin{figure}[H]
	\centering
	\subfigure{\includegraphics[width=0.48\textwidth]{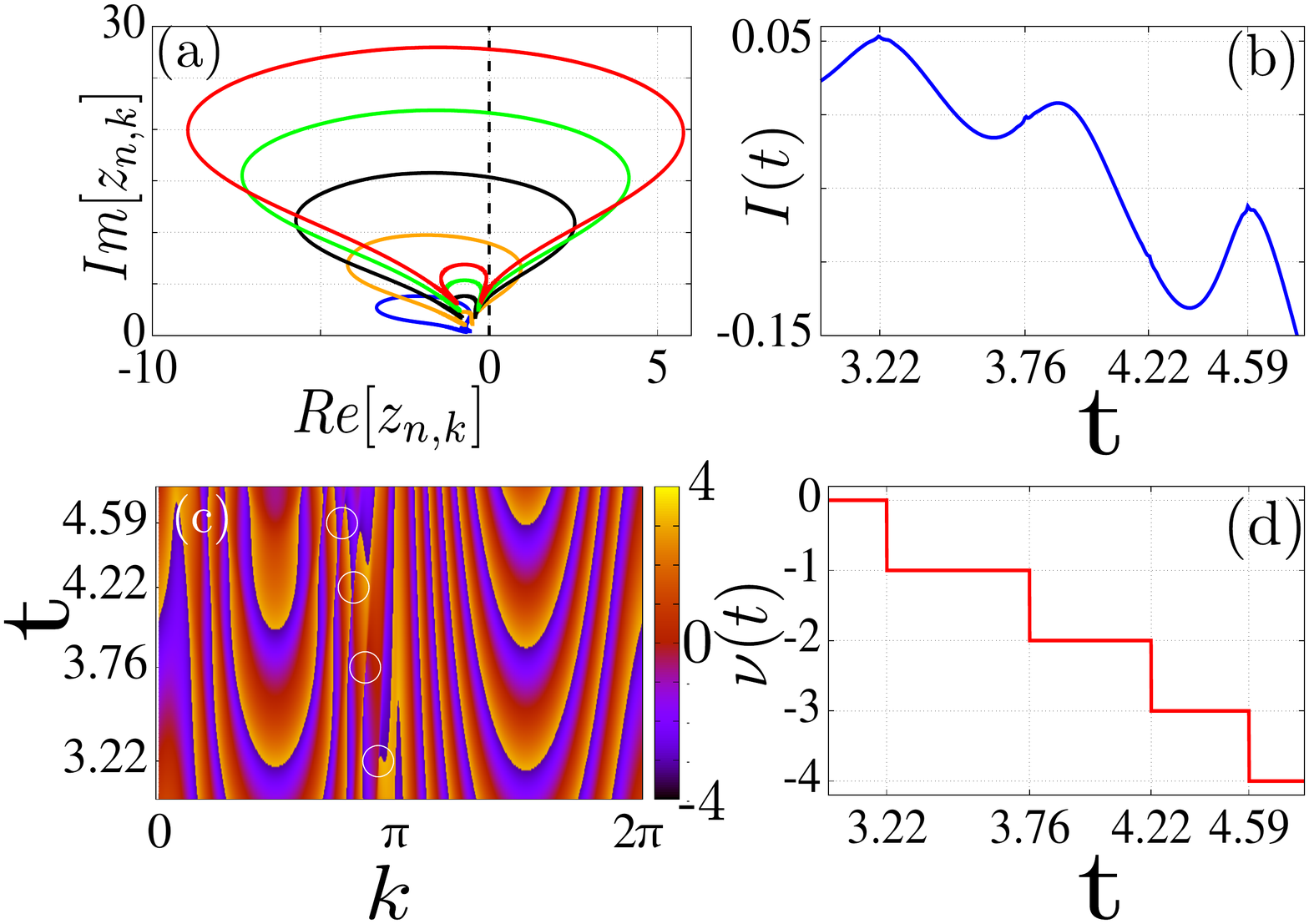}}
	\caption{ We demonstrate (a) the lines of Fisher zeros $z_{n,k}$ with $n=0$ (blue), $\cdots$, $n=4$ (red), as computed from Eq.~(\ref{Fisher2}), (b) nonanalytic nature in the rate function $I(t)$ as obtained from Eq.~(\ref{rate}), (c) the geometric phase  $\Phi_{k}^G(t)$ with time and momenta, and (d) the time evolution of winding number $\nu(t)$, quantified by Eq.~(\ref{wind}) for the case discussed in sec.~\ref{topo_gapped_region}.    
 The Fisher zeros $z_{n,k}$ cross imaginary axis twice leading to the critical times $t_c \approx 3.22,3.76,4.22,4.59,\cdots$ where $\nu(t)$ shows integer jumps. The white circles in (c) highlights the abrupt changes in the geometric phase.  
	The parameters are taken to be ($\mu_{i}$, $\mu_{f}$, $\Delta_{i}$, $\Delta_{f})=(0.1, 0.7, 2.2, 2.2)$. We consider $\beta=1$, $w_{0}=1$, $\phi=\pi/4$, $\gamma_1=0$, and $\gamma_2=1$ for Figs.~\ref{gapped}, \ref{horizontal_gapless}, and \ref{fig:non_hermi_gapless}. 
	}
	\label{gapped}
\end{figure}

We first examine the MSDQPT following the quench within the phase I as shown in  Fig.~\ref{Phase_Diagram} (b). For finite temperature ($\beta=1$), Fisher zeros profile, rate function, geometric phase, and winding number  are depicted in Fig.~\ref{gapped} (a), (b), (c), and (d) referring to the fact that MSDQPT takes place. We notice that $z_{n,k}$ always cross the imaginary axis  except for $n=n_{\rm min}=0$. What we find is that $n_{\rm min}$ increases from zero as $\beta$ increases, but yet indicating to the emergence of MSDQPT for any temperature. Interestingly, the nonanalyticities are not visible macroscopically, however, there exists the singular micro-structures at critical time $t=t_c \approx 3.22,3.76,4.22,4.59,\cdots$  over the oscillating profile. We find abrupt changes in geometric phase $\Phi_{k}^G(t)$, marked by white circles in Fig.~\ref{Phase_Diagram}(c), around the above values of $t$ for $k$ being close to $\pi$. The profile of $\Phi_{k}^G(t)$ looks quite different as compared to the non-Hermitian zero temperature case. The winding number shows step-like jumps at the above critical times. Since $z_{n,k}$ encloses a close loop by crossing the imaginary axis twice, winding number is expected to exhibit both increase and decrease with time. However, we only find decrease in winding number within $3<t<5$ where $z_{n,k}$ crosses the real axis once.


\subsection{Quench within horizontal gapless phase IV}\label{horizontal_gapless_region}

\begin{figure}[H]
	\centering
	\subfigure{\includegraphics[width=0.48\textwidth]{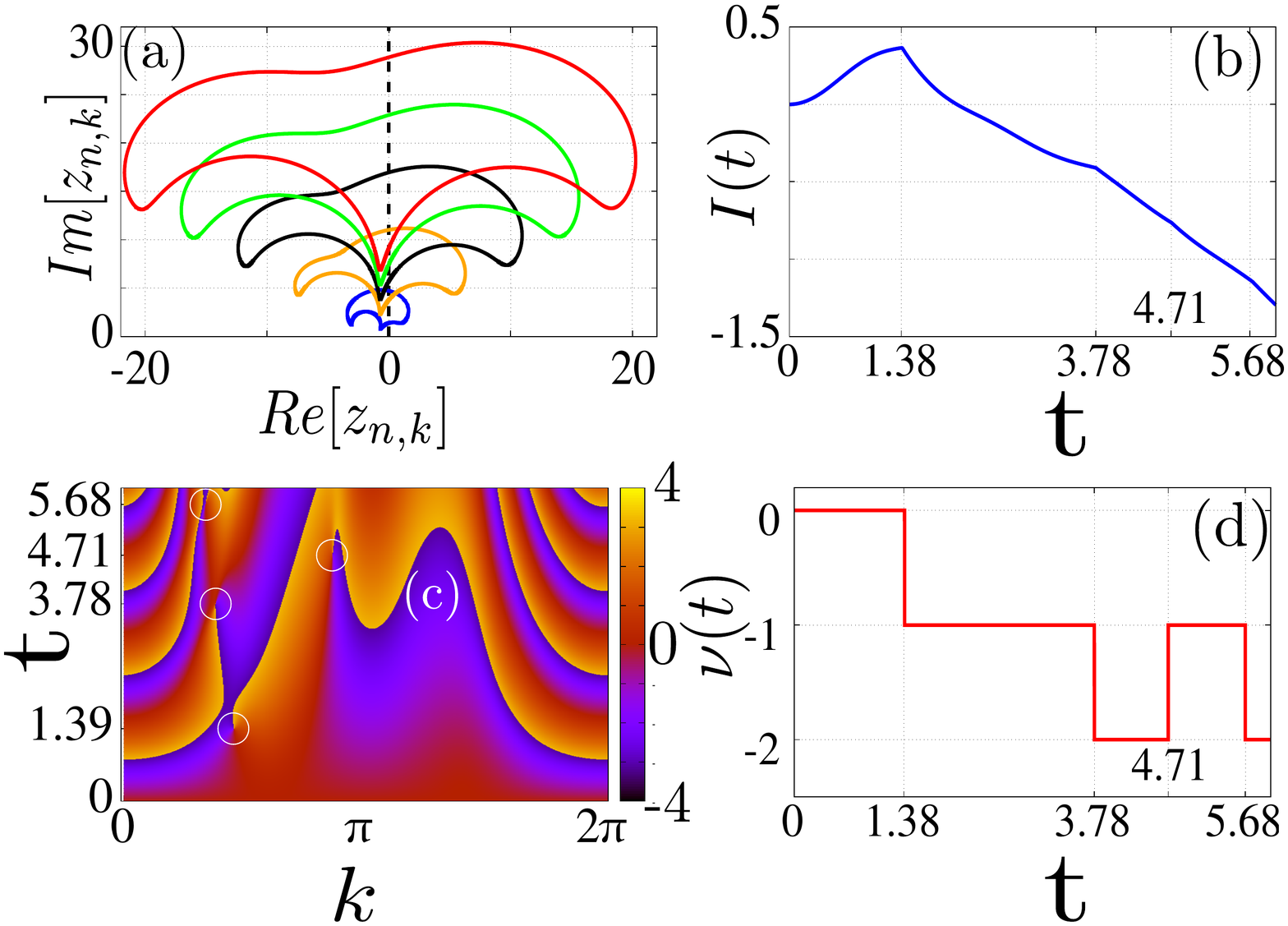}}
	\caption{We repeat Fig.~\ref{gapped} for the case discussed in sec.~\ref{horizontal_gapless_region}. The lines of Fisher zeros, $z_{n,k}$ with $n=0$ (blue), $\cdots$, $n=4$ (red)
  cross imaginary axis twice leading to the nonanalyticities in rate function  at critical times, $t_c\approx 1.38, 3.78, 4.71, 5.68,\cdots$ around which  winding number exhibits integer jumps. The parameters are taken to be ($\mu_{i}$, $\mu_{f}$, $\Delta_{i}$, $\Delta_{f})=(0.1, 0.7, 0.2, 0.2)$.
	}
	\label{horizontal_gapless}
\end{figure}

We now focus on the occurrences of  MSDQPT following the quench inside the gapless phase IV as shown in Fig.~\ref{Phase_Diagram} (b). Figure ~\ref{horizontal_gapless}(a) depicts the lines of Fisher zeroes $z_{n,k}$, crossing imaginary axis twice for all values of $n$. This is in contrast to the previous situation for the quench inside the gapped phase I, presented in Fig.~\ref{gapped}, where MSDQPT only takes place for $n>n_{\rm min}$. The non-analyticities [discontinuous change] in the rate function $I(t)$ [geometric phase] are captured at the critical times $t_c\approx 1.38, 3.78, 4.71, 5.68,\cdots$ in Fig.~\ref{horizontal_gapless}(b) [(c)]. The non-analyticities in the rate function is more clearly visible in the present case as compared to the previous one in Fig.~\ref{gapped} (b). 
The winding number  shows non-monotonic jump profile with time that is caused by the double crossing of imaginary axis by $z_{n,k}$. The important point to note here is that these jumps are always of unit magnitudes unlike the previous zero-temperature case \cite{Mondal22}. The unit jumps are a consequence of the continuous crossing of   Fisher zeros through the imaginary axis that we find in the present case.



\subsection{Quench within vertical gapless phase V}\label{vertical_gapless_region}
\begin{figure}[H]
	\centering
	\subfigure{\includegraphics[width=0.48\textwidth]{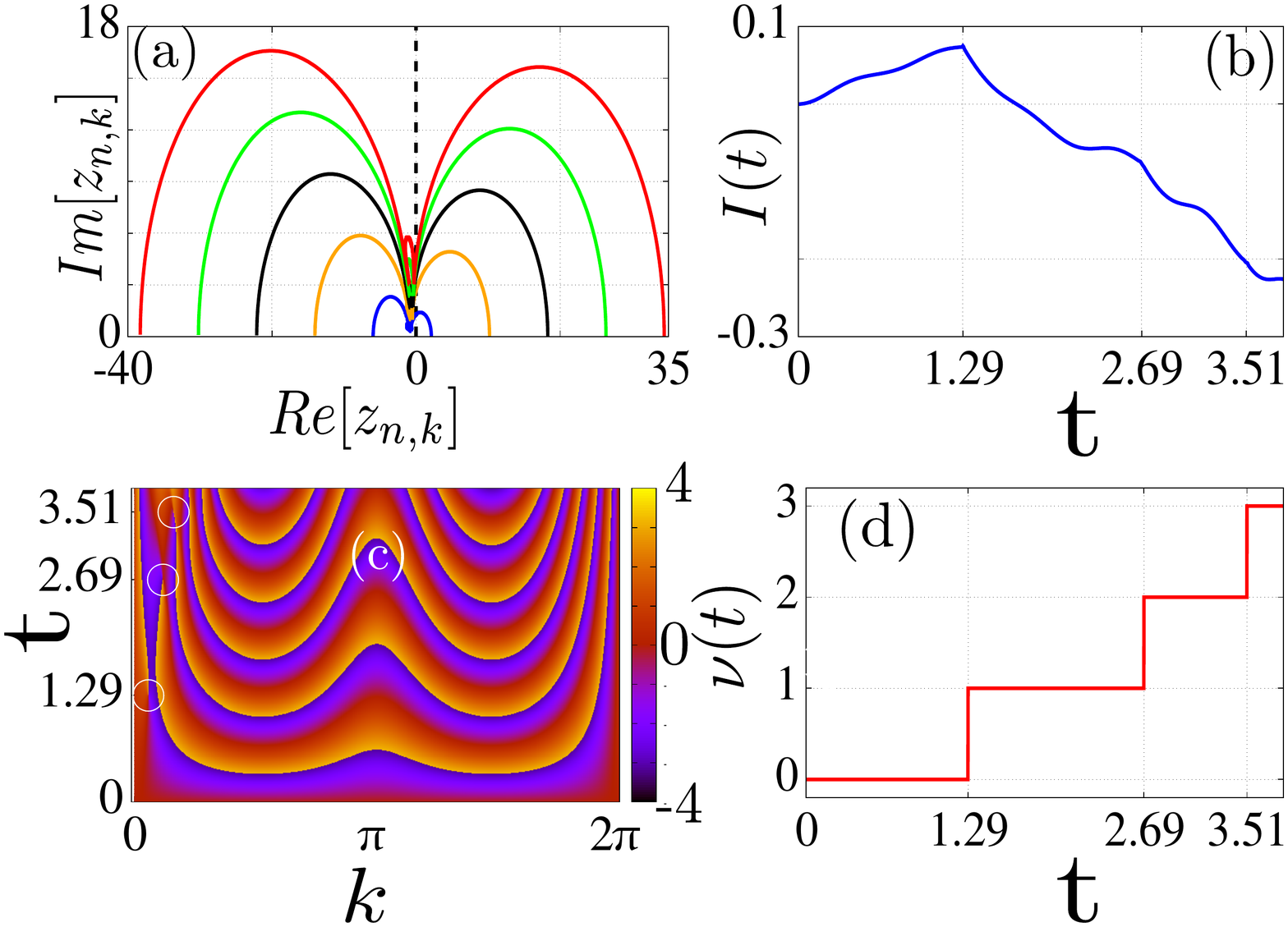}}
	\caption{We repeat Fig.~\ref{gapped} for the case discussed in sec.~\ref{vertical_gapless_region}. The lines of Fisher zeros, $z_{n,k}$ with $n=0$ (blue), $\cdots$, $n=4$ (red) cross imaginary axis once leading to the nonanalyticities in rate function  at critical times, $t_c\approx 1.29, 2.69,3.51,\cdots$ around which  winding number exhibits monotonic integer jumps. All the parameters are taken to be ($\mu_{i}$, $\mu_{f}$, $\Delta_{i}$, $\Delta_{f})=(-1.7, -1.1, 2.2, 2.2)$.
	}
	\label{fig:non_hermi_gapless}
\end{figure}

We now demonstrate the MSDQPT following the quench within the vertical  gapless phase V as presented in  Fig.~\ref{Phase_Diagram} (b). Unlike to the previous cases, we here find that $z_{n,k}$ crosses the imaginary axis once (see Fig.~\ref{fig:non_hermi_gapless}(a)). The nonanalyticities in the rate function are captured with time in   Fig.~\ref{fig:non_hermi_gapless} (b). The geometric phase, shown in Fig.~\ref{fig:non_hermi_gapless} (c), exhibits a similar profile as compared to that of the gapped phase I. The oscillatory profile of geometric phase is a common characteristic of finite-temperature case.  The winding number shows monotonic increase with time due to the single crossing of imaginary axis by $z_{n,k}$. 
The unit jumps in MSDQPT, associated with finite temperature, are in contrast to the half-integer jumps of DQPT corresponding to the zero temperature case \cite{Mondal22}.  We additionally check for $\phi=0$ case  where we also do not find the  half-integer jumps in the  winding number (not shown here).

\subsection{Distinct temperature profile of MSDQPTs for phases I, IV and V }\label{temperature_dependence}
As illustrated above, the MSDQPT takes place in all three phases irrespective of their gap structure. To be precise,  for a given quench amplitude and temperature, there exist multiple critical times $t_c$'s. We here focus on the evolution of the  minimum critical time, referred to as $t_{\rm cm}$, that captures the minimum time taken by the system to  witness the first occurrence of MSDQPT. We numerically study the temperature dependence of $t_{\rm cm}$ such that the phase I, IV and V can be distinguished.

\begin{figure}[H]
	\centering
	\subfigure{\includegraphics[width=0.48\textwidth]{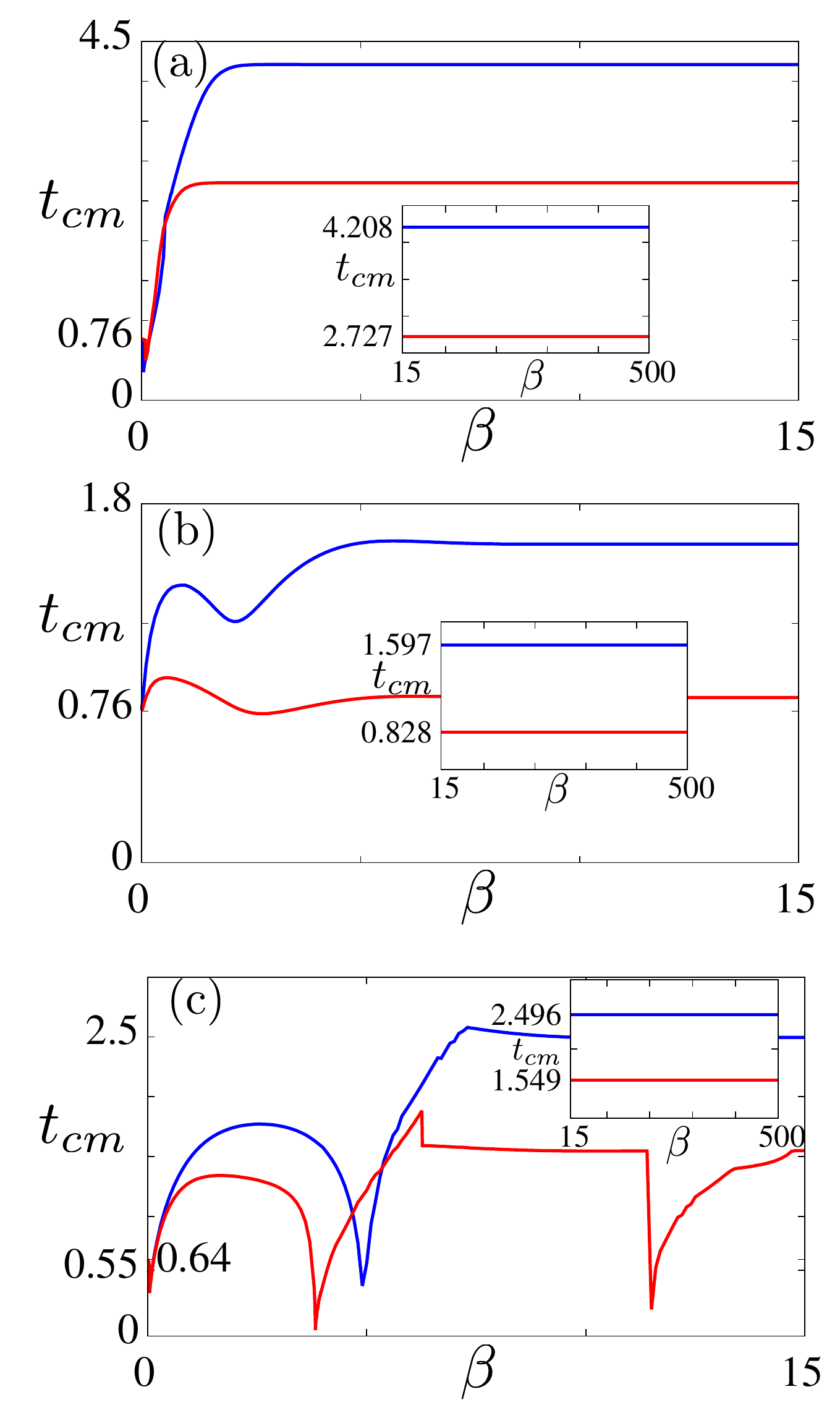}}
	\caption{We plot the minimum time $t_{\rm cm}$, required for MSDQPT to take place, as a function of inverse temperature $\beta$ for quench within phases I, IV, and V in (a), (b), and (c), respectively. For (a), the red [blue] line corresponds to  quench path ($\mu_{i}$, $\mu_{f}$, $\Delta_{i}$, $\Delta_{f})=(-0.7, 0.7, 2.2, 2.2)$ [$(0.1, 0.7, 2.2, 2.2)$]. For (b), the red [blue] line corresponds to  quench path ($\mu_{i}$, $\mu_{f}$, $\Delta_{i}$, $\Delta_{f})=(-0.7, 0.7, 0.2, 0.2)$ [$(0.1, 0.7, 0.2, 0.2)$]. For (c), the red [blue] line corresponds to  quench path ($\mu_{i}$, $\mu_{f}$, $\Delta_{i}$, $\Delta_{f})=(-1.7, -1.1, 2.2, 2.2)$ [$(-1.8, -1.5, 2.2, 2.2)$].  The non-monotonic behavior in (b) and (c) is in complete contrast to that of (a).
    We show the saturation of $t_{\rm cm}$ over a wide range of $\beta \gg1$ as the insets.   The insets show the saturation profile of $t_{\rm cm}$ for $\beta \gg 1$. We consider $w_{0}=1$, $\phi=\pi/4$, $\gamma_1=0$, and $\gamma_2=1$. 
	}
	\label{temp_dep}
\end{figure}

Figure \ref{temp_dep} (a), (b) and (c) depict the temperature profile of  $t_{\rm cm}$ following the large (small) intra-phase quench amplitude, denoted by red (blue) lines, within regions I, IV and V, respectively. The infinite temperature $\beta \to 0$ value of $t_{\rm cm}$ is found to be insensitive to the quench amplitude.
This suggests that MSDQPT is anyway present in the infinite temperature case as long as the quench amplitude is finite. Connecting with the Fig.~\ref{gapped} (a), one can find that $n_{\rm min }$ increases for smaller quench amplitude.
For phase V, MSDQPT takes place early as compared to the phase I and IV in the infinite temperature limit. On the other hand,  
$t_{\rm cm}$ saturates with increasing $\beta$ above a certain value. We now find that 
the zero temperature $\beta \to \infty$ value of $t_{\rm cm}$ strongly depends on the quench amplitude (see insets of Figs. \ref{temp_dep} (a), (b) and (c)). To be precise, MSDQPT appears quickly with time in the zero temperature limit for larger quench amplitude. Interestingly, for the present case, MSDQPT occurs more quickly with time for gapless phases IV and V as compared to the gapped phase I in the limit $\beta \to \infty$.


For the intermediate temperature with finite value of $\beta$, we find non-monotonic behavior of  $t_{\rm cm}$ only for the gapless phases IV and V. In the case of the gapped phase I as shown in Fig.~\ref{temp_dep} (a),  $t_{\rm cm}$ increases almost monotonically from $\beta \to 0$. This is followed by a saturation for  $\beta \to \infty$. However, there exist a small dip around $\beta \approx 0$. The non-Hermiticity induced vertical gapless phase V shows sharp dip  for intermediate values of $\beta$ while a broadened dip is noticed for complex hopping-induced horizontal gapless phase IV (Figs.~\ref{temp_dep} (b) and (c)). Such a dip in   $t_{\rm cm}$ for gapless phases refers to the fact that MSDQPT can even appear early with time as compared to the infinite temperature case. This is in contrast to the behavior of gapped phase where MSDQPT can only appear at later time for any finite temperature as compared to the infinite temperature limit. The details of dip structure of  $t_{\rm cm}$ is expected to depend on the quench amplitudes for a given gapless phase.  
The location of such dips might depend on the details of gapless phase whether it is caused by non-Hermiticity or phase of complex hopping for identical quench amplitude. For example, the relative  locations of such dips on the $\beta$ axis are altered between the phases IV and V.   
However, we emphasize that the detailed future analysis of $t_{\rm cm}$ vs. $\beta$ behavior is yet to be required to comment on their phase-dependent distinct characteristics.

\textcolor{black}{The spectral gap profiles in the phases I, IV and V differ from each other significantly. The above analysis on MSDQPT by varying $\beta$ is able to capture the interplay between the temperature and the gap profile of the eigenstates associated with these phases. In the gapless regions IV and V with finite $T$, the minimum time $t_{\rm cm}$ at which the first   destructive interference takes place is relatively less 
than that for  gapped region I. This can be intimately connected to the  distinct spectral profiles of these phases.   }

\subsection{Discussion}
\label{comments}

The estimation of critical momenta, obtained from Eq.~(\ref{kc}), is hard as far as a closed form is concerned. 
In the non-Hermitian case, one can use non-Bloch form of momentum  to explain the topological properties \cite{YaoPRL2018,Kunst18,helbig2020generalized,Kawabata20b,Ghosh22b}. We can use the same non-Bloch notion to qualitatively predict the critical momenta $k_c$ using the DQPT framework for Hermitian system. The above effective approach is unable to  predict the exact values of the critical momenta, however, one obtains a closed form expression of $k_c$ (see Appendix. \ref{non_bloch}). Importantly, it can explain the emergence of multiple $k_c$'s which is consistently visible for non-Hermitian cases \cite{Mondal22}. The multi-valued nature of $k_c$ also exists for Hermitian system. However, this nature  persists more strongly as the non-Hermiticity allows for additional solutions for $k_c$ as evident from Eq.~(\ref{kc}).


Having demonstrated the results for quenching inside the phases I, IV and V extensively, we comment that MSDQPT is also present for quench inside phases II as well as III. Therefore, all the intra-phase quench leads to MSDQPT in finite temperature when the non-Hermiticity is associated with the superconductivity. By contrast,  the DQPT is not always present for all of the above cases in  the zero temperature \cite{Mondal22}.  
On the other hand, for any inter-phase quench, the MSDQPT is present that we do not show here explicitly. The emergence of DQPT is not intrinsically connected with the crossing of QCP and EP for Hermitian and non-Hermitian systems, respectively \cite{Mondal22,Zhou1} at zero temperature. As shown above, DQPT always present as long as the critical momenta $k_c$ exist. For finite temperature case, 
obtaining the critical momenta is even more probable due to the presence of thermal density matrix instead of the pure quantum state.   The
finite temperature broadening of the  quantum energy levels yields further scope to interact with neighbouring energy levels in addition to the non-Hermiticity. This might lead to the rapid variation of  geometric phases for  different momentum modes  at a given time.   As a result, one can expect to see MSDQPT for all the cases.  
However, we find an exception when we study the intra-phase III quench in presence of non-Hermitian chemical potential only at finite temperature (see Appendix. \ref{nh_chemical_potential}).  We emphasize that the half-integer jumps in winding number for zero temperature are rounded off by the finite temperature where the Fisher zeros do not show any discontinuity over the imaginary axis.

\textcolor{black}{The temperature profiles of MSDQPT in the different phases clearly signal the distinct characteristics of the  post quench evolution of a mixed quantum state (see Fig.~\ref{temp_dep}). The unitary evolution of a pure quantum state results in revival with time referring to the fact that there exist  quantum interferences  between the initial and time evolved state \cite{Rajak14,Rajak17}.  The non-analytic divergences in the rate functions are connected with the  complete destructive interferences. The   absence of such interferences results in  disappearance of the DQPT. 
The evolution of mixed state  LA shows qualitatively similar features as far as the constructive and  destructive interferences are concerned while compared with the  pure state LA.  However,   temperature smoothens the interference patterns and the destructive interferences sustain \cite{Mera18,Cappellaro11,sur2019loschmidt}. In our case, MSDQPT always present means that complete destructive interferences are bound to happen post quench out of the mixed quantum state. For non-Hermitian case,  due to the presence of EPs
 the destructive interferences are more probable \cite{Bergholtz21}. As a result, the non-unitary evolution of mixed state at finite temperature deviates from the initial configuration more often than the zero temperature unitary counterpart. This can be connected to the
thermal behavior of interacting systems where the driven systems traverses the entire phase space. Our findings on the occurence  of MSDQPT essentially refer to the fact that  the non-Hermitian system might not localize in the phase space as opposed to integrable Hermitian system leading to finite temperature thermal phase. A detailed investigation is required in future to study these aspects of DQPTs.    
}

\textcolor{black}{We now discuss the possible experimental connection as far as the model and MSDQPT are concerned.
We know $p$-wave superconductor is naturally unavailable but can be engineered using  proximity effect in Rashba nanowire with $s$-wave superconductivity \cite{das2012zero}.   On the other hand, the non-Hermiticity is more easily realizable in meta-materials as compared to the solid state systems. 
The  non-Hermitian dynamics in ultracold atoms are theoretically proposed to obtain new regime of quantum critical phenomena \cite{Shi23,ashida2017parity,Ashida16}. The PT-symmetric dimerized photonic lattice are experimentally engineered to study the non-Hermitian topological systems \cite{weimann2017topologically}. Moving onto the experimental detection of MSDQPT, we can comment that  the geometric phase for mixed state can be captured by using NMR spectrospy \cite{Du03}.
The DQPT for fermionic many-body states has also been experimentally captured following  time-resolved state tomography in a system of
ultracold atoms in optical lattices \cite{flaschner2018observation}. The quantum logic gates in optical lattices \cite{Brennen99}, where NMR technique can be blended with ultracold atoms, might be instrumental in probing MSDQPT such that the geometric
phase  is measured for a time evolved
mixed state on a Bloch sphere. In short, we believe that meta-material perspective of quantum phenomena could be realized in future to test the theoretical findings.  However, predicting an exact experimental setup is beyond the scope of the present manuscript. }

\section{Conclusions}\label{Sec:V}

Considering $p$-wave superconductor with complex hopping and non-Hermiticity (see Fig.~\ref{Phase_Diagram}), we examine the occurrences of MSDQPT in various gapped and gapless phases. We 
find MSDQPT always exists irrespective of the gap profile of the underlying phases as long as the temperature is non-zero (see Figs. \ref{gapped}, \ref{horizontal_gapless}, and  \ref{fig:non_hermi_gapless}). \textcolor{black}{The
phase boundaries are modified by the particular choice of $\gamma_{1,2}$ in the non-Hermitian case, however,  the qualitative findings whether the MSDQPT appears remain unaltered.}
This is in contrast to the absence of DQPT in the gapped phases at  zero temperature. The half-integer jumps of winding number 
in the zero temperature for gapless phase are washed away at finite temperature.
However, in the gapped phase with finite temperature,
there exist a notion of minimum integer number above 
which the  Fisher zeros crosses the imaginary axis. We do not find any such finite integer number for Fisher zeros under finite temperature in the case of  gapless phases. We analyze the minimum time  $t_{\rm cm}$ required by the system to experience MSDQPT as  a function of the inverse temperature $\beta$ such that we can distinguish the above phases (see Fig.~\ref{temp_dep}).  The non-monotonic (monotonic) nature of  $t_{\rm cm}$ with $\beta$ is noticed for gapless (gapped) phases. There exist finer details in the behavior of $t_{\rm cm}$ with regard to the quench amplitudes through which non-Hermiticity induced and complex hopping induced gapless phases can be differentiated. Our study can successfully bridge between the zero and infinite temperature limits. We provide an effective theoretical framework to qualitatively understand the occurrences of MSDQPT.  However, we stress that 
the exact closed form expression of the critical time for any finite temperature MSDQPT is yet to be examined as a future study. The effect of long-range hopping, various types of disorder can be studied in this context of MSDQPT. 
Provided the experimental advancement on lossy systems 
\cite{Gou20,li2019observation,Zeuner15,weimann2017topologically,Weiwei18,Gao20}, we believe that the present study is experimentally viable.

\subsection*{Acknowledgments}

D.M. acknowledges SAMKHYA: High-Performance Computing Facility provided by Institute of Physics, Bhubaneswar, for numerical computations. We thank to Arijit Saha for useful discussions.
\\

\appendix




	
	\section{Initial density matrix}
	\label{Sec:Initial_density_matrix}

	The initial Hamiltonian is given by
	\begin{equation}
	H_{k,i}=\vect{h}_{k,i} \cdot \vect{\sigma}= h_{k,i}\hat{h}_{k,i}.\vec{\sigma}
	\label{initial_hamiltonian}.
	\end{equation} 
	For temperature $T=\beta^{-1}$, using Eq.~(\ref{initial_hamiltonian}) the initial ($t=0$) density matrix is given by
	\begin{eqnarray}
	\rho_{k}(0)&&=\frac{ e^{- \beta H_{k,i}}}{ \operatorname{Tr} \left[e^{- \beta H_{k,i}}\right]} \nonumber \\
	&&=\frac{ e^{- \beta h_{k,i} \hat{h}_{k,i}.\vec{\sigma}}}{ \operatorname{Tr} \left[e^{- \beta h_{k,i} \hat{h}_{k,i}.\vec{\sigma}}\right]} \nonumber\\
	&&=\frac{\cosh(\beta h_{k,i}) \sigma_{0}-(\hat{h}_{k,i}.\vec{\sigma}) \sinh(\beta h_{k,i})}{\operatorname{Tr} \left[\cosh(\beta h_{k,i}) \sigma_{0}-(\hat{h}_{k,i}.\vec{\sigma}) \sinh(\beta h_{k,i}) \right]} \nonumber\\
	&&=\frac{\cosh(\beta h_{k,i}) \sigma_{0}-(\hat{h}_{k,i}.\vec{\sigma}) \sinh(\beta h_{k,i})}{2 \cosh(\beta h_{k,i})} \nonumber \\
	&&=\frac{1}{2} \left(\sigma_{0}- \tanh(\beta h_{k,i}) (\hat{h}_{k,i}.\vec{\sigma})\right)  \nonumber \\
	&&=\frac{1}{2} \left(\sigma_{0}- m (\hat{h}_{k,i}.\vec{\sigma})\right)
	\label{initial_density_matrix_sup}
	\end{eqnarray}
where, $\sigma_{0}$ is $2 \times 2$ identity matrix, $m=\tanh(\beta h_{k,i})$ and $h_{k,i}=\sqrt{(h_{k,i}^x)^2+(h_{k,i}^y)^2+ (h_{k,i}^z)^2}$. For Hermitian Hamiltonian, $h_{k,i}$ is the positive eigenvalue of $H_{k,i}$ that is always real. On the other hand, for non-Hermitian Hamiltonian, the eigenvalue can be imaginary as well and $h_{k,i}$ corresponds to positive real part of the energy eigenvalue. \\
	
	For infinite temperature with $T \rightarrow \infty$ limit, i.e., $\beta \to 0$, leads to $m=0$. This results in 
	\begin{equation}
	\rho_{k}(0) \arrowvert _{T \rightarrow \infty}=\frac{1}{2} \sigma_{0}.
	\end{equation}
	The above expression exactly matches with previous studies on DQPT with an infinite temperature mixed density matrix at initial time~\cite{Zhou_2021}.

	\section{Loschmidt amplitude}
	\label{Sec:Loschmidt amplitude}
	
	
	The final Hamiltonian is given by
	\begin{equation}
	H_{k,f}=\vect{h}_{k,f} \hspace*{1 mm}. \hspace*{1 mm} \vect{\sigma}=h_{k,f} \hat{h}_{k,f} \hspace*{1 mm}. \hspace*{1 mm} \vec{\sigma}
	\label{final_hamiltonian},
	\end{equation} 
	where, $h_{k,f} $ represents the positive real part of the energy eigenvalue for final Hamiltonian without loss of generality.  The initial thermal density matrix is evolved with the final Hamiltonian $H_{k,f}$ in time  following the sudden quench from $H_{k,i}$.  The  time evolution operator is given by
	\begin{equation}
	U_{k}(t)=e^{-i H_{k,f}t}
	\label{evolution_operator}.
	\end{equation}	
	The LA at zero temperature is given by $g_{k}(t)=\langle \Psi_{k,i}|e^{-i H_{k,f} t}|\Psi_{k,i}\rangle= \operatorname{Tr} \bigg[|\Psi_{k,i}\rangle \langle \Psi_{k,i}|  e^{-i H_{k,f} t}\bigg]$, where initial pure quantum state $|\Psi_{k,i}\rangle$, associated with  $H_{k,i}$, is evolved with $H_{k,f}$. In the case of finite temperature,   the LA thus can be written with the initial thermal density matrix (Eq.~(\ref{initial_density_matrix_sup}) and time evolution operator (Eq.~(\ref{evolution_operator})) as follows ~\cite{Utso_2017}
 \begin{widetext}
	\begin{eqnarray}
	g_{k}(t) && = \operatorname{Tr} \left[\rho_{k}(0) U_{k} (t) \right] \nonumber\\
    && = \operatorname{Tr}\left[\frac{1}{2} \left(\sigma_{0}- m (\hat{h}_{k,i}.\vec{\sigma})\right) e^{-i H_{k,f}t} \right] \nonumber\\
	&&=\operatorname{Tr}\left[\frac{1}{2} \left(\sigma_{0}- m (\hat{h}_{k,i}.\vec{\sigma})\right) \left(\cos(h_{k,f}t)-i (\hat{h}_{k,f}.\vec{\sigma}) \sin(h_{k,f}t) \right) \right] \nonumber\\
	&&= \cos(h_{k,f}t) + i \hspace*{1 mm} m (\hat{h}_{k,i}.\hat{h}_{k,f})  \sin(h_{k,f}t) \nonumber \\
     && =\cos(h_{k,f}t) - i \hspace*{1 mm}  \sin(h_{k,f}t) B_{k}
	\label{Loschmidt2},
	\end{eqnarray}
	where, $B_{k}=-m \frac{\vec{h}_{k,i}.\vec{h}_{k,f}}{ h_{k,i} h_{k,f} }$.\\
\end{widetext}
\textcolor{black}{Note that in DQPT, LA plays the same role as  the partition function for an equilibrium phase transition.}	For the infinite temperature case with $T \rightarrow \infty$, $\beta$ vanishes yielding  $m=0$. This leads to the following
	\begin{equation}
	g_{k}(t) \arrowvert _{T \rightarrow \infty}=\cos(h_{k,f}t).
	\end{equation}
The above expression exactly matches with  previous studies on MSDQPT~\cite{Zhou_2021}.\\
	
\section{Lines of Fisher zeros}
\label{Sec:Fisher}

\textcolor{black}{Similar to 
 the vanishing of the partition function in equilibrium phase transition, here also the lines of Fisher zeros are given by the suppression of LA i.e. $g_{k}(t)=0$. This represents a complete destructive interference between the initial and time evolved states. } From Eq.~(\ref{Loschmidt2}), one can obtain the Fisher zeros as follows 
	\begin{eqnarray}
	&&\cos(h_{k,f}t) - i \hspace*{1 mm}  \sin(h_{k,f}t) B_{k}=0 \nonumber \\
	&& \Rightarrow -i\cot(h_{k,f}t) =  B_{k} \nonumber \\
	&& \Rightarrow \coth(i h_{k,f}t) = B_{k} \nonumber\\
	&& \Rightarrow \coth( h_{k,f} z) = B_{k} \nonumber \\
	&& \Rightarrow z=\frac{1}{h_{k,f}} \coth^{-1}(B_{k}) \nonumber\\
	&& \Rightarrow z=\frac{1}{h_{k,f}} \times \frac{1}{2} \ln \left(\frac{ B_{k}+1}{ B_{k}-1}\right) \nonumber \\
	&& \Rightarrow z= \frac{1}{2 h_{k,f}} \ln (-1) + \frac{1}{2 h_{k,f}} \ln \left(\frac{1 + B_{k}}{1- B_{k}}\right).  \nonumber \\
	\end{eqnarray}
	Hence the general  expression for Fisher zeros $z_{n,k}$ is given by
	\begin{eqnarray}
	&& z_{n,k}=i \left(n+ \frac{1}{2}\right) \frac{\pi}{ h_{k,f}}  + \frac{1}{2 h_{k,f}} \ln \left(\frac{1 + B_{k}}{1- B_{k}}\right), \nonumber \\
	&& \Rightarrow  z_{n,k}=i \left(n+ \frac{1}{2}\right) \frac{\pi}{ h_{k,f}}  + \frac{1}{ h_{k,f}} \tanh^{-1}(B_{k}), \label{Fisher2_sup}
	\end{eqnarray}
	where, $z_{n,k}=it$, and $n \in Z$. \textcolor{black}{Note that $z_{n,k}$ in Eq.~(\ref{Fisher2_sup}) is a complex function, and MSDQPT happens when the lines of Fisher zeros cut the imaginary axis i.e. $\operatorname{Re}[z_{n,k}]=0$. This refers to the fact $\tanh^{-1}(B_{k})/h_{k,f}=0$ with $B_k=0$.}
	
	We again compare with the infinite temperature case i.e,  $T \rightarrow \infty$ limit. Here, $\beta$ becomes zero giving rise to $m=0$, $B_{k}=0$. As a result, we find 
	\begin{equation}
	z_{n,k} \arrowvert _{T \rightarrow \infty}= i \left(n+ \frac{1}{2}\right) \frac{\pi}{ h_{k,f}} .
	\end{equation}
	Note that the above expression exactly matches with the previous findings~\cite{Zhou_2021}.
	
	\section{Critical momenta}
	\label{Sec:kc}
	
	Let us define a new quantity as $C_{k}= \tanh^{-1}(B_{k})$. Hence Eq.~(\ref{Fisher2_sup}) becomes
	\begin{eqnarray}
	z_{n,k}&&=i \left(n+ \frac{1}{2}\right) \frac{\pi}{ h_{k,f}}  + \frac{1}{ h_{k,f}} C_{k}  \nonumber \\
	&&=i \left(n+ \frac{1}{2}\right) \pi \frac{\operatorname{Re}[h_{k,f}] -i \operatorname{Im}[h_{k.f}]}{|h_{k,f}|^{2}} \nonumber \\
 &&+  \frac{\operatorname{Re}[h_{k,f}] -i \operatorname{Im}[h_{k.f}]}{|h_{k,f}|^{2}} \left(\operatorname{Re}[C_{k}] + i \operatorname{Im}[C_{k}] \right). \nonumber
	\end{eqnarray}
	The critical momenta $k_{c}$ is then obtained by solving the equation below for $k_c$
	\begin{eqnarray}
	&&\operatorname{Re}[z_{n,k_{c}}]=0 \nonumber \\
	&& \Rightarrow \pi \left(n + 1/2\right) \operatorname{Im}[h_{k_{c},f}]  + \operatorname{Re}[h_{k_{c},f}] \operatorname{Re}[C_{k_{c}}] \nonumber \\
 &&+  \operatorname{Im}[h_{k_{c},f}] \operatorname{Im}[C_{k_{c}}] =0  \label{kc_sup}.
	\end{eqnarray}
	
	We now reduce the above expression in the case of infinite temperature. For $T \rightarrow \infty$ limit, i.e., $\beta  \to 0$, one can obtain $B_{k}=0$, and $C_{k}=0$. As a result, Eq.~(\ref{kc_sup}) takes the following form
	\begin{equation}
	\operatorname{Im}[h_{k_{c},f}]=0 \label{sup_infinite_temp_kc}.
	\end{equation}
	This is consistent with the previous findings~\cite{Zhou_2021}. 	We would like to point out an important observation here for $\gamma_2, \Delta \ne0$, and $\gamma_1=0$. The above Eq.~(\ref{sup_infinite_temp_kc}) can only yield $k_c=n\pi$ with $n=0,1,2,\cdots$ as the  valid solution. This further indicates that the critical momenta $k_{c}$ are independent of all the model parameters. In order to derive the above solution, we use $z^{1/2}=|z|^{1/4} \exp(i\phi/2)$ with $z=x+i y$, $\phi={\rm arctan}(y/x)$. A complete calculation suggests that $y=0$ is the only solution possible provided $|z| \ne 0$.

	\section{Critical time}
	\label{Sec:tc}

	We here derive the critical time $t_c=-i z_{n,k_{c}}$ corresponding to $k_c$ as given below
	\begin{eqnarray}
	t_{c}=&&\pi \left(n+\frac{1}{2}\right) \frac{\operatorname{Re}[h_{k_{c},f}]}{|h_{k_{c},f}|^{2}} \nonumber \\
 &&+  \frac{\operatorname{Re}[h_{k_{c},f}] \operatorname{Im}[C_{k_{c}}] -\operatorname{Im}[h_{k_{c},f}] \operatorname{Re}[C_{k_{c}}]}{|h_{k_{c},f}|^{2}} \label{tc_sup}.
	\end{eqnarray}
	
	The above general expression in  the case of infinite temperature can be reduced further. Considering  $T \rightarrow \infty$, $\beta \to 0$, i.e., $B_{k}=0$ i.e. $C_{k}=0$, we find 
	\begin{equation}
	t_{c} \arrowvert_{T \rightarrow \infty}=\left(n+\frac{1}{2}\right)\frac{\pi}{\operatorname{Re}[h_{k_{c},f}]}.
	\end{equation}
	This is consistent with the previous findings~\cite{Zhou_2021}. Using the above lines of argument, discussed after Eq.~(\ref{sup_infinite_temp_kc}), one can find that for a fixed $\mu_{f}$ and $\gamma_2, \Delta \ne 0$, $t_{\rm cm}|_{T\rightarrow \infty}$ is independent of quench amplitude. Interestingly, $t_{\rm cm}|_{T\rightarrow \infty}$ can only depend on $\mu_f$ while $\Delta$ dependence is completely absent for the critical momentum $k_c=n\pi$.

	\section{Dynamical phase}
	\label{Sec:dynamical_phase}

	\textcolor{black}{Dynamical phase is nothing but the phase acquired by a quantum state due to the time evolution of the underlying Hamiltonian.} We here illustrate the  dynamical phase for the non-Hermitian system such that $H^{\dagger}\ne H$ \cite{Zhou_2021}.
	Let us denote the right and left eigenvectors as $\ket{\psi_{s}^{r}(k)}$, and $\ket{\psi_{s}^{l}(k)}$ respectively. Here $s=\pm$ denotes two energy bands for the $2$-level systems.  These eigenvectors satisfy the following equations
	\begin{eqnarray}
	&&H(k) \ket{\psi_{s}^{r}(k)} = E_{s}(k) \ket{\psi_{s}^{r}(k)} \label{SE1},\\
	&&H^{\dagger}(k) \ket{\psi_{s}^{l}(k)} = E_{s}^{*}(k) \ket{\psi_{s}^{l}(k)} \label{SE2}.
	\end{eqnarray}
	
	In this representation, the Hamiltonian $H_{k}$ can be expressed as
	\begin{equation}
	H_{k}=\sum_{s=\pm} E_{s}(k) \ket{\psi_{s}^{r}(k)} \bra{\psi_{s}^{l}(k)} \label{Hamiltonian}.
	\end{equation} 
	In the space of these left and right eigenvectors, right and left time evolution operators can be expressed as
	\begin{eqnarray}
	U_{k}^{r}(t)=\sum_{s=\pm} e^{-i E_{s}(k) t} \ket{\psi_{s}^{r}(k)}\bra{\psi_{s}^{l}(k)} \label{rightU}, \\
	U_{k}^{l}(t)=\sum_{s=\pm} e^{-i E_{s}(k) t} \ket{\psi_{s}^{l}(k)}\bra{\psi_{s}^{r}(k)} \label{leftU},
	\end{eqnarray}
	respectively. The bi-orthogonality conditions 
	$\sum_s|\psi^r_{s}(k)\rangle \langle\psi^l_{s}(k)|=\sigma_0$ and 
	$\langle\psi^l_{s}(k)| \psi^r_{s'}(k)\rangle=\delta_{ss'}$ are required to further simplify the expressions. 
	The time evolved density matrix is written as $\rho_{k}(t)=U_{k}^{l \dagger}(t) \rho_{k}(0) U_{k}^{r}(t)$.  The dynamical phase is expressed as follows~\cite{Utso_2017}
 \begin{widetext}
	\begin{eqnarray}
	\Phi_{k}^{\rm dyn}(t) &&= - \int_{0}^{t} dt^{\prime} \operatorname{Re}\left[\frac{\operatorname{Tr}\left[ \rho_{k}(t)  H_{k,f} \right]}{\operatorname{Tr}\left[ \rho_{k}(t) \right]}\right] \nonumber \\
	&& = - \int_{0}^{t} dt^{\prime} \operatorname{Re}\left[\frac{{\operatorname{Tr}}\left[U_{k}^{l \dagger}(t^{\prime}) \rho_{k}(0) U_{k}^{r}(t^{\prime}) H_{k,f} \right]}{\operatorname{Tr}\left[U_{k}^{l \dagger}(t^{\prime}) \rho_{k}(0) U_{k}^{r}(t^{\prime}) \right]}\right] \label{dynamical_main}.
	\end{eqnarray}
 
	Now, using Eq~(\ref{initial_density_matrix_sup}), we obtain 
	
		\begin{eqnarray}
		\Phi_{k}^{\rm dyn}(t) &&= - \int_{0}^{t} dt^{\prime} \operatorname{Re}\left[\frac{\operatorname{Tr}\left[U_{k}^{l\dagger}(t^{\prime}) \times \frac{1}{2} \left(\sigma_{0}- m (\hat{h}_{k,i}.\vec{\sigma})\right) \times U_{k}^{r}(t^{\prime}) H_{k,f} \right]}{\operatorname{Tr}\left[U_{k}^{l \dagger}(t^{\prime}) \times \frac{1}{2} \left(\sigma_{0}- m (\hat{h}_{k,i}.\vec{\sigma})\right) \times U_{k}^{r}(t^{\prime}) \right]}\right] \nonumber\\
		 &&= - \int_{0}^{t} dt^{\prime} \operatorname{Re}\left[\frac{\operatorname{Tr}\left[U_{k}^{l \dagger}(t^{\prime}) \times \frac{1}{2} \left(\sigma_{0}- \frac{m H_{k,i}}{h_{k,i}}  \right) \times U_{k}^{r}(t^{\prime}) H_{k,f} \right]}{\operatorname{Tr}\left[U_{k}^{l \dagger}(t^{\prime}) \times \frac{1}{2} \left(\sigma_{0}- \frac{m H_{k,i}}{h_{k,i}}  \right)\times  U_{k}^{r}(t^{\prime}) \right]}\right]. \nonumber \\
		\label{dynamical_sup}
		\end{eqnarray}
		Now using Eqs.~(\ref{Hamiltonian}), (\ref{rightU}), (\ref{leftU}) 
		\begin{eqnarray}
		&&\operatorname{Tr}\left[U_{k}^{l \dagger}(t^{\prime}) \sigma_{0} U_{k}^{r}(t^{\prime}) H_{k,f} \right] \nonumber \\
		&&=\operatorname{Tr}\left[U_{k}^{l \dagger}(t^{\prime}) U_{k}^{r}(t^{\prime}) H_{k,f} \right] \nonumber\\
		&&=\operatorname{Tr}\left[\sum_{s,s^{\prime},s^{\prime \prime}=\pm} e^{i E_{s,f}^{*}(k) t^{\prime}} e^{-i E_{s^{\prime},f}(k)t^{\prime}} E_{s^{\prime \prime},f}(k) \times \ket{\psi_{s,f}^{r}(k)} \bra{\psi_{s,f}^{l}(k)} \times \ket{\psi_{s^{\prime},f}^{r}(k)} \bra{\psi_{s^{\prime},f}^{l}(k)} \times \ket{\psi_{s^{\prime \prime},f}^{r}(k)} \bra{\psi_{s^{\prime\prime},f }^{l}(k)} \right] \nonumber \\
		&&=\operatorname{Tr}\left[\sum_{s,s^{\prime},s^{\prime \prime}=\pm} e^{i E_{s,f}^{*}(k) t^{\prime}} e^{-i E_{s^{\prime},f}(k)t^{\prime}} E_{s^{\prime \prime},f}(k) \times \ket{\psi_{s,f}^{r}(k)} \bra{\psi_{s^{\prime\prime},f }^{l}(k)}~ \delta_{ss^{\prime}} ~ \delta_{s^{\prime}s^{\prime \prime}}\right] \nonumber \\
		&&=\operatorname{Tr}\left[\sum_{s=\pm} e^{i E_{s,f}^{*}(k) t^{\prime}} e^{-i E_{s,f}(k)t^{\prime}} E_{s,f}(k) \times \ket{\psi_{s,f}^{r}(k)} \bra{\psi_{s,f}^{l}(k)} \right] \nonumber\\
		&&=\operatorname{Tr}\left[\sum_{s=\pm} e^{2 ~\operatorname{Im}[ E_{s,f}(k)] ~t^{\prime}} E_{s,f}(k) \times \ket{\psi_{s,f}^{r}(k)} \bra{\psi_{s,f}^{l}(k)} \right] \nonumber \\
		&&=e^{2~ \operatorname{Im}[h_{k,f}]~ t^{\prime}} h_{k,f}  - e^{-2 ~\operatorname{Im}[h_{k,f}]~ t^{\prime}} h_{k,f} \nonumber \\
		&&= 2~ h_{k,f} \sinh\left(2~ \operatorname{Im}[h_{k,f}]~ t^{\prime}\right)
		\end{eqnarray}
		Similarly,
		\begin{eqnarray}
		&&\operatorname{Tr}\left[U_{k}^{l \dagger}(t^{\prime}) H_{k,i} U_{k}^{r}(t^{\prime}) H_{k,f} \right]=2~h_{k,i}~ h_{k,f} \cosh\left(2~ \operatorname{Im}[h_{k,f}]~ t^{\prime}\right)\\
		&&\operatorname{Tr}\left[U_{k}^{l \dagger}(t^{\prime}) \sigma_{0} U_{k}^{r}(t^{\prime}) \right]=2~ \cosh\left(2~ \operatorname{Im}[h_{k,f}]~ t^{\prime}\right)\\
		&&\operatorname{Tr}\left[U_{k}^{l \dagger}(t^{\prime}) H_{k,i} U_{k}^{r}(t^{\prime}) \right]=2~h_{k,i}~ \sinh\left(2~ \operatorname{Im}[h_{k,f}]~ t^{\prime}\right)
		\end{eqnarray}
		Now combining all the above expression, the    dynamical phase for non-Hermitian system is found to be
		\begin{eqnarray}
		\Phi_{k}^{\rm dyn}(t) &&= - \int_{0}^{t} dt^{\prime} \operatorname{Re} \left[h_{k,f} \frac{\sinh\left(2~ \operatorname{Im}[h_{k,f}]~ t^{\prime}\right)-m~\cosh\left(2~ \operatorname{Im}[h_{k,f}]~ t^{\prime}\right)}{\cosh\left(2~ \operatorname{Im}[h_{k,f}]~ t^{\prime}\right)-m~\sinh\left(2~ \operatorname{Im}[h_{k,f}]~ t^{\prime}\right)} \right] \nonumber \\
		&&= - \int_{0}^{t} dt^{\prime} \operatorname{Re}\left[h_{k,f} \frac{\tanh\left(2~ \operatorname{Im}[h_{k,f}]~ t^{\prime}\right)-m}{1-m~\tanh\left(2~ \operatorname{Im}[h_{k,f}]~ t^{\prime}\right)} \right] \label{dynamical_final}.
		\end{eqnarray}
		In the  infinite temperature $T \rightarrow \infty$ limit, the dynamical phase reads as 
		\begin{equation}
		\Phi_{k}^{\rm dyn}(t) \arrowvert _{T \rightarrow \infty}=- \int_{0}^{t} dt^{\prime} \operatorname{Re}\left[h_{k,f} ~\tanh\left(2~ \operatorname{Im}[h_{k,f}]~ t^{\prime}\right) \right].
		\end{equation}
		The above expression is consistent with earlier findings \cite{Zhou_2021}.
		
	\section{Effective theory for non-Hermitian DQPT}
	\label{non_bloch}
	We here discuss the effective theory for the MSDQPT. We rewrite the non-Hermitian Hamiltonian under consideration 
	\begin{eqnarray}
	{\cal H} _{k}(0,\gamma_2,\phi)&&=  \left(2 \Delta \sin k + \frac{i \gamma_2}{2}\right) \sigma_{y} - \left(2 w_{0} \cos \phi \cos k +\mu \right) \sigma_{z} . 
	\label{Non_Bloch_Hamiltonian}
	\end{eqnarray}
	Replacing `$k$' by `$k+i \kappa$', and saying $e^{k+i \kappa}\equiv x$, we can write the above Hamiltonian as
	\begin{eqnarray}
	{\cal H} _{k}(0,\gamma_2,\phi)&&=  \left(-i \Delta (x-x^{-1}) + \frac{i \gamma_2}{2}\right) \sigma_{y} - \left( w_{0} \cos \phi \hspace*{1 mm} (x + x^{-1}) +\mu \right) \sigma_{z}=\vect{h}_{k}.\vect{\sigma} .
	\label{Non_Bloch_Hamiltonian_change}
	\end{eqnarray}
	The eigenvalues are given by
	\begin{equation}
	E_{\pm}=\pm \sqrt{\left(-i \Delta (x-x^{-1}) + \frac{i \gamma_2}{2}\right)^{2}+\left( w_{0} \cos \phi \hspace*{1 mm} (x + x^{-1}) +\mu \right)^{2}}\label{Eigen_non_bloch}.
	\end{equation}
	Now, $E_{\pm}\rightarrow 0$ limit, we get
	\begin{eqnarray}
	-\left(- \Delta (x^{2}-1) + \frac{ \gamma_2}{2}x\right)^{2}+\left( w_{0} \cos \phi \hspace*{1 mm} (x^{2} + 1) +\mu x \right)^{2}=0
	\label{non_block_criterion}.
	\end{eqnarray}
	Solutions of the above equation are
	\begin{eqnarray}
	&&x_{1}=\frac{\frac{\gamma_2}{2} -\mu -\sqrt{(\mu -\frac{\gamma_2}{2} )^2-4 (w_0 \cos \phi -\Delta ) (\Delta +w_0 \cos \phi )}}{2 (\Delta +w_0 \cos
		\phi )} \nonumber, \\
	&&x_{2}=\frac{\frac{\gamma_2}{2} -\mu +\sqrt{(\mu -\frac{\gamma_2}{2} )^2-4 (w_0 \cos \phi -\Delta ) (\Delta +w_0 \cos \phi )}}{2 (\Delta +w_0 \cos
		\phi )} \nonumber, \\
	&& x_{3}=\frac{\frac{\gamma_2 }{2}+\mu -\sqrt{\frac{\gamma_2 ^2}{4}+\gamma_2  \hspace*{1 mm} \mu +4 \Delta ^2+\mu ^2-2 w_0^2 \cos 2 \phi -2 w_0^2}}{2 (\Delta
		-w_0 \cos \phi )} \nonumber, \\
	&& x_{4}=\frac{\frac{\gamma_2 }{2}+\mu +\sqrt{\frac{\gamma_2 ^2}{4}+\gamma_2 \hspace*{1 mm} \mu +4 \Delta ^2+\mu ^2-2 w_0^2 \cos 2 \phi -2 w_0^2}}{2 (\Delta
		-w_0 \cos \phi )} \label{solution}.
	\end{eqnarray}
 \end{widetext}
	Now, $x_{1} x_{2} x_{3} x_{4}=1$, $x_{1}x_{2}= \frac{- \Delta+w_0 \cos \phi}{\Delta+w_0 \cos \phi}$, and $x_{3}x_{4}= \frac{ \Delta+w_0 \cos \phi}{-\Delta+w_0 \cos \phi}$. Therefore, $x$ can be written as, $x=(x_{1} x_{2} x_{3} x_{4})^{1/4}$, or $x=\sqrt{x_{1} x_{2}}$, or $x=\sqrt{x_{3} x_{4}}$, however, none of the above is a good choice as their final expressions are independent of $\gamma_2$. We hence use $x=\sqrt{x_{1} x_{4}}$ as it depends on $\gamma_2$. This allows us to write the  Eq.~(\ref{Non_Bloch_Hamiltonian_change}) as follows
	\begin{eqnarray}
	&&h_{k}^{y}=-i \Delta (e^{i k}x-e^{-ik}x^{-1}) + \frac{i \gamma_2}{2}, \\
	&&h_{k}^{z}=-[w_{0} \cos \phi \hspace*{1 mm} (e^{i k}x + e^{-ik} x^{-1}) +\mu ]\label{hy_hz_non_bloch}.
	\end{eqnarray}
	
	In order to get the critical momentum, $k_c$ we use the	known DQPT framework for the Hermitian system using $\vec{h}_{k_c,i}.\vec{h}_{k_c,f}$ \cite{Utso_2017} for our case such that 
	\begin{eqnarray}
h_{k_c,i}^y h_{k_c,f}^y + h_{k_c,i}^z h_{k_c,f}^z =0.	
	\end{eqnarray}
	Solving the above equation for a fixed quench amplitude, one can get multiple critical momenta $k_c$ unlike single critical momentum for Hermitian case~\cite{Utso_2017}. This can qualitatively explain the emergence of multiple $k_c$'s for the non-Hermitian case. 
\begin{widetext}	
	\section{Missing MSDQPT for
	non-Hermitian chemical potential}
	\label{nh_chemical_potential}

	\begin{figure}[H]
		\centering
		\subfigure{\includegraphics[width=1.0\textwidth]{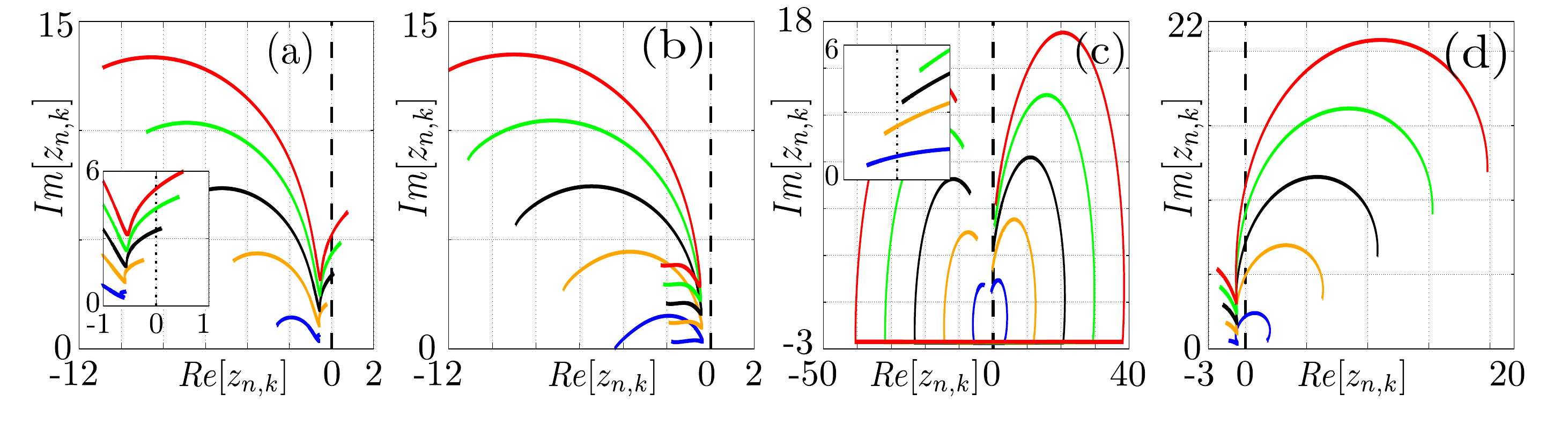}}
		\caption{The lines of Fisher zeroes are plotted with $n=0$ (blue), $\cdots$, $n=4$ (red) for the cases of non-Hermitian chemical potential with $(\gamma_1=1,\gamma_2=0)$. The Fisher zeros 
		for the quench within gapped phase I, III are shown in (a) for ($\mu_{i}$, $\mu_{f}$, $\Delta_{i}$, $\Delta_{f})=(0.1, 0.7, 2.2, 2.2)$, in (b) for ($\mu_{i}$, $\mu_{f}$, $\Delta_{i}$, $\Delta_{f})=(-4.7, -2.1, 2.2, 2.2)$, respectively. The same is shown for gapless phase IV and V in (c) for ($\mu_{i}$, $\mu_{f}$, $\Delta_{i}$, $\Delta_{f})=(0.1, 0.7, 0.2, 0.2)$ and in (d) for ($\mu_{i}$, $\mu_{f}$, $\Delta_{i}$, $\Delta_{f})=(-1.7, -1.1, 2.2, 2.2)$, respectively. We find that there exists  $n_{\rm min}=1$ in (a) above which the Fisher zeros  cross the imaginary axis referring to the occurrence of MSDQPT. The lines of  Fisher zeros do not cross the  imaginary axis for any value of $n$ indicating to the absence of 
	MSDQPT in (b). There exists a 
       $n_{\rm max}=1$ above which the Fisher zeros does not cross the imaginary axis referring to the fact that MSDQPT can only be observed for short times in (c). The lines of Fisher zeros always cross the imaginary axis confirming the occurrences of MSDQPT in (d).  We consider ($w_0,\phi,\gamma_1,\gamma_2,\beta)=(1,\frac{\pi}{4},1,0,1)$. }
		\label{gamma1_fisher}
    \end{figure}
    \end{widetext}	

    We here consider the non-Hermiticity only in the chemical potential i.e., $\gamma_1=1$ and $\gamma_2=0$. Note that the phase diagram changes for lossy chemical potential, however, the phases I, II, III, IV, and V are present similar to the lossy superconductivity (see Fig.~\ref{Phase_Diagram}).
    We study the behavior of the Fisher zeros  for 
    quench within the phases I, III, IV and V, as shown in Fig.~\ref{gamma1_fisher} (a), (b), (c), and (d), respectively.   We find that MSDQPT can exist for phases I, IV and V except for phase III as the  Fisher zeros crosses the imaginary axis in the prior phases but not in the later phase. This is in contrast to the non-Hermitian superconductor case where the MSDQPT always persists irrespective of the phases as long as the temperature is non-zero. On the other hand, for phases I and IV, one can  respectively find $n_{\rm min}$ and  $n_{\rm max}$ for the lines of Fisher zeros, indicating to the fact that MSDQPT is absent below (above) a certain time scale.  This time scale is directly related to $n_{\rm min}$ and  $n_{\rm max}$ for phases I and IV, respectively. For the case of    non-Hermitian superconductor, we do not find any such $n_{\rm max}$ in phase IV, however, we do find $n_{\rm min}$
	for phase I. Therefore, the gapped and gapless phases for non-Hermitian superconductor and non-Hermitian chemical potential do not show identical properties as far as the MSDQPT is concerned. Interestingly, for quench within region III, we find the absence of MSDQPT for the finite temperature non-Hermitian chemical potential similar to the zero temperature  case \cite{Mondal22}. 
However,  likewise the zero temperature  case,  we do not find any discontinuity  in the Fisher zeros in any of the above  cases. \\



\section{\textcolor{black}{MSDQPT in the Hermitian limit}}
\label{sec:hermitian case}
\begin{widetext}
 
\begin{figure}[H]
	\centering
	\subfigure{\includegraphics[width=1.0\textwidth]{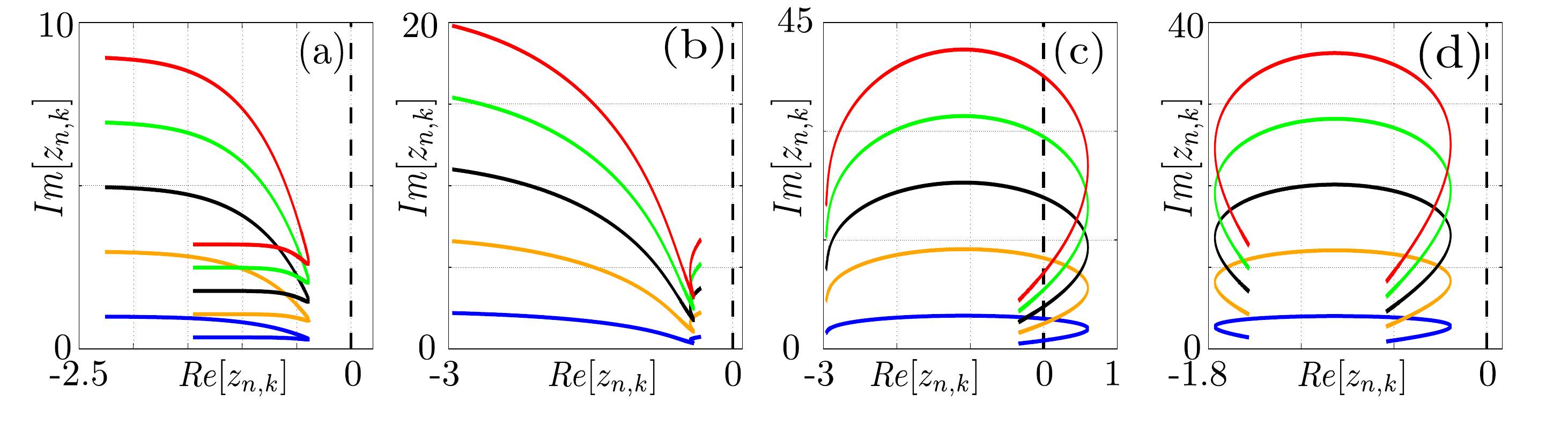}}
	\caption{\textcolor{black}{The lines of Fisher zeros are depicted with $n=0$ (blue), $\cdots$, $n=4$ (red) for Hermitian case $(\gamma_1=0,\gamma_2=0)$. The Fisher zeros 
    for the quench within gapped phase III, and I are shown in (a) for ($\mu_{i}$, $\mu_{f}$, $\Delta_{i}$, $\Delta_{f})=(-5, -3, 2.2, 2.2)$, in (b) for ($\mu_{i}$, $\mu_{f}$, $\Delta_{i}$, $\Delta_{f})=(-0.7, 0.7, 2.2, 2.2)$, respectively. Clearly, there is no crossing over the imaginary axis which confirms the absence of MSDQPT. We do the same for two different quench metric within the gapless phase IV in (c) for ($\mu_{i}$, $\mu_{f}$, $\Delta_{i}$, $\Delta_{f})=(-0.7, 0.7, 0.2, 0.2)$ and in (d) for ($\mu_{i}$, $\mu_{f}$, $\Delta_{i}$, $\Delta_{f})=(-0.3, 0.7, 0.2, 0.2)$, respectively. Here, the lines of Fisher zeros cross the imaginary axis twice~[do not cross the imaginary axis] in (c)~[(d)] suggesting the occurrences of MSDQPT with two types of $k_{c}$~[absence of MSDQPT].  We choose ($w_0,\phi,\beta)=(1,\frac{\pi}{4},1)$.}}
	\label{heritian_fisher}
\end{figure}
\end{widetext}

\textcolor{black}{In this section, we  concentrate on Hermitian limit of our system i.e., $\gamma_{1}=\gamma_{2}=0$. The phase diagram is different for Hermitian case as compared to the non-Hermitian counterpart. The vertical gapless region V vanishes completely, while the horizontal gapless region IV vanishes~(becomes narrower) for $\phi=0$~($\phi=\pi/4$). We are interested in $\phi=\pi/4$ here as it supports extended gapless region where we can observe its effects on MSDQPT following an intra-phase quench. One can rewrite the expressions for the physical quantities like rate function, Fisher zeros, dynamical phases in  the Heritian limit. Note that for Hermitian case energy eigenvalues are real i.e. $\operatorname{Im}[h_{k,i}]=\operatorname{Im}[h_{k,f}]=0$ suggesting $B_{k}$ in Eq.~(\ref{Loschmidt2}), and (\ref{Fisher2_sup}) to be real. This does not result in any  change in the expression for LA $g_{k}(t)$, lines of Fisher zeros $z_{n,k}$, and total phase $\Phi_{k}^{\rm tot}(t)$ while the critical momenta is obtained from 
\begin{eqnarray}
\vec{h}_{k_c,i}.\vec{h}_{k_c,f}=0 \label{eq:hermi_kc}.
\end{eqnarray}
 The critical time is found to be  
 \begin{eqnarray}
 t_{c}=\left(n+\frac{1}{2}\right)\frac{\pi}{h_{k_c,f}} \label{eq:hermi_tc}.
 \end{eqnarray}
However, dynamical phase  has a simple form as
\begin{eqnarray}
\Phi_{k}^{\rm dyn}(t)=-\int_{0}^{t}  dt^{\prime} \hspace*{1.5 mm} m \hspace*{1.5 mm} h_{k,f} =- m \hspace*{1.5 mm} h_{k,f} \hspace*{1.5 mm}t \label{eq:hermi_dynphase}.
\end{eqnarray}
Here, we are interested only in the lines of Fisher zeroes that are  enough to confirm the  occurrences of MSDQPT. We investigate the Fisher zero profiles for quenches within regions III and I which are shown in Fig.~\ref{heritian_fisher} (a) and (b) respectively. The  non-crossing nature conveys  the absence of the MSDQPT for an intra-phase quench inside region I. Therefore, for intra-phase quench within region I,  Fig.~\ref{heritian_fisher} (b) suggests that MSDQPT does not happen. By contrast, $(\gamma_{1},\gamma_{2})=(\neq 0,0)$  and ($0, \neq 0$) as shown in Figs.~\ref{gamma1_fisher} (a) and \ref{fig:non_hermi_gapless} respectively, we find that MSDQPT exists which is a marked difference as compared to the finite temperature Hermitian case. The MSDQPT is absent for intra-phase quench in region III irrespective of the Hermicity of the problem (see Figs.~\ref{heritian_fisher} (a) and \ref{gamma1_fisher} (b)). The Fisher zeros are depicted in Fig.~\ref{heritian_fisher} (c) ((d)) for large~(short) quench mertic within region IV. Interestingly,   the lines of Fisher zeroes cross imaginary axis twice indicating MSDQPT for two types of critical momenta in Fig.~\ref{heritian_fisher} (c). On the other hand, the lines of Fisher zeroes
do not cross imaginary axis referring to the absence of  MSDQPT as demonstrated in Fig.~\ref{heritian_fisher} (d).  Based on the above analysis on the Hermitian case at finite temperature~($\beta=1$), we can comment that the results are similar as obtained for the zero temperature case~\cite{Mondal22}.}

\bibliography{bibfile}{}  

\begin{thebibliography}{113}%
\makeatletter
\providecommand \@ifxundefined [1]{%
 \@ifx{#1\undefined}
}%
\providecommand \@ifnum [1]{%
 \ifnum #1\expandafter \@firstoftwo
 \else \expandafter \@secondoftwo
 \fi
}%
\providecommand \@ifx [1]{%
 \ifx #1\expandafter \@firstoftwo
 \else \expandafter \@secondoftwo
 \fi
}%
\providecommand \natexlab [1]{#1}%
\providecommand \enquote  [1]{``#1''}%
\providecommand \bibnamefont  [1]{#1}%
\providecommand \bibfnamefont [1]{#1}%
\providecommand \citenamefont [1]{#1}%
\providecommand \href@noop [0]{\@secondoftwo}%
\providecommand \href [0]{\begingroup \@sanitize@url \@href}%
\providecommand \@href[1]{\@@startlink{#1}\@@href}%
\providecommand \@@href[1]{\endgroup#1\@@endlink}%
\providecommand \@sanitize@url [0]{\catcode `\\12\catcode `\$12\catcode
  `\&12\catcode `\#12\catcode `\^12\catcode `\_12\catcode `\%12\relax}%
\providecommand \@@startlink[1]{}%
\providecommand \@@endlink[0]{}%
\providecommand \url  [0]{\begingroup\@sanitize@url \@url }%
\providecommand \@url [1]{\endgroup\@href {#1}{\urlprefix }}%
\providecommand \urlprefix  [0]{URL }%
\providecommand \Eprint [0]{\href }%
\providecommand \doibase [0]{http://dx.doi.org/}%
\providecommand \selectlanguage [0]{\@gobble}%
\providecommand \bibinfo  [0]{\@secondoftwo}%
\providecommand \bibfield  [0]{\@secondoftwo}%
\providecommand \translation [1]{[#1]}%
\providecommand \BibitemOpen [0]{}%
\providecommand \bibitemStop [0]{}%
\providecommand \bibitemNoStop [0]{.\EOS\space}%
\providecommand \EOS [0]{\spacefactor3000\relax}%
\providecommand \BibitemShut  [1]{\csname bibitem#1\endcsname}%
\let\auto@bib@innerbib\@empty
\bibitem [{\citenamefont {Fisher}(1967)}]{fisher1967theory}%
  \BibitemOpen
  \bibfield  {author} {\bibinfo {author} {\bibfnamefont {M.~E.}\ \bibnamefont
  {Fisher}},\ }\bibfield  {title} {\enquote {\bibinfo {title} {The theory of
  equilibrium critical phenomena},}\ }\href@noop {} {\bibfield  {journal}
  {\bibinfo  {journal} {Reports on progress in physics}\ }\textbf {\bibinfo
  {volume} {30}},\ \bibinfo {pages} {615} (\bibinfo {year} {1967})}\BibitemShut
  {NoStop}%
\bibitem [{\citenamefont {Yang}\ and\ \citenamefont
  {Lee}(1952)}]{PhysRev.87.404}%
  \BibitemOpen
  \bibfield  {author} {\bibinfo {author} {\bibfnamefont {C.~N.}\ \bibnamefont
  {Yang}}\ and\ \bibinfo {author} {\bibfnamefont {T.~D.}\ \bibnamefont {Lee}},\
  }\bibfield  {title} {\enquote {\bibinfo {title} {Statistical theory of
  equations of state and phase transitions. i. theory of condensation},}\
  }\href {\doibase 10.1103/PhysRev.87.404} {\bibfield  {journal} {\bibinfo
  {journal} {Phys. Rev.}\ }\textbf {\bibinfo {volume} {87}},\ \bibinfo {pages}
  {404--409} (\bibinfo {year} {1952})}\BibitemShut {NoStop}%
\bibitem [{\citenamefont {Lee}\ and\ \citenamefont
  {Yang}(1952)}]{PhysRev.87.410}%
  \BibitemOpen
  \bibfield  {author} {\bibinfo {author} {\bibfnamefont {T.~D.}\ \bibnamefont
  {Lee}}\ and\ \bibinfo {author} {\bibfnamefont {C.~N.}\ \bibnamefont {Yang}},\
  }\bibfield  {title} {\enquote {\bibinfo {title} {Statistical theory of
  equations of state and phase transitions. ii. lattice gas and ising model},}\
  }\href {\doibase 10.1103/PhysRev.87.410} {\bibfield  {journal} {\bibinfo
  {journal} {Phys. Rev.}\ }\textbf {\bibinfo {volume} {87}},\ \bibinfo {pages}
  {410--419} (\bibinfo {year} {1952})}\BibitemShut {NoStop}%
\bibitem [{\citenamefont {Heyl}\ \emph {et~al.}(2013)\citenamefont {Heyl},
  \citenamefont {Polkovnikov},\ and\ \citenamefont {Kehrein}}]{Heyl13}%
  \BibitemOpen
  \bibfield  {author} {\bibinfo {author} {\bibfnamefont {M.}~\bibnamefont
  {Heyl}}, \bibinfo {author} {\bibfnamefont {A.}~\bibnamefont {Polkovnikov}}, \
  and\ \bibinfo {author} {\bibfnamefont {S.}~\bibnamefont {Kehrein}},\
  }\bibfield  {title} {\enquote {\bibinfo {title} {Dynamical quantum phase
  transitions in the transverse-field ising model},}\ }\href {\doibase
  10.1103/PhysRevLett.110.135704} {\bibfield  {journal} {\bibinfo  {journal}
  {Phys. Rev. Lett.}\ }\textbf {\bibinfo {volume} {110}},\ \bibinfo {pages}
  {135704} (\bibinfo {year} {2013})}\BibitemShut {NoStop}%
\bibitem [{\citenamefont {Kriel}\ \emph {et~al.}(2014)\citenamefont {Kriel},
  \citenamefont {Karrasch},\ and\ \citenamefont
  {Kehrein}}]{PhysRevB.90.125106}%
  \BibitemOpen
  \bibfield  {author} {\bibinfo {author} {\bibfnamefont {J.~N.}\ \bibnamefont
  {Kriel}}, \bibinfo {author} {\bibfnamefont {C.}~\bibnamefont {Karrasch}}, \
  and\ \bibinfo {author} {\bibfnamefont {S.}~\bibnamefont {Kehrein}},\
  }\bibfield  {title} {\enquote {\bibinfo {title} {Dynamical quantum phase
  transitions in the axial next-nearest-neighbor ising chain},}\ }\href
  {\doibase 10.1103/PhysRevB.90.125106} {\bibfield  {journal} {\bibinfo
  {journal} {Phys. Rev. B}\ }\textbf {\bibinfo {volume} {90}},\ \bibinfo
  {pages} {125106} (\bibinfo {year} {2014})}\BibitemShut {NoStop}%
\bibitem [{\citenamefont {Karrasch}\ and\ \citenamefont
  {Schuricht}(2013)}]{PhysRevB.87.195104}%
  \BibitemOpen
  \bibfield  {author} {\bibinfo {author} {\bibfnamefont {C.}~\bibnamefont
  {Karrasch}}\ and\ \bibinfo {author} {\bibfnamefont {D.}~\bibnamefont
  {Schuricht}},\ }\bibfield  {title} {\enquote {\bibinfo {title} {Dynamical
  phase transitions after quenches in nonintegrable models},}\ }\href {\doibase
  10.1103/PhysRevB.87.195104} {\bibfield  {journal} {\bibinfo  {journal} {Phys.
  Rev. B}\ }\textbf {\bibinfo {volume} {87}},\ \bibinfo {pages} {195104}
  (\bibinfo {year} {2013})}\BibitemShut {NoStop}%
\bibitem [{\citenamefont {Heyl}(2015)}]{Heyl15}%
  \BibitemOpen
  \bibfield  {author} {\bibinfo {author} {\bibfnamefont {M.}~\bibnamefont
  {Heyl}},\ }\bibfield  {title} {\enquote {\bibinfo {title} {Scaling and
  universality at dynamical quantum phase transitions},}\ }\href {\doibase
  10.1103/PhysRevLett.115.140602} {\bibfield  {journal} {\bibinfo  {journal}
  {Phys. Rev. Lett.}\ }\textbf {\bibinfo {volume} {115}},\ \bibinfo {pages}
  {140602} (\bibinfo {year} {2015})}\BibitemShut {NoStop}%
\bibitem [{\citenamefont {Heyl}(2018)}]{Heyl_2018}%
  \BibitemOpen
  \bibfield  {author} {\bibinfo {author} {\bibfnamefont {M.}~\bibnamefont
  {Heyl}},\ }\bibfield  {title} {\enquote {\bibinfo {title} {Dynamical quantum
  phase transitions: a review},}\ }\href {\doibase 10.1088/1361-6633/aaaf9a}
  {\bibfield  {journal} {\bibinfo  {journal} {Reports on Progress in Physics}\
  }\textbf {\bibinfo {volume} {81}},\ \bibinfo {pages} {054001} (\bibinfo
  {year} {2018})}\BibitemShut {NoStop}%
\bibitem [{\citenamefont {Canovi}\ \emph {et~al.}(2014)\citenamefont {Canovi},
  \citenamefont {Werner},\ and\ \citenamefont {Eckstein}}]{Canovi_PRL}%
  \BibitemOpen
  \bibfield  {author} {\bibinfo {author} {\bibfnamefont {E.}~\bibnamefont
  {Canovi}}, \bibinfo {author} {\bibfnamefont {P.}~\bibnamefont {Werner}}, \
  and\ \bibinfo {author} {\bibfnamefont {M.}~\bibnamefont {Eckstein}},\
  }\bibfield  {title} {\enquote {\bibinfo {title} {First-order dynamical phase
  transitions},}\ }\href {\doibase 10.1103/PhysRevLett.113.265702} {\bibfield
  {journal} {\bibinfo  {journal} {Phys. Rev. Lett.}\ }\textbf {\bibinfo
  {volume} {113}},\ \bibinfo {pages} {265702} (\bibinfo {year}
  {2014})}\BibitemShut {NoStop}%
\bibitem [{\citenamefont {Bhattacharya}\ \emph {et~al.}(2017)\citenamefont
  {Bhattacharya}, \citenamefont {Bandyopadhyay},\ and\ \citenamefont
  {Dutta}}]{Utso_2017}%
  \BibitemOpen
  \bibfield  {author} {\bibinfo {author} {\bibfnamefont {U.}~\bibnamefont
  {Bhattacharya}}, \bibinfo {author} {\bibfnamefont {S.}~\bibnamefont
  {Bandyopadhyay}}, \ and\ \bibinfo {author} {\bibfnamefont {A.}~\bibnamefont
  {Dutta}},\ }\bibfield  {title} {\enquote {\bibinfo {title} {Mixed state
  dynamical quantum phase transitions},}\ }\href {\doibase
  10.1103/PhysRevB.96.180303} {\bibfield  {journal} {\bibinfo  {journal} {Phys.
  Rev. B}\ }\textbf {\bibinfo {volume} {96}},\ \bibinfo {pages} {180303}
  (\bibinfo {year} {2017})}\BibitemShut {NoStop}%
\bibitem [{\citenamefont {Jafari}\ \emph {et~al.}(2019)\citenamefont {Jafari},
  \citenamefont {Johannesson}, \citenamefont {Langari},\ and\ \citenamefont
  {Martin-Delgado}}]{Jafari_19feb}%
  \BibitemOpen
  \bibfield  {author} {\bibinfo {author} {\bibfnamefont {R.}~\bibnamefont
  {Jafari}}, \bibinfo {author} {\bibfnamefont {H.}~\bibnamefont {Johannesson}},
  \bibinfo {author} {\bibfnamefont {A.}~\bibnamefont {Langari}}, \ and\
  \bibinfo {author} {\bibfnamefont {M.~A.}\ \bibnamefont {Martin-Delgado}},\
  }\bibfield  {title} {\enquote {\bibinfo {title} {Quench dynamics and
  zero-energy modes: The case of the creutz model},}\ }\href {\doibase
  10.1103/PhysRevB.99.054302} {\bibfield  {journal} {\bibinfo  {journal} {Phys.
  Rev. B}\ }\textbf {\bibinfo {volume} {99}},\ \bibinfo {pages} {054302}
  (\bibinfo {year} {2019})}\BibitemShut {NoStop}%
\bibitem [{\citenamefont {Vajna}\ and\ \citenamefont {D\'ora}(2014)}]{Vajna14}%
  \BibitemOpen
  \bibfield  {author} {\bibinfo {author} {\bibfnamefont {S.}~\bibnamefont
  {Vajna}}\ and\ \bibinfo {author} {\bibfnamefont {B.}~\bibnamefont {D\'ora}},\
  }\bibfield  {title} {\enquote {\bibinfo {title} {Disentangling dynamical
  phase transitions from equilibrium phase transitions},}\ }\href {\doibase
  10.1103/PhysRevB.89.161105} {\bibfield  {journal} {\bibinfo  {journal} {Phys.
  Rev. B}\ }\textbf {\bibinfo {volume} {89}},\ \bibinfo {pages} {161105}
  (\bibinfo {year} {2014})}\BibitemShut {NoStop}%
\bibitem [{\citenamefont {Vajna}\ and\ \citenamefont
  {D\'ora}(2015{\natexlab{a}})}]{Dora2015}%
  \BibitemOpen
  \bibfield  {author} {\bibinfo {author} {\bibfnamefont {S.}~\bibnamefont
  {Vajna}}\ and\ \bibinfo {author} {\bibfnamefont {B.}~\bibnamefont {D\'ora}},\
  }\bibfield  {title} {\enquote {\bibinfo {title} {Topological classification
  of dynamical phase transitions},}\ }\href {\doibase
  10.1103/PhysRevB.91.155127} {\bibfield  {journal} {\bibinfo  {journal} {Phys.
  Rev. B}\ }\textbf {\bibinfo {volume} {91}},\ \bibinfo {pages} {155127}
  (\bibinfo {year} {2015}{\natexlab{a}})}\BibitemShut {NoStop}%
\bibitem [{\citenamefont {Khatun}\ and\ \citenamefont
  {Bhattacharjee}(2019)}]{Somendra_PRL}%
  \BibitemOpen
  \bibfield  {author} {\bibinfo {author} {\bibfnamefont {A.}~\bibnamefont
  {Khatun}}\ and\ \bibinfo {author} {\bibfnamefont {S.~M.}\ \bibnamefont
  {Bhattacharjee}},\ }\bibfield  {title} {\enquote {\bibinfo {title}
  {Boundaries and unphysical fixed points in dynamical quantum phase
  transitions},}\ }\href {\doibase 10.1103/PhysRevLett.123.160603} {\bibfield
  {journal} {\bibinfo  {journal} {Phys. Rev. Lett.}\ }\textbf {\bibinfo
  {volume} {123}},\ \bibinfo {pages} {160603} (\bibinfo {year}
  {2019})}\BibitemShut {NoStop}%
\bibitem [{\citenamefont {Nag}(2016)}]{Nag16}%
  \BibitemOpen
  \bibfield  {author} {\bibinfo {author} {\bibfnamefont {T.}~\bibnamefont
  {Nag}},\ }\bibfield  {title} {\enquote {\bibinfo {title} {Excess energy and
  decoherence factor of a qubit coupled to a one-dimensional periodically
  driven spin chain},}\ }\href {\doibase 10.1103/PhysRevE.93.062119} {\bibfield
   {journal} {\bibinfo  {journal} {Phys. Rev. E}\ }\textbf {\bibinfo {volume}
  {93}},\ \bibinfo {pages} {062119} (\bibinfo {year} {2016})}\BibitemShut
  {NoStop}%
\bibitem [{\citenamefont {Suzuki}\ \emph {et~al.}(2016)\citenamefont {Suzuki},
  \citenamefont {Nag},\ and\ \citenamefont {Dutta}}]{Suzuki16}%
  \BibitemOpen
  \bibfield  {author} {\bibinfo {author} {\bibfnamefont {S.}~\bibnamefont
  {Suzuki}}, \bibinfo {author} {\bibfnamefont {T.}~\bibnamefont {Nag}}, \ and\
  \bibinfo {author} {\bibfnamefont {A.}~\bibnamefont {Dutta}},\ }\bibfield
  {title} {\enquote {\bibinfo {title} {Dynamics of decoherence: Universal
  scaling of the decoherence factor},}\ }\href {\doibase
  10.1103/PhysRevA.93.012112} {\bibfield  {journal} {\bibinfo  {journal} {Phys.
  Rev. A}\ }\textbf {\bibinfo {volume} {93}},\ \bibinfo {pages} {012112}
  (\bibinfo {year} {2016})}\BibitemShut {NoStop}%
\bibitem [{\citenamefont {Sachdeva}\ \emph {et~al.}(2014)\citenamefont
  {Sachdeva}, \citenamefont {Nag}, \citenamefont {Agarwal},\ and\ \citenamefont
  {Dutta}}]{Sachdeva14}%
  \BibitemOpen
  \bibfield  {author} {\bibinfo {author} {\bibfnamefont {R.}~\bibnamefont
  {Sachdeva}}, \bibinfo {author} {\bibfnamefont {T.}~\bibnamefont {Nag}},
  \bibinfo {author} {\bibfnamefont {A.}~\bibnamefont {Agarwal}}, \ and\
  \bibinfo {author} {\bibfnamefont {A.}~\bibnamefont {Dutta}},\ }\bibfield
  {title} {\enquote {\bibinfo {title} {Finite-time interaction quench in a
  luttinger liquid},}\ }\href {\doibase 10.1103/PhysRevB.90.045421} {\bibfield
  {journal} {\bibinfo  {journal} {Phys. Rev. B}\ }\textbf {\bibinfo {volume}
  {90}},\ \bibinfo {pages} {045421} (\bibinfo {year} {2014})}\BibitemShut
  {NoStop}%
\bibitem [{\citenamefont {Nag}\ \emph {et~al.}(2012)\citenamefont {Nag},
  \citenamefont {Divakaran},\ and\ \citenamefont {Dutta}}]{Nag12}%
  \BibitemOpen
  \bibfield  {author} {\bibinfo {author} {\bibfnamefont {T.}~\bibnamefont
  {Nag}}, \bibinfo {author} {\bibfnamefont {U.}~\bibnamefont {Divakaran}}, \
  and\ \bibinfo {author} {\bibfnamefont {A.}~\bibnamefont {Dutta}},\ }\bibfield
   {title} {\enquote {\bibinfo {title} {Scaling of the decoherence factor of a
  qubit coupled to a spin chain driven across quantum critical points},}\
  }\href {\doibase 10.1103/PhysRevB.86.020401} {\bibfield  {journal} {\bibinfo
  {journal} {Phys. Rev. B}\ }\textbf {\bibinfo {volume} {86}},\ \bibinfo
  {pages} {020401} (\bibinfo {year} {2012})}\BibitemShut {NoStop}%
\bibitem [{\citenamefont {Quan}\ \emph {et~al.}(2006)\citenamefont {Quan},
  \citenamefont {Song}, \citenamefont {Liu}, \citenamefont {Zanardi},\ and\
  \citenamefont {Sun}}]{PhysRevLett.96.140604}%
  \BibitemOpen
  \bibfield  {author} {\bibinfo {author} {\bibfnamefont {H.~T.}\ \bibnamefont
  {Quan}}, \bibinfo {author} {\bibfnamefont {Z.}~\bibnamefont {Song}}, \bibinfo
  {author} {\bibfnamefont {X.~F.}\ \bibnamefont {Liu}}, \bibinfo {author}
  {\bibfnamefont {P.}~\bibnamefont {Zanardi}}, \ and\ \bibinfo {author}
  {\bibfnamefont {C.~P.}\ \bibnamefont {Sun}},\ }\bibfield  {title} {\enquote
  {\bibinfo {title} {Decay of loschmidt echo enhanced by quantum
  criticality},}\ }\href {\doibase 10.1103/PhysRevLett.96.140604} {\bibfield
  {journal} {\bibinfo  {journal} {Phys. Rev. Lett.}\ }\textbf {\bibinfo
  {volume} {96}},\ \bibinfo {pages} {140604} (\bibinfo {year}
  {2006})}\BibitemShut {NoStop}%
\bibitem [{\citenamefont {Cucchietti}\ \emph {et~al.}(2003)\citenamefont
  {Cucchietti}, \citenamefont {Dalvit}, \citenamefont {Paz},\ and\
  \citenamefont {Zurek}}]{Cucchietti03}%
  \BibitemOpen
  \bibfield  {author} {\bibinfo {author} {\bibfnamefont {F.~M.}\ \bibnamefont
  {Cucchietti}}, \bibinfo {author} {\bibfnamefont {D.~A.~R.}\ \bibnamefont
  {Dalvit}}, \bibinfo {author} {\bibfnamefont {J.~P.}\ \bibnamefont {Paz}}, \
  and\ \bibinfo {author} {\bibfnamefont {W.~H.}\ \bibnamefont {Zurek}},\
  }\bibfield  {title} {\enquote {\bibinfo {title} {Decoherence and the
  loschmidt echo},}\ }\href {\doibase 10.1103/PhysRevLett.91.210403} {\bibfield
   {journal} {\bibinfo  {journal} {Phys. Rev. Lett.}\ }\textbf {\bibinfo
  {volume} {91}},\ \bibinfo {pages} {210403} (\bibinfo {year}
  {2003})}\BibitemShut {NoStop}%
\bibitem [{\citenamefont {Jafari}\ and\ \citenamefont
  {Johannesson}(2017)}]{Jafari17}%
  \BibitemOpen
  \bibfield  {author} {\bibinfo {author} {\bibfnamefont {R.}~\bibnamefont
  {Jafari}}\ and\ \bibinfo {author} {\bibfnamefont {H.}~\bibnamefont
  {Johannesson}},\ }\bibfield  {title} {\enquote {\bibinfo {title} {Loschmidt
  echo revivals: Critical and noncritical},}\ }\href {\doibase
  10.1103/PhysRevLett.118.015701} {\bibfield  {journal} {\bibinfo  {journal}
  {Phys. Rev. Lett.}\ }\textbf {\bibinfo {volume} {118}},\ \bibinfo {pages}
  {015701} (\bibinfo {year} {2017})}\BibitemShut {NoStop}%
\bibitem [{\citenamefont {Uhrich}\ \emph {et~al.}(2020)\citenamefont {Uhrich},
  \citenamefont {Defenu}, \citenamefont {Jafari},\ and\ \citenamefont
  {Halimeh}}]{Uhrich20}%
  \BibitemOpen
  \bibfield  {author} {\bibinfo {author} {\bibfnamefont {P.}~\bibnamefont
  {Uhrich}}, \bibinfo {author} {\bibfnamefont {N.}~\bibnamefont {Defenu}},
  \bibinfo {author} {\bibfnamefont {R.}~\bibnamefont {Jafari}}, \ and\ \bibinfo
  {author} {\bibfnamefont {J.~C.}\ \bibnamefont {Halimeh}},\ }\bibfield
  {title} {\enquote {\bibinfo {title} {Out-of-equilibrium phase diagram of
  long-range superconductors},}\ }\href {\doibase 10.1103/PhysRevB.101.245148}
  {\bibfield  {journal} {\bibinfo  {journal} {Phys. Rev. B}\ }\textbf {\bibinfo
  {volume} {101}},\ \bibinfo {pages} {245148} (\bibinfo {year}
  {2020})}\BibitemShut {NoStop}%
\bibitem [{\citenamefont {Schmitt}\ and\ \citenamefont
  {Kehrein}(2015)}]{Schmitt15}%
  \BibitemOpen
  \bibfield  {author} {\bibinfo {author} {\bibfnamefont {M.}~\bibnamefont
  {Schmitt}}\ and\ \bibinfo {author} {\bibfnamefont {S.}~\bibnamefont
  {Kehrein}},\ }\bibfield  {title} {\enquote {\bibinfo {title} {Dynamical
  quantum phase transitions in the kitaev honeycomb model},}\ }\href {\doibase
  10.1103/PhysRevB.92.075114} {\bibfield  {journal} {\bibinfo  {journal} {Phys.
  Rev. B}\ }\textbf {\bibinfo {volume} {92}},\ \bibinfo {pages} {075114}
  (\bibinfo {year} {2015})}\BibitemShut {NoStop}%
\bibitem [{\citenamefont {Halimeh}\ and\ \citenamefont
  {Zauner-Stauber}(2017)}]{Halimeh17}%
  \BibitemOpen
  \bibfield  {author} {\bibinfo {author} {\bibfnamefont {J.~C.}\ \bibnamefont
  {Halimeh}}\ and\ \bibinfo {author} {\bibfnamefont {V.}~\bibnamefont
  {Zauner-Stauber}},\ }\bibfield  {title} {\enquote {\bibinfo {title}
  {Dynamical phase diagram of quantum spin chains with long-range
  interactions},}\ }\href {\doibase 10.1103/PhysRevB.96.134427} {\bibfield
  {journal} {\bibinfo  {journal} {Phys. Rev. B}\ }\textbf {\bibinfo {volume}
  {96}},\ \bibinfo {pages} {134427} (\bibinfo {year} {2017})}\BibitemShut
  {NoStop}%
\bibitem [{\citenamefont {\ifmmode \check{Z}\else
  \v{Z}\fi{}unkovi\ifmmode~\check{c}\else \v{c}\fi{}}\ \emph
  {et~al.}(2018)\citenamefont {\ifmmode \check{Z}\else
  \v{Z}\fi{}unkovi\ifmmode~\check{c}\else \v{c}\fi{}}, \citenamefont {Heyl},
  \citenamefont {Knap},\ and\ \citenamefont {Silva}}]{Silva18}%
  \BibitemOpen
  \bibfield  {author} {\bibinfo {author} {\bibfnamefont {B.}~\bibnamefont
  {\ifmmode \check{Z}\else \v{Z}\fi{}unkovi\ifmmode~\check{c}\else
  \v{c}\fi{}}}, \bibinfo {author} {\bibfnamefont {M.}~\bibnamefont {Heyl}},
  \bibinfo {author} {\bibfnamefont {M.}~\bibnamefont {Knap}}, \ and\ \bibinfo
  {author} {\bibfnamefont {A.}~\bibnamefont {Silva}},\ }\bibfield  {title}
  {\enquote {\bibinfo {title} {Dynamical quantum phase transitions in spin
  chains with long-range interactions: Merging different concepts of
  nonequilibrium criticality},}\ }\href {\doibase
  10.1103/PhysRevLett.120.130601} {\bibfield  {journal} {\bibinfo  {journal}
  {Phys. Rev. Lett.}\ }\textbf {\bibinfo {volume} {120}},\ \bibinfo {pages}
  {130601} (\bibinfo {year} {2018})}\BibitemShut {NoStop}%
\bibitem [{\citenamefont {Halimeh}\ \emph {et~al.}(2020)\citenamefont
  {Halimeh}, \citenamefont {Van~Damme}, \citenamefont {Zauner-Stauber},\ and\
  \citenamefont {Vanderstraeten}}]{Halimeh20c}%
  \BibitemOpen
  \bibfield  {author} {\bibinfo {author} {\bibfnamefont {J.~C.}\ \bibnamefont
  {Halimeh}}, \bibinfo {author} {\bibfnamefont {M.}~\bibnamefont {Van~Damme}},
  \bibinfo {author} {\bibfnamefont {V.}~\bibnamefont {Zauner-Stauber}}, \ and\
  \bibinfo {author} {\bibfnamefont {L.}~\bibnamefont {Vanderstraeten}},\
  }\bibfield  {title} {\enquote {\bibinfo {title} {Quasiparticle origin of
  dynamical quantum phase transitions},}\ }\href {\doibase
  10.1103/PhysRevResearch.2.033111} {\bibfield  {journal} {\bibinfo  {journal}
  {Phys. Rev. Research}\ }\textbf {\bibinfo {volume} {2}},\ \bibinfo {pages}
  {033111} (\bibinfo {year} {2020})}\BibitemShut {NoStop}%
\bibitem [{\citenamefont {Hashizume}\ \emph {et~al.}(2022)\citenamefont
  {Hashizume}, \citenamefont {McCulloch},\ and\ \citenamefont
  {Halimeh}}]{Hashizume2022}%
  \BibitemOpen
  \bibfield  {author} {\bibinfo {author} {\bibfnamefont {T.}~\bibnamefont
  {Hashizume}}, \bibinfo {author} {\bibfnamefont {I.~P.}\ \bibnamefont
  {McCulloch}}, \ and\ \bibinfo {author} {\bibfnamefont {J.~C.}\ \bibnamefont
  {Halimeh}},\ }\bibfield  {title} {\enquote {\bibinfo {title} {Dynamical phase
  transitions in the two-dimensional transverse-field ising model},}\ }\href
  {\doibase 10.1103/PhysRevResearch.4.013250} {\bibfield  {journal} {\bibinfo
  {journal} {Phys. Rev. Research}\ }\textbf {\bibinfo {volume} {4}},\ \bibinfo
  {pages} {013250} (\bibinfo {year} {2022})}\BibitemShut {NoStop}%
\bibitem [{\citenamefont {Lang}\ \emph
  {et~al.}(2018{\natexlab{a}})\citenamefont {Lang}, \citenamefont {Frank},\
  and\ \citenamefont {Halimeh}}]{Lang_2018}%
  \BibitemOpen
  \bibfield  {author} {\bibinfo {author} {\bibfnamefont {J.}~\bibnamefont
  {Lang}}, \bibinfo {author} {\bibfnamefont {B.}~\bibnamefont {Frank}}, \ and\
  \bibinfo {author} {\bibfnamefont {J.~C.}\ \bibnamefont {Halimeh}},\
  }\bibfield  {title} {\enquote {\bibinfo {title} {Concurrence of dynamical
  phase transitions at finite temperature in the fully connected
  transverse-field ising model},}\ }\href {\doibase 10.1103/PhysRevB.97.174401}
  {\bibfield  {journal} {\bibinfo  {journal} {Phys. Rev. B}\ }\textbf {\bibinfo
  {volume} {97}},\ \bibinfo {pages} {174401} (\bibinfo {year}
  {2018}{\natexlab{a}})}\BibitemShut {NoStop}%
\bibitem [{\citenamefont {Homrighausen}\ \emph {et~al.}(2017)\citenamefont
  {Homrighausen}, \citenamefont {Abeling}, \citenamefont {Zauner-Stauber},\
  and\ \citenamefont {Halimeh}}]{Homrighausen_17}%
  \BibitemOpen
  \bibfield  {author} {\bibinfo {author} {\bibfnamefont {I.}~\bibnamefont
  {Homrighausen}}, \bibinfo {author} {\bibfnamefont {N.~O.}\ \bibnamefont
  {Abeling}}, \bibinfo {author} {\bibfnamefont {V.}~\bibnamefont
  {Zauner-Stauber}}, \ and\ \bibinfo {author} {\bibfnamefont {J.~C.}\
  \bibnamefont {Halimeh}},\ }\bibfield  {title} {\enquote {\bibinfo {title}
  {Anomalous dynamical phase in quantum spin chains with long-range
  interactions},}\ }\href {\doibase 10.1103/PhysRevB.96.104436} {\bibfield
  {journal} {\bibinfo  {journal} {Phys. Rev. B}\ }\textbf {\bibinfo {volume}
  {96}},\ \bibinfo {pages} {104436} (\bibinfo {year} {2017})}\BibitemShut
  {NoStop}%
\bibitem [{\citenamefont {Rossi}\ and\ \citenamefont
  {Dolcini}(2022)}]{PhysRevB.106.045410}%
  \BibitemOpen
  \bibfield  {author} {\bibinfo {author} {\bibfnamefont {L.}~\bibnamefont
  {Rossi}}\ and\ \bibinfo {author} {\bibfnamefont {F.}~\bibnamefont
  {Dolcini}},\ }\bibfield  {title} {\enquote {\bibinfo {title} {Nonlinear
  current and dynamical quantum phase transitions in the flux-quenched
  su-schrieffer-heeger model},}\ }\href {\doibase 10.1103/PhysRevB.106.045410}
  {\bibfield  {journal} {\bibinfo  {journal} {Phys. Rev. B}\ }\textbf {\bibinfo
  {volume} {106}},\ \bibinfo {pages} {045410} (\bibinfo {year}
  {2022})}\BibitemShut {NoStop}%
\bibitem [{\citenamefont {Mishra}\ \emph {et~al.}(2020)\citenamefont {Mishra},
  \citenamefont {Jafari},\ and\ \citenamefont {Akbari}}]{mishra2020disordered}%
  \BibitemOpen
  \bibfield  {author} {\bibinfo {author} {\bibfnamefont {U.}~\bibnamefont
  {Mishra}}, \bibinfo {author} {\bibfnamefont {R.}~\bibnamefont {Jafari}}, \
  and\ \bibinfo {author} {\bibfnamefont {A.}~\bibnamefont {Akbari}},\
  }\bibfield  {title} {\enquote {\bibinfo {title} {Disordered kitaev chain with
  long-range pairing: Loschmidt echo revivals and dynamical phase
  transitions},}\ }\href@noop {} {\bibfield  {journal} {\bibinfo  {journal}
  {Journal of Physics A: Mathematical and Theoretical}\ }\textbf {\bibinfo
  {volume} {53}},\ \bibinfo {pages} {375301} (\bibinfo {year}
  {2020})}\BibitemShut {NoStop}%
\bibitem [{\citenamefont {Divakaran}\ \emph {et~al.}(2016)\citenamefont
  {Divakaran}, \citenamefont {Sharma},\ and\ \citenamefont
  {Dutta}}]{Divakaran16}%
  \BibitemOpen
  \bibfield  {author} {\bibinfo {author} {\bibfnamefont {U.}~\bibnamefont
  {Divakaran}}, \bibinfo {author} {\bibfnamefont {S.}~\bibnamefont {Sharma}}, \
  and\ \bibinfo {author} {\bibfnamefont {A.}~\bibnamefont {Dutta}},\ }\bibfield
   {title} {\enquote {\bibinfo {title} {Tuning the presence of dynamical phase
  transitions in a generalized $xy$ spin chain},}\ }\href {\doibase
  10.1103/PhysRevE.93.052133} {\bibfield  {journal} {\bibinfo  {journal} {Phys.
  Rev. E}\ }\textbf {\bibinfo {volume} {93}},\ \bibinfo {pages} {052133}
  (\bibinfo {year} {2016})}\BibitemShut {NoStop}%
\bibitem [{\citenamefont {Sharma}\ \emph {et~al.}(2016)\citenamefont {Sharma},
  \citenamefont {Divakaran}, \citenamefont {Polkovnikov},\ and\ \citenamefont
  {Dutta}}]{SS}%
  \BibitemOpen
  \bibfield  {author} {\bibinfo {author} {\bibfnamefont {S.}~\bibnamefont
  {Sharma}}, \bibinfo {author} {\bibfnamefont {U.}~\bibnamefont {Divakaran}},
  \bibinfo {author} {\bibfnamefont {A.}~\bibnamefont {Polkovnikov}}, \ and\
  \bibinfo {author} {\bibfnamefont {A.}~\bibnamefont {Dutta}},\ }\bibfield
  {title} {\enquote {\bibinfo {title} {Slow quenches in a quantum ising chain:
  Dynamical phase transitions and topology},}\ }\href {\doibase
  10.1103/PhysRevB.93.144306} {\bibfield  {journal} {\bibinfo  {journal} {Phys.
  Rev. B}\ }\textbf {\bibinfo {volume} {93}},\ \bibinfo {pages} {144306}
  (\bibinfo {year} {2016})}\BibitemShut {NoStop}%
\bibitem [{\citenamefont {Sharma}\ \emph {et~al.}(2015)\citenamefont {Sharma},
  \citenamefont {Suzuki},\ and\ \citenamefont {Dutta}}]{PhysRevB.92.104306}%
  \BibitemOpen
  \bibfield  {author} {\bibinfo {author} {\bibfnamefont {S.}~\bibnamefont
  {Sharma}}, \bibinfo {author} {\bibfnamefont {S.}~\bibnamefont {Suzuki}}, \
  and\ \bibinfo {author} {\bibfnamefont {A.}~\bibnamefont {Dutta}},\ }\bibfield
   {title} {\enquote {\bibinfo {title} {Quenches and dynamical phase
  transitions in a nonintegrable quantum ising model},}\ }\href {\doibase
  10.1103/PhysRevB.92.104306} {\bibfield  {journal} {\bibinfo  {journal} {Phys.
  Rev. B}\ }\textbf {\bibinfo {volume} {92}},\ \bibinfo {pages} {104306}
  (\bibinfo {year} {2015})}\BibitemShut {NoStop}%
\bibitem [{\citenamefont {Dutta}\ and\ \citenamefont {Dutta}(2017)}]{Dutta17}%
  \BibitemOpen
  \bibfield  {author} {\bibinfo {author} {\bibfnamefont {A.}~\bibnamefont
  {Dutta}}\ and\ \bibinfo {author} {\bibfnamefont {A.}~\bibnamefont {Dutta}},\
  }\bibfield  {title} {\enquote {\bibinfo {title} {Probing the role of
  long-range interactions in the dynamics of a long-range kitaev chain},}\
  }\href {\doibase 10.1103/PhysRevB.96.125113} {\bibfield  {journal} {\bibinfo
  {journal} {Phys. Rev. B}\ }\textbf {\bibinfo {volume} {96}},\ \bibinfo
  {pages} {125113} (\bibinfo {year} {2017})}\BibitemShut {NoStop}%
\bibitem [{\citenamefont {Zamani}\ \emph {et~al.}(2020)\citenamefont {Zamani},
  \citenamefont {Jafari},\ and\ \citenamefont {Langari}}]{Zamani20}%
  \BibitemOpen
  \bibfield  {author} {\bibinfo {author} {\bibfnamefont {S.}~\bibnamefont
  {Zamani}}, \bibinfo {author} {\bibfnamefont {R.}~\bibnamefont {Jafari}}, \
  and\ \bibinfo {author} {\bibfnamefont {A.}~\bibnamefont {Langari}},\
  }\bibfield  {title} {\enquote {\bibinfo {title} {Floquet dynamical quantum
  phase transition in the extended xy model: Nonadiabatic to adiabatic
  topological transition},}\ }\href {\doibase 10.1103/PhysRevB.102.144306}
  {\bibfield  {journal} {\bibinfo  {journal} {Phys. Rev. B}\ }\textbf {\bibinfo
  {volume} {102}},\ \bibinfo {pages} {144306} (\bibinfo {year}
  {2020})}\BibitemShut {NoStop}%
\bibitem [{\citenamefont {Jafari}\ and\ \citenamefont
  {Akbari}(2021)}]{Jafari21}%
  \BibitemOpen
  \bibfield  {author} {\bibinfo {author} {\bibfnamefont {R.}~\bibnamefont
  {Jafari}}\ and\ \bibinfo {author} {\bibfnamefont {A.}~\bibnamefont
  {Akbari}},\ }\bibfield  {title} {\enquote {\bibinfo {title} {Floquet
  dynamical phase transition and entanglement spectrum},}\ }\href {\doibase
  10.1103/PhysRevA.103.012204} {\bibfield  {journal} {\bibinfo  {journal}
  {Phys. Rev. A}\ }\textbf {\bibinfo {volume} {103}},\ \bibinfo {pages}
  {012204} (\bibinfo {year} {2021})}\BibitemShut {NoStop}%
\bibitem [{\citenamefont {Jafari}\ \emph {et~al.}(2022)\citenamefont {Jafari},
  \citenamefont {Akbari}, \citenamefont {Mishra},\ and\ \citenamefont
  {Johannesson}}]{Jafari22}%
  \BibitemOpen
  \bibfield  {author} {\bibinfo {author} {\bibfnamefont {R.}~\bibnamefont
  {Jafari}}, \bibinfo {author} {\bibfnamefont {A.}~\bibnamefont {Akbari}},
  \bibinfo {author} {\bibfnamefont {U.}~\bibnamefont {Mishra}}, \ and\ \bibinfo
  {author} {\bibfnamefont {H.}~\bibnamefont {Johannesson}},\ }\bibfield
  {title} {\enquote {\bibinfo {title} {Floquet dynamical quantum phase
  transitions under synchronized periodic driving},}\ }\href {\doibase
  10.1103/PhysRevB.105.094311} {\bibfield  {journal} {\bibinfo  {journal}
  {Phys. Rev. B}\ }\textbf {\bibinfo {volume} {105}},\ \bibinfo {pages}
  {094311} (\bibinfo {year} {2022})}\BibitemShut {NoStop}%
\bibitem [{\citenamefont {Naji}\ \emph
  {et~al.}(2022{\natexlab{a}})\citenamefont {Naji}, \citenamefont {Jafari},
  \citenamefont {Zhou},\ and\ \citenamefont {Langari}}]{Naji22}%
  \BibitemOpen
  \bibfield  {author} {\bibinfo {author} {\bibfnamefont {J.}~\bibnamefont
  {Naji}}, \bibinfo {author} {\bibfnamefont {R.}~\bibnamefont {Jafari}},
  \bibinfo {author} {\bibfnamefont {L.}~\bibnamefont {Zhou}}, \ and\ \bibinfo
  {author} {\bibfnamefont {A.}~\bibnamefont {Langari}},\ }\bibfield  {title}
  {\enquote {\bibinfo {title} {Engineering floquet dynamical quantum phase
  transitions},}\ }\href {\doibase 10.1103/PhysRevB.106.094314} {\bibfield
  {journal} {\bibinfo  {journal} {Phys. Rev. B}\ }\textbf {\bibinfo {volume}
  {106}},\ \bibinfo {pages} {094314} (\bibinfo {year}
  {2022}{\natexlab{a}})}\BibitemShut {NoStop}%
\bibitem [{\citenamefont {Yang}\ \emph {et~al.}(2019)\citenamefont {Yang},
  \citenamefont {Zhou}, \citenamefont {Ma}, \citenamefont {Kong}, \citenamefont
  {Wang}, \citenamefont {Qin}, \citenamefont {Rong}, \citenamefont {Wang},
  \citenamefont {Shi}, \citenamefont {Gong},\ and\ \citenamefont
  {Du}}]{Yang19}%
  \BibitemOpen
  \bibfield  {author} {\bibinfo {author} {\bibfnamefont {K.}~\bibnamefont
  {Yang}}, \bibinfo {author} {\bibfnamefont {L.}~\bibnamefont {Zhou}}, \bibinfo
  {author} {\bibfnamefont {W.}~\bibnamefont {Ma}}, \bibinfo {author}
  {\bibfnamefont {X.}~\bibnamefont {Kong}}, \bibinfo {author} {\bibfnamefont
  {P.}~\bibnamefont {Wang}}, \bibinfo {author} {\bibfnamefont {X.}~\bibnamefont
  {Qin}}, \bibinfo {author} {\bibfnamefont {X.}~\bibnamefont {Rong}}, \bibinfo
  {author} {\bibfnamefont {Y.}~\bibnamefont {Wang}}, \bibinfo {author}
  {\bibfnamefont {F.}~\bibnamefont {Shi}}, \bibinfo {author} {\bibfnamefont
  {J.}~\bibnamefont {Gong}}, \ and\ \bibinfo {author} {\bibfnamefont
  {J.}~\bibnamefont {Du}},\ }\bibfield  {title} {\enquote {\bibinfo {title}
  {Floquet dynamical quantum phase transitions},}\ }\href {\doibase
  10.1103/PhysRevB.100.085308} {\bibfield  {journal} {\bibinfo  {journal}
  {Phys. Rev. B}\ }\textbf {\bibinfo {volume} {100}},\ \bibinfo {pages}
  {085308} (\bibinfo {year} {2019})}\BibitemShut {NoStop}%
\bibitem [{\citenamefont {Zhou}\ and\ \citenamefont
  {Du}(2021{\natexlab{a}})}]{zhou2021floquet}%
  \BibitemOpen
  \bibfield  {author} {\bibinfo {author} {\bibfnamefont {L.}~\bibnamefont
  {Zhou}}\ and\ \bibinfo {author} {\bibfnamefont {Q.}~\bibnamefont {Du}},\
  }\bibfield  {title} {\enquote {\bibinfo {title} {Floquet dynamical quantum
  phase transitions in periodically quenched systems},}\ }\href@noop {}
  {\bibfield  {journal} {\bibinfo  {journal} {Journal of Physics: Condensed
  Matter}\ }\textbf {\bibinfo {volume} {33}},\ \bibinfo {pages} {345403}
  (\bibinfo {year} {2021}{\natexlab{a}})}\BibitemShut {NoStop}%
\bibitem [{\citenamefont {Palmai}(2015)}]{Palmai_2015}%
  \BibitemOpen
  \bibfield  {author} {\bibinfo {author} {\bibfnamefont {T.}~\bibnamefont
  {Palmai}},\ }\bibfield  {title} {\enquote {\bibinfo {title} {Edge exponents
  in work statistics out of equilibrium and dynamical phase transitions from
  scattering theory in one-dimensional gapped systems},}\ }\href {\doibase
  10.1103/PhysRevB.92.235433} {\bibfield  {journal} {\bibinfo  {journal} {Phys.
  Rev. B}\ }\textbf {\bibinfo {volume} {92}},\ \bibinfo {pages} {235433}
  (\bibinfo {year} {2015})}\BibitemShut {NoStop}%
\bibitem [{\citenamefont {Andraschko}\ and\ \citenamefont
  {Sirker}(2014)}]{Andraschko}%
  \BibitemOpen
  \bibfield  {author} {\bibinfo {author} {\bibfnamefont {F.}~\bibnamefont
  {Andraschko}}\ and\ \bibinfo {author} {\bibfnamefont {J.}~\bibnamefont
  {Sirker}},\ }\bibfield  {title} {\enquote {\bibinfo {title} {Dynamical
  quantum phase transitions and the loschmidt echo: A transfer matrix
  approach},}\ }\href {\doibase 10.1103/PhysRevB.89.125120} {\bibfield
  {journal} {\bibinfo  {journal} {Phys. Rev. B}\ }\textbf {\bibinfo {volume}
  {89}},\ \bibinfo {pages} {125120} (\bibinfo {year} {2014})}\BibitemShut
  {NoStop}%
\bibitem [{\citenamefont {Modak}\ and\ \citenamefont
  {Rakshit}(2021)}]{Modak21}%
  \BibitemOpen
  \bibfield  {author} {\bibinfo {author} {\bibfnamefont {R.}~\bibnamefont
  {Modak}}\ and\ \bibinfo {author} {\bibfnamefont {D.}~\bibnamefont
  {Rakshit}},\ }\bibfield  {title} {\enquote {\bibinfo {title} {Many-body
  dynamical phase transition in a quasiperiodic potential},}\ }\href {\doibase
  10.1103/PhysRevB.103.224310} {\bibfield  {journal} {\bibinfo  {journal}
  {Phys. Rev. B}\ }\textbf {\bibinfo {volume} {103}},\ \bibinfo {pages}
  {224310} (\bibinfo {year} {2021})}\BibitemShut {NoStop}%
\bibitem [{\citenamefont {Abdi}(2019)}]{Abdi19}%
  \BibitemOpen
  \bibfield  {author} {\bibinfo {author} {\bibfnamefont {M.}~\bibnamefont
  {Abdi}},\ }\bibfield  {title} {\enquote {\bibinfo {title} {Dynamical quantum
  phase transition in bose-einstein condensates},}\ }\href {\doibase
  10.1103/PhysRevB.100.184310} {\bibfield  {journal} {\bibinfo  {journal}
  {Phys. Rev. B}\ }\textbf {\bibinfo {volume} {100}},\ \bibinfo {pages}
  {184310} (\bibinfo {year} {2019})}\BibitemShut {NoStop}%
\bibitem [{\citenamefont {Syed}\ \emph {et~al.}(2021)\citenamefont {Syed},
  \citenamefont {Enss},\ and\ \citenamefont {Defenu}}]{Seyd_21}%
  \BibitemOpen
  \bibfield  {author} {\bibinfo {author} {\bibfnamefont {M.}~\bibnamefont
  {Syed}}, \bibinfo {author} {\bibfnamefont {T.}~\bibnamefont {Enss}}, \ and\
  \bibinfo {author} {\bibfnamefont {N.}~\bibnamefont {Defenu}},\ }\bibfield
  {title} {\enquote {\bibinfo {title} {Dynamical quantum phase transition in a
  bosonic system with long-range interactions},}\ }\href {\doibase
  10.1103/PhysRevB.103.064306} {\bibfield  {journal} {\bibinfo  {journal}
  {Phys. Rev. B}\ }\textbf {\bibinfo {volume} {103}},\ \bibinfo {pages}
  {064306} (\bibinfo {year} {2021})}\BibitemShut {NoStop}%
\bibitem [{\citenamefont {Stumper}\ \emph {et~al.}(2022)\citenamefont
  {Stumper}, \citenamefont {Thoss},\ and\ \citenamefont {Okamoto}}]{Stumper22}%
  \BibitemOpen
  \bibfield  {author} {\bibinfo {author} {\bibfnamefont {S.}~\bibnamefont
  {Stumper}}, \bibinfo {author} {\bibfnamefont {M.}~\bibnamefont {Thoss}}, \
  and\ \bibinfo {author} {\bibfnamefont {J.}~\bibnamefont {Okamoto}},\
  }\bibfield  {title} {\enquote {\bibinfo {title} {Interaction-driven dynamical
  quantum phase transitions in a strongly correlated bosonic system},}\ }\href
  {\doibase 10.1103/PhysRevResearch.4.013002} {\bibfield  {journal} {\bibinfo
  {journal} {Phys. Rev. Research}\ }\textbf {\bibinfo {volume} {4}},\ \bibinfo
  {pages} {013002} (\bibinfo {year} {2022})}\BibitemShut {NoStop}%
\bibitem [{\citenamefont {Kosior}\ \emph {et~al.}(2018)\citenamefont {Kosior},
  \citenamefont {Syrwid},\ and\ \citenamefont {Sacha}}]{Kosior18}%
  \BibitemOpen
  \bibfield  {author} {\bibinfo {author} {\bibfnamefont {A.}~\bibnamefont
  {Kosior}}, \bibinfo {author} {\bibfnamefont {A.}~\bibnamefont {Syrwid}}, \
  and\ \bibinfo {author} {\bibfnamefont {K.}~\bibnamefont {Sacha}},\ }\bibfield
   {title} {\enquote {\bibinfo {title} {Dynamical quantum phase transitions in
  systems with broken continuous time and space translation symmetries},}\
  }\href {\doibase 10.1103/PhysRevA.98.023612} {\bibfield  {journal} {\bibinfo
  {journal} {Phys. Rev. A}\ }\textbf {\bibinfo {volume} {98}},\ \bibinfo
  {pages} {023612} (\bibinfo {year} {2018})}\BibitemShut {NoStop}%
\bibitem [{\citenamefont {Kosior}\ and\ \citenamefont
  {Sacha}(2018)}]{Kosior18b}%
  \BibitemOpen
  \bibfield  {author} {\bibinfo {author} {\bibfnamefont {A.}~\bibnamefont
  {Kosior}}\ and\ \bibinfo {author} {\bibfnamefont {K.}~\bibnamefont {Sacha}},\
  }\bibfield  {title} {\enquote {\bibinfo {title} {Dynamical quantum phase
  transitions in discrete time crystals},}\ }\href {\doibase
  10.1103/PhysRevA.97.053621} {\bibfield  {journal} {\bibinfo  {journal} {Phys.
  Rev. A}\ }\textbf {\bibinfo {volume} {97}},\ \bibinfo {pages} {053621}
  (\bibinfo {year} {2018})}\BibitemShut {NoStop}%
\bibitem [{\citenamefont {Bergholtz}\ and\ \citenamefont
  {Budich}(2019)}]{Bergholtz19}%
  \BibitemOpen
  \bibfield  {author} {\bibinfo {author} {\bibfnamefont {E.~J.}\ \bibnamefont
  {Bergholtz}}\ and\ \bibinfo {author} {\bibfnamefont {J.~C.}\ \bibnamefont
  {Budich}},\ }\bibfield  {title} {\enquote {\bibinfo {title} {Non-hermitian
  weyl physics in topological insulator ferromagnet junctions},}\ }\href
  {\doibase 10.1103/PhysRevResearch.1.012003} {\bibfield  {journal} {\bibinfo
  {journal} {Phys. Rev. Research}\ }\textbf {\bibinfo {volume} {1}},\ \bibinfo
  {pages} {012003} (\bibinfo {year} {2019})}\BibitemShut {NoStop}%
\bibitem [{\citenamefont {Yang}\ \emph {et~al.}(2021)\citenamefont {Yang},
  \citenamefont {Morampudi},\ and\ \citenamefont {Bergholtz}}]{Yang21}%
  \BibitemOpen
  \bibfield  {author} {\bibinfo {author} {\bibfnamefont {K.}~\bibnamefont
  {Yang}}, \bibinfo {author} {\bibfnamefont {S.~C.}\ \bibnamefont {Morampudi}},
  \ and\ \bibinfo {author} {\bibfnamefont {E.~J.}\ \bibnamefont {Bergholtz}},\
  }\bibfield  {title} {\enquote {\bibinfo {title} {Exceptional spin liquids
  from couplings to the environment},}\ }\href {\doibase
  10.1103/PhysRevLett.126.077201} {\bibfield  {journal} {\bibinfo  {journal}
  {Phys. Rev. Lett.}\ }\textbf {\bibinfo {volume} {126}},\ \bibinfo {pages}
  {077201} (\bibinfo {year} {2021})}\BibitemShut {NoStop}%
\bibitem [{\citenamefont {Kozii}\ and\ \citenamefont
  {Fu}(2017)}]{kozii2017non}%
  \BibitemOpen
  \bibfield  {author} {\bibinfo {author} {\bibfnamefont {V.}~\bibnamefont
  {Kozii}}\ and\ \bibinfo {author} {\bibfnamefont {L.}~\bibnamefont {Fu}},\
  }\bibfield  {title} {\enquote {\bibinfo {title} {Non-hermitian topological
  theory of finite-lifetime quasiparticles: prediction of bulk fermi arc due to
  exceptional point},}\ }\href@noop {} {\bibfield  {journal} {\bibinfo
  {journal} {arXiv preprint arXiv:1708.05841}\ } (\bibinfo {year}
  {2017})}\BibitemShut {NoStop}%
\bibitem [{\citenamefont {Yoshida}\ \emph {et~al.}(2018)\citenamefont
  {Yoshida}, \citenamefont {Peters},\ and\ \citenamefont
  {Kawakami}}]{Yoshida18}%
  \BibitemOpen
  \bibfield  {author} {\bibinfo {author} {\bibfnamefont {T.}~\bibnamefont
  {Yoshida}}, \bibinfo {author} {\bibfnamefont {R.}~\bibnamefont {Peters}}, \
  and\ \bibinfo {author} {\bibfnamefont {N.}~\bibnamefont {Kawakami}},\
  }\bibfield  {title} {\enquote {\bibinfo {title} {Non-hermitian perspective of
  the band structure in heavy-fermion systems},}\ }\href {\doibase
  10.1103/PhysRevB.98.035141} {\bibfield  {journal} {\bibinfo  {journal} {Phys.
  Rev. B}\ }\textbf {\bibinfo {volume} {98}},\ \bibinfo {pages} {035141}
  (\bibinfo {year} {2018})}\BibitemShut {NoStop}%
\bibitem [{\citenamefont {Shen}\ \emph {et~al.}(2018)\citenamefont {Shen},
  \citenamefont {Zhen},\ and\ \citenamefont {Fu}}]{Shen18}%
  \BibitemOpen
  \bibfield  {author} {\bibinfo {author} {\bibfnamefont {H.}~\bibnamefont
  {Shen}}, \bibinfo {author} {\bibfnamefont {B.}~\bibnamefont {Zhen}}, \ and\
  \bibinfo {author} {\bibfnamefont {L.}~\bibnamefont {Fu}},\ }\bibfield
  {title} {\enquote {\bibinfo {title} {Topological band theory for
  non-hermitian hamiltonians},}\ }\href {\doibase
  10.1103/PhysRevLett.120.146402} {\bibfield  {journal} {\bibinfo  {journal}
  {Phys. Rev. Lett.}\ }\textbf {\bibinfo {volume} {120}},\ \bibinfo {pages}
  {146402} (\bibinfo {year} {2018})}\BibitemShut {NoStop}%
\bibitem [{\citenamefont {Bergholtz}\ \emph {et~al.}(2021)\citenamefont
  {Bergholtz}, \citenamefont {Budich},\ and\ \citenamefont
  {Kunst}}]{Bergholtz21}%
  \BibitemOpen
  \bibfield  {author} {\bibinfo {author} {\bibfnamefont {E.~J.}\ \bibnamefont
  {Bergholtz}}, \bibinfo {author} {\bibfnamefont {J.~C.}\ \bibnamefont
  {Budich}}, \ and\ \bibinfo {author} {\bibfnamefont {F.~K.}\ \bibnamefont
  {Kunst}},\ }\bibfield  {title} {\enquote {\bibinfo {title} {Exceptional
  topology of non-hermitian systems},}\ }\href {\doibase
  10.1103/RevModPhys.93.015005} {\bibfield  {journal} {\bibinfo  {journal}
  {Rev. Mod. Phys.}\ }\textbf {\bibinfo {volume} {93}},\ \bibinfo {pages}
  {015005} (\bibinfo {year} {2021})}\BibitemShut {NoStop}%
\bibitem [{\citenamefont {Ghatak}\ and\ \citenamefont
  {Das}(2019)}]{ghatak2019new}%
  \BibitemOpen
  \bibfield  {author} {\bibinfo {author} {\bibfnamefont {A.}~\bibnamefont
  {Ghatak}}\ and\ \bibinfo {author} {\bibfnamefont {T.}~\bibnamefont {Das}},\
  }\bibfield  {title} {\enquote {\bibinfo {title} {New topological invariants
  in non-hermitian systems},}\ }\href@noop {} {\bibfield  {journal} {\bibinfo
  {journal} {Journal of Physics: Condensed Matter}\ }\textbf {\bibinfo {volume}
  {31}},\ \bibinfo {pages} {263001} (\bibinfo {year} {2019})}\BibitemShut
  {NoStop}%
\bibitem [{\citenamefont {Ashida}\ \emph {et~al.}(2020)\citenamefont {Ashida},
  \citenamefont {Gong},\ and\ \citenamefont {Ueda}}]{ashida2020non}%
  \BibitemOpen
  \bibfield  {author} {\bibinfo {author} {\bibfnamefont {Y.}~\bibnamefont
  {Ashida}}, \bibinfo {author} {\bibfnamefont {Z.}~\bibnamefont {Gong}}, \ and\
  \bibinfo {author} {\bibfnamefont {M.}~\bibnamefont {Ueda}},\ }\bibfield
  {title} {\enquote {\bibinfo {title} {Non-hermitian physics},}\ }\href@noop {}
  {\bibfield  {journal} {\bibinfo  {journal} {Advances in Physics}\ }\textbf
  {\bibinfo {volume} {69}},\ \bibinfo {pages} {249--435} (\bibinfo {year}
  {2020})}\BibitemShut {NoStop}%
\bibitem [{\citenamefont {Kawabata}\ \emph {et~al.}(2019)\citenamefont
  {Kawabata}, \citenamefont {Shiozaki}, \citenamefont {Ueda},\ and\
  \citenamefont {Sato}}]{Kawabata19}%
  \BibitemOpen
  \bibfield  {author} {\bibinfo {author} {\bibfnamefont {K.}~\bibnamefont
  {Kawabata}}, \bibinfo {author} {\bibfnamefont {K.}~\bibnamefont {Shiozaki}},
  \bibinfo {author} {\bibfnamefont {M.}~\bibnamefont {Ueda}}, \ and\ \bibinfo
  {author} {\bibfnamefont {M.}~\bibnamefont {Sato}},\ }\bibfield  {title}
  {\enquote {\bibinfo {title} {Symmetry and topology in non-hermitian
  physics},}\ }\href {\doibase 10.1103/PhysRevX.9.041015} {\bibfield  {journal}
  {\bibinfo  {journal} {Phys. Rev. X}\ }\textbf {\bibinfo {volume} {9}},\
  \bibinfo {pages} {041015} (\bibinfo {year} {2019})}\BibitemShut {NoStop}%
\bibitem [{\citenamefont {Yoshida}\ \emph {et~al.}(2020)\citenamefont
  {Yoshida}, \citenamefont {Mizoguchi},\ and\ \citenamefont
  {Hatsugai}}]{Yoshida20}%
  \BibitemOpen
  \bibfield  {author} {\bibinfo {author} {\bibfnamefont {T.}~\bibnamefont
  {Yoshida}}, \bibinfo {author} {\bibfnamefont {T.}~\bibnamefont {Mizoguchi}},
  \ and\ \bibinfo {author} {\bibfnamefont {Y.}~\bibnamefont {Hatsugai}},\
  }\bibfield  {title} {\enquote {\bibinfo {title} {Mirror skin effect and its
  electric circuit simulation},}\ }\href {\doibase
  10.1103/PhysRevResearch.2.022062} {\bibfield  {journal} {\bibinfo  {journal}
  {Phys. Rev. Research}\ }\textbf {\bibinfo {volume} {2}},\ \bibinfo {pages}
  {022062} (\bibinfo {year} {2020})}\BibitemShut {NoStop}%
\bibitem [{\citenamefont {Yoshida}\ \emph {et~al.}(2019)\citenamefont
  {Yoshida}, \citenamefont {Peters}, \citenamefont {Kawakami},\ and\
  \citenamefont {Hatsugai}}]{Yoshida19}%
  \BibitemOpen
  \bibfield  {author} {\bibinfo {author} {\bibfnamefont {T.}~\bibnamefont
  {Yoshida}}, \bibinfo {author} {\bibfnamefont {R.}~\bibnamefont {Peters}},
  \bibinfo {author} {\bibfnamefont {N.}~\bibnamefont {Kawakami}}, \ and\
  \bibinfo {author} {\bibfnamefont {Y.}~\bibnamefont {Hatsugai}},\ }\bibfield
  {title} {\enquote {\bibinfo {title} {Symmetry-protected exceptional rings in
  two-dimensional correlated systems with chiral symmetry},}\ }\href {\doibase
  10.1103/PhysRevB.99.121101} {\bibfield  {journal} {\bibinfo  {journal} {Phys.
  Rev. B}\ }\textbf {\bibinfo {volume} {99}},\ \bibinfo {pages} {121101}
  (\bibinfo {year} {2019})}\BibitemShut {NoStop}%
\bibitem [{\citenamefont {Yoshida}(2021)}]{Yoshida21}%
  \BibitemOpen
  \bibfield  {author} {\bibinfo {author} {\bibfnamefont {T.}~\bibnamefont
  {Yoshida}},\ }\bibfield  {title} {\enquote {\bibinfo {title} {Real-space
  dynamical mean field theory study of non-hermitian skin effect for correlated
  systems: Analysis based on pseudospectrum},}\ }\href {\doibase
  10.1103/PhysRevB.103.125145} {\bibfield  {journal} {\bibinfo  {journal}
  {Phys. Rev. B}\ }\textbf {\bibinfo {volume} {103}},\ \bibinfo {pages}
  {125145} (\bibinfo {year} {2021})}\BibitemShut {NoStop}%
\bibitem [{\citenamefont {Vajna}\ and\ \citenamefont
  {D\'ora}(2015{\natexlab{b}})}]{Vajna_15}%
  \BibitemOpen
  \bibfield  {author} {\bibinfo {author} {\bibfnamefont {S.}~\bibnamefont
  {Vajna}}\ and\ \bibinfo {author} {\bibfnamefont {B.}~\bibnamefont {D\'ora}},\
  }\bibfield  {title} {\enquote {\bibinfo {title} {Topological classification
  of dynamical phase transitions},}\ }\href {\doibase
  10.1103/PhysRevB.91.155127} {\bibfield  {journal} {\bibinfo  {journal} {Phys.
  Rev. B}\ }\textbf {\bibinfo {volume} {91}},\ \bibinfo {pages} {155127}
  (\bibinfo {year} {2015}{\natexlab{b}})}\BibitemShut {NoStop}%
\bibitem [{\citenamefont {Budich}\ and\ \citenamefont {Heyl}(2016)}]{Budich1}%
  \BibitemOpen
  \bibfield  {author} {\bibinfo {author} {\bibfnamefont {J.~C.}\ \bibnamefont
  {Budich}}\ and\ \bibinfo {author} {\bibfnamefont {M.}~\bibnamefont {Heyl}},\
  }\bibfield  {title} {\enquote {\bibinfo {title} {Dynamical topological order
  parameters far from equilibrium},}\ }\href {\doibase
  10.1103/PhysRevB.93.085416} {\bibfield  {journal} {\bibinfo  {journal} {Phys.
  Rev. B}\ }\textbf {\bibinfo {volume} {93}},\ \bibinfo {pages} {085416}
  (\bibinfo {year} {2016})}\BibitemShut {NoStop}%
\bibitem [{\citenamefont {Zhou}\ \emph {et~al.}(2018)\citenamefont {Zhou},
  \citenamefont {Wang}, \citenamefont {Wang},\ and\ \citenamefont
  {Gong}}]{Zhou1}%
  \BibitemOpen
  \bibfield  {author} {\bibinfo {author} {\bibfnamefont {L.}~\bibnamefont
  {Zhou}}, \bibinfo {author} {\bibfnamefont {Q.-h.}\ \bibnamefont {Wang}},
  \bibinfo {author} {\bibfnamefont {H.}~\bibnamefont {Wang}}, \ and\ \bibinfo
  {author} {\bibfnamefont {J.}~\bibnamefont {Gong}},\ }\bibfield  {title}
  {\enquote {\bibinfo {title} {Dynamical quantum phase transitions in
  non-hermitian lattices},}\ }\href {\doibase 10.1103/PhysRevA.98.022129}
  {\bibfield  {journal} {\bibinfo  {journal} {Phys. Rev. A}\ }\textbf {\bibinfo
  {volume} {98}},\ \bibinfo {pages} {022129} (\bibinfo {year}
  {2018})}\BibitemShut {NoStop}%
\bibitem [{\citenamefont {Zhou}\ and\ \citenamefont
  {Du}(2021{\natexlab{b}})}]{Zhou_2021}%
  \BibitemOpen
  \bibfield  {author} {\bibinfo {author} {\bibfnamefont {L.}~\bibnamefont
  {Zhou}}\ and\ \bibinfo {author} {\bibfnamefont {Q.}~\bibnamefont {Du}},\
  }\bibfield  {title} {\enquote {\bibinfo {title} {Non-hermitian topological
  phases and dynamical quantum phase transitions: a generic connection},}\
  }\href {\doibase 10.1088/1367-2630/ac0574} {\bibfield  {journal} {\bibinfo
  {journal} {New Journal of Physics}\ }\textbf {\bibinfo {volume} {23}},\
  \bibinfo {pages} {063041} (\bibinfo {year} {2021}{\natexlab{b}})}\BibitemShut
  {NoStop}%
\bibitem [{\citenamefont {Naji}\ \emph
  {et~al.}(2022{\natexlab{b}})\citenamefont {Naji}, \citenamefont {Jafari},
  \citenamefont {Jafari},\ and\ \citenamefont {Akbari}}]{Naji_PRA}%
  \BibitemOpen
  \bibfield  {author} {\bibinfo {author} {\bibfnamefont {J.}~\bibnamefont
  {Naji}}, \bibinfo {author} {\bibfnamefont {M.}~\bibnamefont {Jafari}},
  \bibinfo {author} {\bibfnamefont {R.}~\bibnamefont {Jafari}}, \ and\ \bibinfo
  {author} {\bibfnamefont {A.}~\bibnamefont {Akbari}},\ }\bibfield  {title}
  {\enquote {\bibinfo {title} {Dissipative floquet dynamical quantum phase
  transition},}\ }\href {\doibase 10.1103/PhysRevA.105.022220} {\bibfield
  {journal} {\bibinfo  {journal} {Phys. Rev. A}\ }\textbf {\bibinfo {volume}
  {105}},\ \bibinfo {pages} {022220} (\bibinfo {year}
  {2022}{\natexlab{b}})}\BibitemShut {NoStop}%
\bibitem [{\citenamefont {Hamazaki}(2021)}]{Hamazaki_2021}%
  \BibitemOpen
  \bibfield  {author} {\bibinfo {author} {\bibfnamefont {R.}~\bibnamefont
  {Hamazaki}},\ }\bibfield  {title} {\enquote {\bibinfo {title} {Exceptional
  dynamical quantum phase transitions in periodically driven systems},}\ }\href
  {\doibase 10.1038/s41467-021-25355-3} {\bibfield  {journal} {\bibinfo
  {journal} {Nature Communications}\ }\textbf {\bibinfo {volume} {12}},\
  \bibinfo {pages} {5108} (\bibinfo {year} {2021})}\BibitemShut {NoStop}%
\bibitem [{\citenamefont {Mondal}\ and\ \citenamefont {Nag}(2022)}]{Mondal22}%
  \BibitemOpen
  \bibfield  {author} {\bibinfo {author} {\bibfnamefont {D.}~\bibnamefont
  {Mondal}}\ and\ \bibinfo {author} {\bibfnamefont {T.}~\bibnamefont {Nag}},\
  }\bibfield  {title} {\enquote {\bibinfo {title} {Anomaly in the dynamical
  quantum phase transition in a non-hermitian system with extended gapless
  phases},}\ }\href {\doibase 10.1103/PhysRevB.106.054308} {\bibfield
  {journal} {\bibinfo  {journal} {Phys. Rev. B}\ }\textbf {\bibinfo {volume}
  {106}},\ \bibinfo {pages} {054308} (\bibinfo {year} {2022})}\BibitemShut
  {NoStop}%
\bibitem [{\citenamefont {Jurcevic}\ \emph {et~al.}(2017)\citenamefont
  {Jurcevic}, \citenamefont {Shen}, \citenamefont {Hauke}, \citenamefont
  {Maier}, \citenamefont {Brydges}, \citenamefont {Hempel}, \citenamefont
  {Lanyon}, \citenamefont {Heyl}, \citenamefont {Blatt},\ and\ \citenamefont
  {Roos}}]{PhysRevLett.119.080501}%
  \BibitemOpen
  \bibfield  {author} {\bibinfo {author} {\bibfnamefont {P.}~\bibnamefont
  {Jurcevic}}, \bibinfo {author} {\bibfnamefont {H.}~\bibnamefont {Shen}},
  \bibinfo {author} {\bibfnamefont {P.}~\bibnamefont {Hauke}}, \bibinfo
  {author} {\bibfnamefont {C.}~\bibnamefont {Maier}}, \bibinfo {author}
  {\bibfnamefont {T.}~\bibnamefont {Brydges}}, \bibinfo {author} {\bibfnamefont
  {C.}~\bibnamefont {Hempel}}, \bibinfo {author} {\bibfnamefont {B.~P.}\
  \bibnamefont {Lanyon}}, \bibinfo {author} {\bibfnamefont {M.}~\bibnamefont
  {Heyl}}, \bibinfo {author} {\bibfnamefont {R.}~\bibnamefont {Blatt}}, \ and\
  \bibinfo {author} {\bibfnamefont {C.~F.}\ \bibnamefont {Roos}},\ }\bibfield
  {title} {\enquote {\bibinfo {title} {Direct observation of dynamical quantum
  phase transitions in an interacting many-body system},}\ }\href {\doibase
  10.1103/PhysRevLett.119.080501} {\bibfield  {journal} {\bibinfo  {journal}
  {Phys. Rev. Lett.}\ }\textbf {\bibinfo {volume} {119}},\ \bibinfo {pages}
  {080501} (\bibinfo {year} {2017})}\BibitemShut {NoStop}%
\bibitem [{\citenamefont {Nie}\ \emph {et~al.}(2020)\citenamefont {Nie},
  \citenamefont {Wei}, \citenamefont {Chen}, \citenamefont {Zhang},
  \citenamefont {Zhao}, \citenamefont {Qiu}, \citenamefont {Tian},
  \citenamefont {Ji}, \citenamefont {Xin}, \citenamefont {Lu},\ and\
  \citenamefont {Li}}]{Nie20}%
  \BibitemOpen
  \bibfield  {author} {\bibinfo {author} {\bibfnamefont {X.}~\bibnamefont
  {Nie}}, \bibinfo {author} {\bibfnamefont {B.-B.}\ \bibnamefont {Wei}},
  \bibinfo {author} {\bibfnamefont {X.}~\bibnamefont {Chen}}, \bibinfo {author}
  {\bibfnamefont {Z.}~\bibnamefont {Zhang}}, \bibinfo {author} {\bibfnamefont
  {X.}~\bibnamefont {Zhao}}, \bibinfo {author} {\bibfnamefont {C.}~\bibnamefont
  {Qiu}}, \bibinfo {author} {\bibfnamefont {Y.}~\bibnamefont {Tian}}, \bibinfo
  {author} {\bibfnamefont {Y.}~\bibnamefont {Ji}}, \bibinfo {author}
  {\bibfnamefont {T.}~\bibnamefont {Xin}}, \bibinfo {author} {\bibfnamefont
  {D.}~\bibnamefont {Lu}}, \ and\ \bibinfo {author} {\bibfnamefont
  {J.}~\bibnamefont {Li}},\ }\bibfield  {title} {\enquote {\bibinfo {title}
  {Experimental observation of equilibrium and dynamical quantum phase
  transitions via out-of-time-ordered correlators},}\ }\href {\doibase
  10.1103/PhysRevLett.124.250601} {\bibfield  {journal} {\bibinfo  {journal}
  {Phys. Rev. Lett.}\ }\textbf {\bibinfo {volume} {124}},\ \bibinfo {pages}
  {250601} (\bibinfo {year} {2020})}\BibitemShut {NoStop}%
\bibitem [{\citenamefont {Fl{\"a}schner}\ \emph {et~al.}(2018)\citenamefont
  {Fl{\"a}schner}, \citenamefont {Vogel}, \citenamefont {Tarnowski},
  \citenamefont {Rem}, \citenamefont {L{\"u}hmann}, \citenamefont {Heyl},
  \citenamefont {Budich}, \citenamefont {Mathey}, \citenamefont {Sengstock},\
  and\ \citenamefont {Weitenberg}}]{flaschner2018observation}%
  \BibitemOpen
  \bibfield  {author} {\bibinfo {author} {\bibfnamefont {N.}~\bibnamefont
  {Fl{\"a}schner}}, \bibinfo {author} {\bibfnamefont {D.}~\bibnamefont
  {Vogel}}, \bibinfo {author} {\bibfnamefont {M.}~\bibnamefont {Tarnowski}},
  \bibinfo {author} {\bibfnamefont {B.}~\bibnamefont {Rem}}, \bibinfo {author}
  {\bibfnamefont {D.-S.}\ \bibnamefont {L{\"u}hmann}}, \bibinfo {author}
  {\bibfnamefont {M.}~\bibnamefont {Heyl}}, \bibinfo {author} {\bibfnamefont
  {J.}~\bibnamefont {Budich}}, \bibinfo {author} {\bibfnamefont
  {L.}~\bibnamefont {Mathey}}, \bibinfo {author} {\bibfnamefont
  {K.}~\bibnamefont {Sengstock}}, \ and\ \bibinfo {author} {\bibfnamefont
  {C.}~\bibnamefont {Weitenberg}},\ }\bibfield  {title} {\enquote {\bibinfo
  {title} {Observation of dynamical vortices after quenches in a system with
  topology},}\ }\href@noop {} {\bibfield  {journal} {\bibinfo  {journal}
  {Nature Physics}\ }\textbf {\bibinfo {volume} {14}},\ \bibinfo {pages}
  {265--268} (\bibinfo {year} {2018})}\BibitemShut {NoStop}%
\bibitem [{\citenamefont {Gou}\ \emph {et~al.}(2020)\citenamefont {Gou},
  \citenamefont {Chen}, \citenamefont {Xie}, \citenamefont {Xiao},
  \citenamefont {Deng}, \citenamefont {Gadway}, \citenamefont {Yi},\ and\
  \citenamefont {Yan}}]{Gou20}%
  \BibitemOpen
  \bibfield  {author} {\bibinfo {author} {\bibfnamefont {W.}~\bibnamefont
  {Gou}}, \bibinfo {author} {\bibfnamefont {T.}~\bibnamefont {Chen}}, \bibinfo
  {author} {\bibfnamefont {D.}~\bibnamefont {Xie}}, \bibinfo {author}
  {\bibfnamefont {T.}~\bibnamefont {Xiao}}, \bibinfo {author} {\bibfnamefont
  {T.-S.}\ \bibnamefont {Deng}}, \bibinfo {author} {\bibfnamefont
  {B.}~\bibnamefont {Gadway}}, \bibinfo {author} {\bibfnamefont
  {W.}~\bibnamefont {Yi}}, \ and\ \bibinfo {author} {\bibfnamefont
  {B.}~\bibnamefont {Yan}},\ }\bibfield  {title} {\enquote {\bibinfo {title}
  {Tunable nonreciprocal quantum transport through a dissipative aharonov-bohm
  ring in ultracold atoms},}\ }\href {\doibase 10.1103/PhysRevLett.124.070402}
  {\bibfield  {journal} {\bibinfo  {journal} {Phys. Rev. Lett.}\ }\textbf
  {\bibinfo {volume} {124}},\ \bibinfo {pages} {070402} (\bibinfo {year}
  {2020})}\BibitemShut {NoStop}%
\bibitem [{\citenamefont {Li}\ \emph {et~al.}(2019)\citenamefont {Li},
  \citenamefont {Harter}, \citenamefont {Liu}, \citenamefont {de~Melo},
  \citenamefont {Joglekar},\ and\ \citenamefont {Luo}}]{li2019observation}%
  \BibitemOpen
  \bibfield  {author} {\bibinfo {author} {\bibfnamefont {J.}~\bibnamefont
  {Li}}, \bibinfo {author} {\bibfnamefont {A.~K.}\ \bibnamefont {Harter}},
  \bibinfo {author} {\bibfnamefont {J.}~\bibnamefont {Liu}}, \bibinfo {author}
  {\bibfnamefont {L.}~\bibnamefont {de~Melo}}, \bibinfo {author} {\bibfnamefont
  {Y.~N.}\ \bibnamefont {Joglekar}}, \ and\ \bibinfo {author} {\bibfnamefont
  {L.}~\bibnamefont {Luo}},\ }\bibfield  {title} {\enquote {\bibinfo {title}
  {Observation of parity-time symmetry breaking transitions in a dissipative
  floquet system of ultracold atoms},}\ }\href@noop {} {\bibfield  {journal}
  {\bibinfo  {journal} {Nature communications}\ }\textbf {\bibinfo {volume}
  {10}},\ \bibinfo {pages} {855} (\bibinfo {year} {2019})}\BibitemShut
  {NoStop}%
\bibitem [{\citenamefont {Zeuner}\ \emph {et~al.}(2015)\citenamefont {Zeuner},
  \citenamefont {Rechtsman}, \citenamefont {Plotnik}, \citenamefont {Lumer},
  \citenamefont {Nolte}, \citenamefont {Rudner}, \citenamefont {Segev},\ and\
  \citenamefont {Szameit}}]{Zeuner15}%
  \BibitemOpen
  \bibfield  {author} {\bibinfo {author} {\bibfnamefont {J.~M.}\ \bibnamefont
  {Zeuner}}, \bibinfo {author} {\bibfnamefont {M.~C.}\ \bibnamefont
  {Rechtsman}}, \bibinfo {author} {\bibfnamefont {Y.}~\bibnamefont {Plotnik}},
  \bibinfo {author} {\bibfnamefont {Y.}~\bibnamefont {Lumer}}, \bibinfo
  {author} {\bibfnamefont {S.}~\bibnamefont {Nolte}}, \bibinfo {author}
  {\bibfnamefont {M.~S.}\ \bibnamefont {Rudner}}, \bibinfo {author}
  {\bibfnamefont {M.}~\bibnamefont {Segev}}, \ and\ \bibinfo {author}
  {\bibfnamefont {A.}~\bibnamefont {Szameit}},\ }\bibfield  {title} {\enquote
  {\bibinfo {title} {Observation of a topological transition in the bulk of a
  non-hermitian system},}\ }\href {\doibase 10.1103/PhysRevLett.115.040402}
  {\bibfield  {journal} {\bibinfo  {journal} {Phys. Rev. Lett.}\ }\textbf
  {\bibinfo {volume} {115}},\ \bibinfo {pages} {040402} (\bibinfo {year}
  {2015})}\BibitemShut {NoStop}%
\bibitem [{\citenamefont {Weimann}\ \emph {et~al.}(2017)\citenamefont
  {Weimann}, \citenamefont {Kremer}, \citenamefont {Plotnik}, \citenamefont
  {Lumer}, \citenamefont {Nolte}, \citenamefont {Makris}, \citenamefont
  {Segev}, \citenamefont {Rechtsman},\ and\ \citenamefont
  {Szameit}}]{weimann2017topologically}%
  \BibitemOpen
  \bibfield  {author} {\bibinfo {author} {\bibfnamefont {S.}~\bibnamefont
  {Weimann}}, \bibinfo {author} {\bibfnamefont {M.}~\bibnamefont {Kremer}},
  \bibinfo {author} {\bibfnamefont {Y.}~\bibnamefont {Plotnik}}, \bibinfo
  {author} {\bibfnamefont {Y.}~\bibnamefont {Lumer}}, \bibinfo {author}
  {\bibfnamefont {S.}~\bibnamefont {Nolte}}, \bibinfo {author} {\bibfnamefont
  {K.~G.}\ \bibnamefont {Makris}}, \bibinfo {author} {\bibfnamefont
  {M.}~\bibnamefont {Segev}}, \bibinfo {author} {\bibfnamefont {M.~C.}\
  \bibnamefont {Rechtsman}}, \ and\ \bibinfo {author} {\bibfnamefont
  {A.}~\bibnamefont {Szameit}},\ }\bibfield  {title} {\enquote {\bibinfo
  {title} {Topologically protected bound states in photonic
  parity--time-symmetric crystals},}\ }\href@noop {} {\bibfield  {journal}
  {\bibinfo  {journal} {Nature materials}\ }\textbf {\bibinfo {volume} {16}},\
  \bibinfo {pages} {433--438} (\bibinfo {year} {2017})}\BibitemShut {NoStop}%
\bibitem [{\citenamefont {Zhu}\ \emph {et~al.}(2018)\citenamefont {Zhu},
  \citenamefont {Fang}, \citenamefont {Li}, \citenamefont {Sun}, \citenamefont
  {Li}, \citenamefont {Jing},\ and\ \citenamefont {Chen}}]{Weiwei18}%
  \BibitemOpen
  \bibfield  {author} {\bibinfo {author} {\bibfnamefont {W.}~\bibnamefont
  {Zhu}}, \bibinfo {author} {\bibfnamefont {X.}~\bibnamefont {Fang}}, \bibinfo
  {author} {\bibfnamefont {D.}~\bibnamefont {Li}}, \bibinfo {author}
  {\bibfnamefont {Y.}~\bibnamefont {Sun}}, \bibinfo {author} {\bibfnamefont
  {Y.}~\bibnamefont {Li}}, \bibinfo {author} {\bibfnamefont {Y.}~\bibnamefont
  {Jing}}, \ and\ \bibinfo {author} {\bibfnamefont {H.}~\bibnamefont {Chen}},\
  }\bibfield  {title} {\enquote {\bibinfo {title} {Simultaneous observation of
  a topological edge state and exceptional point in an open and non-hermitian
  acoustic system},}\ }\href {\doibase 10.1103/PhysRevLett.121.124501}
  {\bibfield  {journal} {\bibinfo  {journal} {Phys. Rev. Lett.}\ }\textbf
  {\bibinfo {volume} {121}},\ \bibinfo {pages} {124501} (\bibinfo {year}
  {2018})}\BibitemShut {NoStop}%
\bibitem [{\citenamefont {Gao}\ \emph {et~al.}(2020)\citenamefont {Gao},
  \citenamefont {Xue}, \citenamefont {Wang}, \citenamefont {Gu}, \citenamefont
  {Liu}, \citenamefont {Zhu},\ and\ \citenamefont {Zhang}}]{Gao20}%
  \BibitemOpen
  \bibfield  {author} {\bibinfo {author} {\bibfnamefont {H.}~\bibnamefont
  {Gao}}, \bibinfo {author} {\bibfnamefont {H.}~\bibnamefont {Xue}}, \bibinfo
  {author} {\bibfnamefont {Q.}~\bibnamefont {Wang}}, \bibinfo {author}
  {\bibfnamefont {Z.}~\bibnamefont {Gu}}, \bibinfo {author} {\bibfnamefont
  {T.}~\bibnamefont {Liu}}, \bibinfo {author} {\bibfnamefont {J.}~\bibnamefont
  {Zhu}}, \ and\ \bibinfo {author} {\bibfnamefont {B.}~\bibnamefont {Zhang}},\
  }\bibfield  {title} {\enquote {\bibinfo {title} {Observation of topological
  edge states induced solely by non-hermiticity in an acoustic crystal},}\
  }\href {\doibase 10.1103/PhysRevB.101.180303} {\bibfield  {journal} {\bibinfo
   {journal} {Phys. Rev. B}\ }\textbf {\bibinfo {volume} {101}},\ \bibinfo
  {pages} {180303} (\bibinfo {year} {2020})}\BibitemShut {NoStop}%
\bibitem [{\citenamefont {Zanardi}\ \emph {et~al.}(2007)\citenamefont
  {Zanardi}, \citenamefont {Quan}, \citenamefont {Wang},\ and\ \citenamefont
  {Sun}}]{Zanardi07}%
  \BibitemOpen
  \bibfield  {author} {\bibinfo {author} {\bibfnamefont {P.}~\bibnamefont
  {Zanardi}}, \bibinfo {author} {\bibfnamefont {H.~T.}\ \bibnamefont {Quan}},
  \bibinfo {author} {\bibfnamefont {X.}~\bibnamefont {Wang}}, \ and\ \bibinfo
  {author} {\bibfnamefont {C.~P.}\ \bibnamefont {Sun}},\ }\bibfield  {title}
  {\enquote {\bibinfo {title} {Mixed-state fidelity and quantum criticality at
  finite temperature},}\ }\href {\doibase 10.1103/PhysRevA.75.032109}
  {\bibfield  {journal} {\bibinfo  {journal} {Phys. Rev. A}\ }\textbf {\bibinfo
  {volume} {75}},\ \bibinfo {pages} {032109} (\bibinfo {year}
  {2007})}\BibitemShut {NoStop}%
\bibitem [{\citenamefont {Liang}\ \emph {et~al.}(2019)\citenamefont {Liang},
  \citenamefont {Yeh}, \citenamefont {Mendon{\c{c}}a}, \citenamefont {Teh},
  \citenamefont {Reid},\ and\ \citenamefont {Drummond}}]{liang2019quantum}%
  \BibitemOpen
  \bibfield  {author} {\bibinfo {author} {\bibfnamefont {Y.-C.}\ \bibnamefont
  {Liang}}, \bibinfo {author} {\bibfnamefont {Y.-H.}\ \bibnamefont {Yeh}},
  \bibinfo {author} {\bibfnamefont {P.~E.}\ \bibnamefont {Mendon{\c{c}}a}},
  \bibinfo {author} {\bibfnamefont {R.~Y.}\ \bibnamefont {Teh}}, \bibinfo
  {author} {\bibfnamefont {M.~D.}\ \bibnamefont {Reid}}, \ and\ \bibinfo
  {author} {\bibfnamefont {P.~D.}\ \bibnamefont {Drummond}},\ }\bibfield
  {title} {\enquote {\bibinfo {title} {Quantum fidelity measures for mixed
  states},}\ }\href@noop {} {\bibfield  {journal} {\bibinfo  {journal} {Reports
  on Progress in Physics}\ }\textbf {\bibinfo {volume} {82}},\ \bibinfo {pages}
  {076001} (\bibinfo {year} {2019})}\BibitemShut {NoStop}%
\bibitem [{\citenamefont {Gu}(2010)}]{gu2010fidelity}%
  \BibitemOpen
  \bibfield  {author} {\bibinfo {author} {\bibfnamefont {S.-J.}\ \bibnamefont
  {Gu}},\ }\bibfield  {title} {\enquote {\bibinfo {title} {Fidelity approach to
  quantum phase transitions},}\ }\href@noop {} {\bibfield  {journal} {\bibinfo
  {journal} {International Journal of Modern Physics B}\ }\textbf {\bibinfo
  {volume} {24}},\ \bibinfo {pages} {4371--4458} (\bibinfo {year}
  {2010})}\BibitemShut {NoStop}%
\bibitem [{\citenamefont {Quan}\ and\ \citenamefont
  {Cucchietti}(2009)}]{Quan09}%
  \BibitemOpen
  \bibfield  {author} {\bibinfo {author} {\bibfnamefont {H.~T.}\ \bibnamefont
  {Quan}}\ and\ \bibinfo {author} {\bibfnamefont {F.~M.}\ \bibnamefont
  {Cucchietti}},\ }\bibfield  {title} {\enquote {\bibinfo {title} {Quantum
  fidelity and thermal phase transitions},}\ }\href {\doibase
  10.1103/PhysRevE.79.031101} {\bibfield  {journal} {\bibinfo  {journal} {Phys.
  Rev. E}\ }\textbf {\bibinfo {volume} {79}},\ \bibinfo {pages} {031101}
  (\bibinfo {year} {2009})}\BibitemShut {NoStop}%
\bibitem [{\citenamefont {Lang}\ \emph
  {et~al.}(2018{\natexlab{b}})\citenamefont {Lang}, \citenamefont {Frank},\
  and\ \citenamefont {Halimeh}}]{Lang18}%
  \BibitemOpen
  \bibfield  {author} {\bibinfo {author} {\bibfnamefont {J.}~\bibnamefont
  {Lang}}, \bibinfo {author} {\bibfnamefont {B.}~\bibnamefont {Frank}}, \ and\
  \bibinfo {author} {\bibfnamefont {J.~C.}\ \bibnamefont {Halimeh}},\
  }\bibfield  {title} {\enquote {\bibinfo {title} {Concurrence of dynamical
  phase transitions at finite temperature in the fully connected
  transverse-field ising model},}\ }\href {\doibase 10.1103/PhysRevB.97.174401}
  {\bibfield  {journal} {\bibinfo  {journal} {Phys. Rev. B}\ }\textbf {\bibinfo
  {volume} {97}},\ \bibinfo {pages} {174401} (\bibinfo {year}
  {2018}{\natexlab{b}})}\BibitemShut {NoStop}%
\bibitem [{\citenamefont {Lang}\ \emph
  {et~al.}(2018{\natexlab{c}})\citenamefont {Lang}, \citenamefont {Frank},\
  and\ \citenamefont {Halimeh}}]{Lang18b}%
  \BibitemOpen
  \bibfield  {author} {\bibinfo {author} {\bibfnamefont {J.}~\bibnamefont
  {Lang}}, \bibinfo {author} {\bibfnamefont {B.}~\bibnamefont {Frank}}, \ and\
  \bibinfo {author} {\bibfnamefont {J.~C.}\ \bibnamefont {Halimeh}},\
  }\bibfield  {title} {\enquote {\bibinfo {title} {Dynamical quantum phase
  transitions: A geometric picture},}\ }\href {\doibase
  10.1103/PhysRevLett.121.130603} {\bibfield  {journal} {\bibinfo  {journal}
  {Phys. Rev. Lett.}\ }\textbf {\bibinfo {volume} {121}},\ \bibinfo {pages}
  {130603} (\bibinfo {year} {2018}{\natexlab{c}})}\BibitemShut {NoStop}%
\bibitem [{\citenamefont {Mera}\ \emph {et~al.}(2018)\citenamefont {Mera},
  \citenamefont {Vlachou}, \citenamefont {Paunkovi\ifmmode~\acute{c}\else
  \'{c}\fi{}}, \citenamefont {Vieira},\ and\ \citenamefont {Viyuela}}]{Mera18}%
  \BibitemOpen
  \bibfield  {author} {\bibinfo {author} {\bibfnamefont {B.}~\bibnamefont
  {Mera}}, \bibinfo {author} {\bibfnamefont {C.}~\bibnamefont {Vlachou}},
  \bibinfo {author} {\bibfnamefont {N.}~\bibnamefont
  {Paunkovi\ifmmode~\acute{c}\else \'{c}\fi{}}}, \bibinfo {author}
  {\bibfnamefont {V.~R.}\ \bibnamefont {Vieira}}, \ and\ \bibinfo {author}
  {\bibfnamefont {O.}~\bibnamefont {Viyuela}},\ }\bibfield  {title} {\enquote
  {\bibinfo {title} {Dynamical phase transitions at finite temperature from
  fidelity and interferometric loschmidt echo induced metrics},}\ }\href
  {\doibase 10.1103/PhysRevB.97.094110} {\bibfield  {journal} {\bibinfo
  {journal} {Phys. Rev. B}\ }\textbf {\bibinfo {volume} {97}},\ \bibinfo
  {pages} {094110} (\bibinfo {year} {2018})}\BibitemShut {NoStop}%
\bibitem [{\citenamefont {Bandyopadhyay}\ \emph {et~al.}(2018)\citenamefont
  {Bandyopadhyay}, \citenamefont {Laha}, \citenamefont {Bhattacharya},\ and\
  \citenamefont {Dutta}}]{Bandyopadhyay18}%
  \BibitemOpen
  \bibfield  {author} {\bibinfo {author} {\bibfnamefont {S.}~\bibnamefont
  {Bandyopadhyay}}, \bibinfo {author} {\bibfnamefont {S.}~\bibnamefont {Laha}},
  \bibinfo {author} {\bibfnamefont {U.}~\bibnamefont {Bhattacharya}}, \ and\
  \bibinfo {author} {\bibfnamefont {A.}~\bibnamefont {Dutta}},\ }\bibfield
  {title} {\enquote {\bibinfo {title} {Exploring the possibilities of dynamical
  quantum phase transitions in the presence of a markovian bath},}\ }\href@noop
  {} {\bibfield  {journal} {\bibinfo  {journal} {Scientific Reports}\ }\textbf
  {\bibinfo {volume} {8}},\ \bibinfo {pages} {1--15} (\bibinfo {year}
  {2018})}\BibitemShut {NoStop}%
\bibitem [{\citenamefont {Hou}\ \emph {et~al.}(2020)\citenamefont {Hou},
  \citenamefont {Gao}, \citenamefont {Guo}, \citenamefont {He}, \citenamefont
  {Liu},\ and\ \citenamefont {Chien}}]{Hou20}%
  \BibitemOpen
  \bibfield  {author} {\bibinfo {author} {\bibfnamefont {X.-Y.}\ \bibnamefont
  {Hou}}, \bibinfo {author} {\bibfnamefont {Q.-C.}\ \bibnamefont {Gao}},
  \bibinfo {author} {\bibfnamefont {H.}~\bibnamefont {Guo}}, \bibinfo {author}
  {\bibfnamefont {Y.}~\bibnamefont {He}}, \bibinfo {author} {\bibfnamefont
  {T.}~\bibnamefont {Liu}}, \ and\ \bibinfo {author} {\bibfnamefont {C.-C.}\
  \bibnamefont {Chien}},\ }\bibfield  {title} {\enquote {\bibinfo {title}
  {Ubiquity of zeros of the loschmidt amplitude for mixed states in different
  physical processes and its implication},}\ }\href {\doibase
  10.1103/PhysRevB.102.104305} {\bibfield  {journal} {\bibinfo  {journal}
  {Phys. Rev. B}\ }\textbf {\bibinfo {volume} {102}},\ \bibinfo {pages}
  {104305} (\bibinfo {year} {2020})}\BibitemShut {NoStop}%
\bibitem [{\citenamefont {Hou}\ \emph {et~al.}(2022)\citenamefont {Hou},
  \citenamefont {Gao}, \citenamefont {Guo},\ and\ \citenamefont
  {Chien}}]{Hou22}%
  \BibitemOpen
  \bibfield  {author} {\bibinfo {author} {\bibfnamefont {X.-Y.}\ \bibnamefont
  {Hou}}, \bibinfo {author} {\bibfnamefont {Q.-C.}\ \bibnamefont {Gao}},
  \bibinfo {author} {\bibfnamefont {H.}~\bibnamefont {Guo}}, \ and\ \bibinfo
  {author} {\bibfnamefont {C.-C.}\ \bibnamefont {Chien}},\ }\bibfield  {title}
  {\enquote {\bibinfo {title} {Metamorphic dynamical quantum phase transition
  in double-quench processes at finite temperatures},}\ }\href {\doibase
  10.1103/PhysRevB.106.014301} {\bibfield  {journal} {\bibinfo  {journal}
  {Phys. Rev. B}\ }\textbf {\bibinfo {volume} {106}},\ \bibinfo {pages}
  {014301} (\bibinfo {year} {2022})}\BibitemShut {NoStop}%
\bibitem [{\citenamefont {Lang}\ \emph
  {et~al.}(2018{\natexlab{d}})\citenamefont {Lang}, \citenamefont {Chen},
  \citenamefont {Hong},\ and\ \citenamefont {Fan}}]{Lang18c}%
  \BibitemOpen
  \bibfield  {author} {\bibinfo {author} {\bibfnamefont {H.}~\bibnamefont
  {Lang}}, \bibinfo {author} {\bibfnamefont {Y.}~\bibnamefont {Chen}}, \bibinfo
  {author} {\bibfnamefont {Q.}~\bibnamefont {Hong}}, \ and\ \bibinfo {author}
  {\bibfnamefont {H.}~\bibnamefont {Fan}},\ }\bibfield  {title} {\enquote
  {\bibinfo {title} {Dynamical quantum phase transition for mixed states in
  open systems},}\ }\href {\doibase 10.1103/PhysRevB.98.134310} {\bibfield
  {journal} {\bibinfo  {journal} {Phys. Rev. B}\ }\textbf {\bibinfo {volume}
  {98}},\ \bibinfo {pages} {134310} (\bibinfo {year}
  {2018}{\natexlab{d}})}\BibitemShut {NoStop}%
\bibitem [{\citenamefont {Zhou}\ and\ \citenamefont
  {Du}(2021{\natexlab{c}})}]{zhou2021non}%
  \BibitemOpen
  \bibfield  {author} {\bibinfo {author} {\bibfnamefont {L.}~\bibnamefont
  {Zhou}}\ and\ \bibinfo {author} {\bibfnamefont {Q.}~\bibnamefont {Du}},\
  }\bibfield  {title} {\enquote {\bibinfo {title} {Non-hermitian topological
  phases and dynamical quantum phase transitions: a generic connection},}\
  }\href@noop {} {\bibfield  {journal} {\bibinfo  {journal} {New Journal of
  Physics}\ }\textbf {\bibinfo {volume} {23}},\ \bibinfo {pages} {063041}
  (\bibinfo {year} {2021}{\natexlab{c}})}\BibitemShut {NoStop}%
\bibitem [{\citenamefont {Sedlmayr}\ \emph {et~al.}(2018)\citenamefont
  {Sedlmayr}, \citenamefont {Fleischhauer},\ and\ \citenamefont
  {Sirker}}]{Sedlmayr18}%
  \BibitemOpen
  \bibfield  {author} {\bibinfo {author} {\bibfnamefont {N.}~\bibnamefont
  {Sedlmayr}}, \bibinfo {author} {\bibfnamefont {M.}~\bibnamefont
  {Fleischhauer}}, \ and\ \bibinfo {author} {\bibfnamefont {J.}~\bibnamefont
  {Sirker}},\ }\bibfield  {title} {\enquote {\bibinfo {title} {Fate of
  dynamical phase transitions at finite temperatures and in open systems},}\
  }\href {\doibase 10.1103/PhysRevB.97.045147} {\bibfield  {journal} {\bibinfo
  {journal} {Phys. Rev. B}\ }\textbf {\bibinfo {volume} {97}},\ \bibinfo
  {pages} {045147} (\bibinfo {year} {2018})}\BibitemShut {NoStop}%
\bibitem [{\citenamefont {Sj\"oqvist}\ \emph {et~al.}(2000)\citenamefont
  {Sj\"oqvist}, \citenamefont {Pati}, \citenamefont {Ekert}, \citenamefont
  {Anandan}, \citenamefont {Ericsson}, \citenamefont {Oi},\ and\ \citenamefont
  {Vedral}}]{Erik00}%
  \BibitemOpen
  \bibfield  {author} {\bibinfo {author} {\bibfnamefont {E.}~\bibnamefont
  {Sj\"oqvist}}, \bibinfo {author} {\bibfnamefont {A.~K.}\ \bibnamefont
  {Pati}}, \bibinfo {author} {\bibfnamefont {A.}~\bibnamefont {Ekert}},
  \bibinfo {author} {\bibfnamefont {J.~S.}\ \bibnamefont {Anandan}}, \bibinfo
  {author} {\bibfnamefont {M.}~\bibnamefont {Ericsson}}, \bibinfo {author}
  {\bibfnamefont {D.~K.~L.}\ \bibnamefont {Oi}}, \ and\ \bibinfo {author}
  {\bibfnamefont {V.}~\bibnamefont {Vedral}},\ }\bibfield  {title} {\enquote
  {\bibinfo {title} {Geometric phases for mixed states in interferometry},}\
  }\href {\doibase 10.1103/PhysRevLett.85.2845} {\bibfield  {journal} {\bibinfo
   {journal} {Phys. Rev. Lett.}\ }\textbf {\bibinfo {volume} {85}},\ \bibinfo
  {pages} {2845--2849} (\bibinfo {year} {2000})}\BibitemShut {NoStop}%
\bibitem [{\citenamefont {Kitaev}(2001)}]{kitaev2001unpaired}%
  \BibitemOpen
  \bibfield  {author} {\bibinfo {author} {\bibfnamefont {A.~Y.}\ \bibnamefont
  {Kitaev}},\ }\bibfield  {title} {\enquote {\bibinfo {title} {Unpaired
  majorana fermions in quantum wires},}\ }\href@noop {} {\bibfield  {journal}
  {\bibinfo  {journal} {Physics-uspekhi}\ }\textbf {\bibinfo {volume} {44}},\
  \bibinfo {pages} {131} (\bibinfo {year} {2001})}\BibitemShut {NoStop}%
\bibitem [{\citenamefont {DeGottardi}\ \emph
  {et~al.}(2013{\natexlab{a}})\citenamefont {DeGottardi}, \citenamefont {Sen},\
  and\ \citenamefont {Vishveshwara}}]{DeGo1}%
  \BibitemOpen
  \bibfield  {author} {\bibinfo {author} {\bibfnamefont {W.}~\bibnamefont
  {DeGottardi}}, \bibinfo {author} {\bibfnamefont {D.}~\bibnamefont {Sen}}, \
  and\ \bibinfo {author} {\bibfnamefont {S.}~\bibnamefont {Vishveshwara}},\
  }\bibfield  {title} {\enquote {\bibinfo {title} {Majorana fermions in
  superconducting 1d systems having periodic, quasiperiodic, and disordered
  potentials},}\ }\href {\doibase 10.1103/PhysRevLett.110.146404} {\bibfield
  {journal} {\bibinfo  {journal} {Phys. Rev. Lett.}\ }\textbf {\bibinfo
  {volume} {110}},\ \bibinfo {pages} {146404} (\bibinfo {year}
  {2013}{\natexlab{a}})}\BibitemShut {NoStop}%
\bibitem [{\citenamefont {DeGottardi}\ \emph
  {et~al.}(2013{\natexlab{b}})\citenamefont {DeGottardi}, \citenamefont
  {Thakurathi}, \citenamefont {Vishveshwara},\ and\ \citenamefont
  {Sen}}]{Manisha1}%
  \BibitemOpen
  \bibfield  {author} {\bibinfo {author} {\bibfnamefont {W.}~\bibnamefont
  {DeGottardi}}, \bibinfo {author} {\bibfnamefont {M.}~\bibnamefont
  {Thakurathi}}, \bibinfo {author} {\bibfnamefont {S.}~\bibnamefont
  {Vishveshwara}}, \ and\ \bibinfo {author} {\bibfnamefont {D.}~\bibnamefont
  {Sen}},\ }\bibfield  {title} {\enquote {\bibinfo {title} {Majorana fermions
  in superconducting wires: Effects of long-range hopping, broken time-reversal
  symmetry, and potential landscapes},}\ }\href {\doibase
  10.1103/PhysRevB.88.165111} {\bibfield  {journal} {\bibinfo  {journal} {Phys.
  Rev. B}\ }\textbf {\bibinfo {volume} {88}},\ \bibinfo {pages} {165111}
  (\bibinfo {year} {2013}{\natexlab{b}})}\BibitemShut {NoStop}%
\bibitem [{\citenamefont {Rajak}\ \emph {et~al.}(2014)\citenamefont {Rajak},
  \citenamefont {Nag},\ and\ \citenamefont {Dutta}}]{Rajak1}%
  \BibitemOpen
  \bibfield  {author} {\bibinfo {author} {\bibfnamefont {A.}~\bibnamefont
  {Rajak}}, \bibinfo {author} {\bibfnamefont {T.}~\bibnamefont {Nag}}, \ and\
  \bibinfo {author} {\bibfnamefont {A.}~\bibnamefont {Dutta}},\ }\bibfield
  {title} {\enquote {\bibinfo {title} {Possibility of adiabatic transport of a
  majorana edge state through an extended gapless region},}\ }\href {\doibase
  10.1103/PhysRevE.90.042107} {\bibfield  {journal} {\bibinfo  {journal} {Phys.
  Rev. E}\ }\textbf {\bibinfo {volume} {90}},\ \bibinfo {pages} {042107}
  (\bibinfo {year} {2014})}\BibitemShut {NoStop}%
\bibitem [{\citenamefont {Shi}\ and\ \citenamefont
  {Song}(2022)}]{shi2022topological}%
  \BibitemOpen
  \bibfield  {author} {\bibinfo {author} {\bibfnamefont {Y.}~\bibnamefont
  {Shi}}\ and\ \bibinfo {author} {\bibfnamefont {Z.}~\bibnamefont {Song}},\
  }\bibfield  {title} {\enquote {\bibinfo {title} {Topological phase in kitaev
  chain with spatially separated pairing processes},}\ }\href@noop {}
  {\bibfield  {journal} {\bibinfo  {journal} {arXiv preprint arXiv:2211.07920}\
  } (\bibinfo {year} {2022})}\BibitemShut {NoStop}%
\bibitem [{\citenamefont {Li}\ \emph {et~al.}(2020)\citenamefont {Li},
  \citenamefont {Lee}, \citenamefont {Mu},\ and\ \citenamefont
  {Gong}}]{li2020critical}%
  \BibitemOpen
  \bibfield  {author} {\bibinfo {author} {\bibfnamefont {L.}~\bibnamefont
  {Li}}, \bibinfo {author} {\bibfnamefont {C.~H.}\ \bibnamefont {Lee}},
  \bibinfo {author} {\bibfnamefont {S.}~\bibnamefont {Mu}}, \ and\ \bibinfo
  {author} {\bibfnamefont {J.}~\bibnamefont {Gong}},\ }\bibfield  {title}
  {\enquote {\bibinfo {title} {Critical non-hermitian skin effect},}\
  }\href@noop {} {\bibfield  {journal} {\bibinfo  {journal} {Nature
  communications}\ }\textbf {\bibinfo {volume} {11}},\ \bibinfo {pages} {5491}
  (\bibinfo {year} {2020})}\BibitemShut {NoStop}%
\bibitem [{\citenamefont {Yuce}(2016)}]{Yuce16}%
  \BibitemOpen
  \bibfield  {author} {\bibinfo {author} {\bibfnamefont {C.}~\bibnamefont
  {Yuce}},\ }\bibfield  {title} {\enquote {\bibinfo {title} {Majorana edge
  modes with gain and loss},}\ }\href {\doibase 10.1103/PhysRevA.93.062130}
  {\bibfield  {journal} {\bibinfo  {journal} {Phys. Rev. A}\ }\textbf {\bibinfo
  {volume} {93}},\ \bibinfo {pages} {062130} (\bibinfo {year}
  {2016})}\BibitemShut {NoStop}%
\bibitem [{\citenamefont {Yao}\ and\ \citenamefont {Wang}(2018)}]{YaoPRL2018}%
  \BibitemOpen
  \bibfield  {author} {\bibinfo {author} {\bibfnamefont {S.}~\bibnamefont
  {Yao}}\ and\ \bibinfo {author} {\bibfnamefont {Z.}~\bibnamefont {Wang}},\
  }\bibfield  {title} {\enquote {\bibinfo {title} {Edge states and topological
  invariants of non-hermitian systems},}\ }\href {\doibase
  10.1103/PhysRevLett.121.086803} {\bibfield  {journal} {\bibinfo  {journal}
  {Phys. Rev. Lett.}\ }\textbf {\bibinfo {volume} {121}},\ \bibinfo {pages}
  {086803} (\bibinfo {year} {2018})}\BibitemShut {NoStop}%
\bibitem [{\citenamefont {Kunst}\ \emph {et~al.}(2018)\citenamefont {Kunst},
  \citenamefont {Edvardsson}, \citenamefont {Budich},\ and\ \citenamefont
  {Bergholtz}}]{Kunst18}%
  \BibitemOpen
  \bibfield  {author} {\bibinfo {author} {\bibfnamefont {F.~K.}\ \bibnamefont
  {Kunst}}, \bibinfo {author} {\bibfnamefont {E.}~\bibnamefont {Edvardsson}},
  \bibinfo {author} {\bibfnamefont {J.~C.}\ \bibnamefont {Budich}}, \ and\
  \bibinfo {author} {\bibfnamefont {E.~J.}\ \bibnamefont {Bergholtz}},\
  }\bibfield  {title} {\enquote {\bibinfo {title} {Biorthogonal bulk-boundary
  correspondence in non-hermitian systems},}\ }\href {\doibase
  10.1103/PhysRevLett.121.026808} {\bibfield  {journal} {\bibinfo  {journal}
  {Phys. Rev. Lett.}\ }\textbf {\bibinfo {volume} {121}},\ \bibinfo {pages}
  {026808} (\bibinfo {year} {2018})}\BibitemShut {NoStop}%
\bibitem [{\citenamefont {Helbig}\ \emph {et~al.}(2020)\citenamefont {Helbig},
  \citenamefont {Hofmann}, \citenamefont {Imhof}, \citenamefont {Abdelghany},
  \citenamefont {Kiessling}, \citenamefont {Molenkamp}, \citenamefont {Lee},
  \citenamefont {Szameit}, \citenamefont {Greiter},\ and\ \citenamefont
  {Thomale}}]{helbig2020generalized}%
  \BibitemOpen
  \bibfield  {author} {\bibinfo {author} {\bibfnamefont {T.}~\bibnamefont
  {Helbig}}, \bibinfo {author} {\bibfnamefont {T.}~\bibnamefont {Hofmann}},
  \bibinfo {author} {\bibfnamefont {S.}~\bibnamefont {Imhof}}, \bibinfo
  {author} {\bibfnamefont {M.}~\bibnamefont {Abdelghany}}, \bibinfo {author}
  {\bibfnamefont {T.}~\bibnamefont {Kiessling}}, \bibinfo {author}
  {\bibfnamefont {L.~W.}\ \bibnamefont {Molenkamp}}, \bibinfo {author}
  {\bibfnamefont {C.~H.}\ \bibnamefont {Lee}}, \bibinfo {author} {\bibfnamefont
  {A.}~\bibnamefont {Szameit}}, \bibinfo {author} {\bibfnamefont
  {M.}~\bibnamefont {Greiter}}, \ and\ \bibinfo {author} {\bibfnamefont
  {R.}~\bibnamefont {Thomale}},\ }\bibfield  {title} {\enquote {\bibinfo
  {title} {Generalized bulk--boundary correspondence in non-hermitian
  topolectrical circuits},}\ }\href {\doibase 10.1038/s41567-020-0922-9}
  {\bibfield  {journal} {\bibinfo  {journal} {Nature Physics}\ }\textbf
  {\bibinfo {volume} {16}},\ \bibinfo {pages} {747--750} (\bibinfo {year}
  {2020})}\BibitemShut {NoStop}%
\bibitem [{\citenamefont {Kawabata}\ \emph {et~al.}(2020)\citenamefont
  {Kawabata}, \citenamefont {Sato},\ and\ \citenamefont
  {Shiozaki}}]{Kawabata20b}%
  \BibitemOpen
  \bibfield  {author} {\bibinfo {author} {\bibfnamefont {K.}~\bibnamefont
  {Kawabata}}, \bibinfo {author} {\bibfnamefont {M.}~\bibnamefont {Sato}}, \
  and\ \bibinfo {author} {\bibfnamefont {K.}~\bibnamefont {Shiozaki}},\
  }\bibfield  {title} {\enquote {\bibinfo {title} {Higher-order non-hermitian
  skin effect},}\ }\href {\doibase 10.1103/PhysRevB.102.205118} {\bibfield
  {journal} {\bibinfo  {journal} {Phys. Rev. B}\ }\textbf {\bibinfo {volume}
  {102}},\ \bibinfo {pages} {205118} (\bibinfo {year} {2020})}\BibitemShut
  {NoStop}%
\bibitem [{\citenamefont {Ghosh}\ and\ \citenamefont {Nag}(2022)}]{Ghosh22b}%
  \BibitemOpen
  \bibfield  {author} {\bibinfo {author} {\bibfnamefont {A.~K.}\ \bibnamefont
  {Ghosh}}\ and\ \bibinfo {author} {\bibfnamefont {T.}~\bibnamefont {Nag}},\
  }\bibfield  {title} {\enquote {\bibinfo {title} {Non-hermitian higher-order
  topological superconductors in two dimensions: Statics and dynamics},}\
  }\href {\doibase 10.1103/PhysRevB.106.L140303} {\bibfield  {journal}
  {\bibinfo  {journal} {Phys. Rev. B}\ }\textbf {\bibinfo {volume} {106}},\
  \bibinfo {pages} {L140303} (\bibinfo {year} {2022})}\BibitemShut {NoStop}%
\bibitem [{\citenamefont {Rajak}\ and\ \citenamefont {Dutta}(2014)}]{Rajak14}%
  \BibitemOpen
  \bibfield  {author} {\bibinfo {author} {\bibfnamefont {A.}~\bibnamefont
  {Rajak}}\ and\ \bibinfo {author} {\bibfnamefont {A.}~\bibnamefont {Dutta}},\
  }\bibfield  {title} {\enquote {\bibinfo {title} {Survival probability of an
  edge majorana in a one-dimensional $p$-wave superconducting chain under
  sudden quenching of parameters},}\ }\href {\doibase
  10.1103/PhysRevE.89.042125} {\bibfield  {journal} {\bibinfo  {journal} {Phys.
  Rev. E}\ }\textbf {\bibinfo {volume} {89}},\ \bibinfo {pages} {042125}
  (\bibinfo {year} {2014})}\BibitemShut {NoStop}%
\bibitem [{\citenamefont {Rajak}\ and\ \citenamefont {Nag}(2017)}]{Rajak17}%
  \BibitemOpen
  \bibfield  {author} {\bibinfo {author} {\bibfnamefont {A.}~\bibnamefont
  {Rajak}}\ and\ \bibinfo {author} {\bibfnamefont {T.}~\bibnamefont {Nag}},\
  }\bibfield  {title} {\enquote {\bibinfo {title} {Survival probability in a
  quenched majorana chain with an impurity},}\ }\href {\doibase
  10.1103/PhysRevE.96.022136} {\bibfield  {journal} {\bibinfo  {journal} {Phys.
  Rev. E}\ }\textbf {\bibinfo {volume} {96}},\ \bibinfo {pages} {022136}
  (\bibinfo {year} {2017})}\BibitemShut {NoStop}%
\bibitem [{\citenamefont {Cappellaro}\ \emph {et~al.}(2011)\citenamefont
  {Cappellaro}, \citenamefont {Viola},\ and\ \citenamefont
  {Ramanathan}}]{Cappellaro11}%
  \BibitemOpen
  \bibfield  {author} {\bibinfo {author} {\bibfnamefont {P.}~\bibnamefont
  {Cappellaro}}, \bibinfo {author} {\bibfnamefont {L.}~\bibnamefont {Viola}}, \
  and\ \bibinfo {author} {\bibfnamefont {C.}~\bibnamefont {Ramanathan}},\
  }\bibfield  {title} {\enquote {\bibinfo {title} {Coherent-state transfer via
  highly mixed quantum spin chains},}\ }\href {\doibase
  10.1103/PhysRevA.83.032304} {\bibfield  {journal} {\bibinfo  {journal} {Phys.
  Rev. A}\ }\textbf {\bibinfo {volume} {83}},\ \bibinfo {pages} {032304}
  (\bibinfo {year} {2011})}\BibitemShut {NoStop}%
\bibitem [{\citenamefont {Sur}\ and\ \citenamefont
  {Subrahmanyam}(2019)}]{sur2019loschmidt}%
  \BibitemOpen
  \bibfield  {author} {\bibinfo {author} {\bibfnamefont {S.}~\bibnamefont
  {Sur}}\ and\ \bibinfo {author} {\bibfnamefont {V.}~\bibnamefont
  {Subrahmanyam}},\ }\bibfield  {title} {\enquote {\bibinfo {title} {Loschmidt
  echo of local dynamical processes in integrable and non integrable spin
  chains},}\ }\href@noop {} {\bibfield  {journal} {\bibinfo  {journal} {Journal
  of Physics A: Mathematical and Theoretical}\ }\textbf {\bibinfo {volume}
  {52}},\ \bibinfo {pages} {345301} (\bibinfo {year} {2019})}\BibitemShut
  {NoStop}%
\bibitem [{\citenamefont {Das}\ \emph {et~al.}(2012)\citenamefont {Das},
  \citenamefont {Ronen}, \citenamefont {Most}, \citenamefont {Oreg},
  \citenamefont {Heiblum},\ and\ \citenamefont {Shtrikman}}]{das2012zero}%
  \BibitemOpen
  \bibfield  {author} {\bibinfo {author} {\bibfnamefont {A.}~\bibnamefont
  {Das}}, \bibinfo {author} {\bibfnamefont {Y.}~\bibnamefont {Ronen}}, \bibinfo
  {author} {\bibfnamefont {Y.}~\bibnamefont {Most}}, \bibinfo {author}
  {\bibfnamefont {Y.}~\bibnamefont {Oreg}}, \bibinfo {author} {\bibfnamefont
  {M.}~\bibnamefont {Heiblum}}, \ and\ \bibinfo {author} {\bibfnamefont
  {H.}~\bibnamefont {Shtrikman}},\ }\bibfield  {title} {\enquote {\bibinfo
  {title} {Zero-bias peaks and splitting in an al--inas nanowire topological
  superconductor as a signature of majorana fermions},}\ }\href@noop {}
  {\bibfield  {journal} {\bibinfo  {journal} {Nature Physics}\ }\textbf
  {\bibinfo {volume} {8}},\ \bibinfo {pages} {887--895} (\bibinfo {year}
  {2012})}\BibitemShut {NoStop}%
\bibitem [{\citenamefont {Shi}\ and\ \citenamefont {Song}(2023)}]{Shi23}%
  \BibitemOpen
  \bibfield  {author} {\bibinfo {author} {\bibfnamefont {Y.~B.}\ \bibnamefont
  {Shi}}\ and\ \bibinfo {author} {\bibfnamefont {Z.}~\bibnamefont {Song}},\
  }\bibfield  {title} {\enquote {\bibinfo {title} {Topological phase in a
  kitaev chain with spatially separated pairing processes},}\ }\href {\doibase
  10.1103/PhysRevB.107.125110} {\bibfield  {journal} {\bibinfo  {journal}
  {Phys. Rev. B}\ }\textbf {\bibinfo {volume} {107}},\ \bibinfo {pages}
  {125110} (\bibinfo {year} {2023})}\BibitemShut {NoStop}%
\bibitem [{\citenamefont {Ashida}\ \emph {et~al.}(2017)\citenamefont {Ashida},
  \citenamefont {Furukawa},\ and\ \citenamefont {Ueda}}]{ashida2017parity}%
  \BibitemOpen
  \bibfield  {author} {\bibinfo {author} {\bibfnamefont {Y.}~\bibnamefont
  {Ashida}}, \bibinfo {author} {\bibfnamefont {S.}~\bibnamefont {Furukawa}}, \
  and\ \bibinfo {author} {\bibfnamefont {M.}~\bibnamefont {Ueda}},\ }\bibfield
  {title} {\enquote {\bibinfo {title} {Parity-time-symmetric quantum critical
  phenomena},}\ }\href@noop {} {\bibfield  {journal} {\bibinfo  {journal}
  {Nature communications}\ }\textbf {\bibinfo {volume} {8}},\ \bibinfo {pages}
  {15791} (\bibinfo {year} {2017})}\BibitemShut {NoStop}%
\bibitem [{\citenamefont {Ashida}\ \emph {et~al.}(2016)\citenamefont {Ashida},
  \citenamefont {Furukawa},\ and\ \citenamefont {Ueda}}]{Ashida16}%
  \BibitemOpen
  \bibfield  {author} {\bibinfo {author} {\bibfnamefont {Y.}~\bibnamefont
  {Ashida}}, \bibinfo {author} {\bibfnamefont {S.}~\bibnamefont {Furukawa}}, \
  and\ \bibinfo {author} {\bibfnamefont {M.}~\bibnamefont {Ueda}},\ }\bibfield
  {title} {\enquote {\bibinfo {title} {Quantum critical behavior influenced by
  measurement backaction in ultracold gases},}\ }\href {\doibase
  10.1103/PhysRevA.94.053615} {\bibfield  {journal} {\bibinfo  {journal} {Phys.
  Rev. A}\ }\textbf {\bibinfo {volume} {94}},\ \bibinfo {pages} {053615}
  (\bibinfo {year} {2016})}\BibitemShut {NoStop}%
\bibitem [{\citenamefont {Du}\ \emph {et~al.}(2003)\citenamefont {Du},
  \citenamefont {Zou}, \citenamefont {Shi}, \citenamefont {Kwek}, \citenamefont
  {Pan}, \citenamefont {Oh}, \citenamefont {Ekert}, \citenamefont {Oi},\ and\
  \citenamefont {Ericsson}}]{Du03}%
  \BibitemOpen
  \bibfield  {author} {\bibinfo {author} {\bibfnamefont {J.}~\bibnamefont
  {Du}}, \bibinfo {author} {\bibfnamefont {P.}~\bibnamefont {Zou}}, \bibinfo
  {author} {\bibfnamefont {M.}~\bibnamefont {Shi}}, \bibinfo {author}
  {\bibfnamefont {L.~C.}\ \bibnamefont {Kwek}}, \bibinfo {author}
  {\bibfnamefont {J.-W.}\ \bibnamefont {Pan}}, \bibinfo {author} {\bibfnamefont
  {C.~H.}\ \bibnamefont {Oh}}, \bibinfo {author} {\bibfnamefont
  {A.}~\bibnamefont {Ekert}}, \bibinfo {author} {\bibfnamefont {D.~K.~L.}\
  \bibnamefont {Oi}}, \ and\ \bibinfo {author} {\bibfnamefont {M.}~\bibnamefont
  {Ericsson}},\ }\bibfield  {title} {\enquote {\bibinfo {title} {Observation of
  geometric phases for mixed states using nmr interferometry},}\ }\href
  {\doibase 10.1103/PhysRevLett.91.100403} {\bibfield  {journal} {\bibinfo
  {journal} {Phys. Rev. Lett.}\ }\textbf {\bibinfo {volume} {91}},\ \bibinfo
  {pages} {100403} (\bibinfo {year} {2003})}\BibitemShut {NoStop}%
\bibitem [{\citenamefont {Brennen}\ \emph {et~al.}(1999)\citenamefont
  {Brennen}, \citenamefont {Caves}, \citenamefont {Jessen},\ and\ \citenamefont
  {Deutsch}}]{Brennen99}%
  \BibitemOpen
  \bibfield  {author} {\bibinfo {author} {\bibfnamefont {G.~K.}\ \bibnamefont
  {Brennen}}, \bibinfo {author} {\bibfnamefont {C.~M.}\ \bibnamefont {Caves}},
  \bibinfo {author} {\bibfnamefont {P.~S.}\ \bibnamefont {Jessen}}, \ and\
  \bibinfo {author} {\bibfnamefont {I.~H.}\ \bibnamefont {Deutsch}},\
  }\bibfield  {title} {\enquote {\bibinfo {title} {Quantum logic gates in
  optical lattices},}\ }\href {\doibase 10.1103/PhysRevLett.82.1060} {\bibfield
   {journal} {\bibinfo  {journal} {Phys. Rev. Lett.}\ }\textbf {\bibinfo
  {volume} {82}},\ \bibinfo {pages} {1060--1063} (\bibinfo {year}
  {1999})}\BibitemShut {NoStop}%
\end{thebibliography}%


\end{document}